\documentclass[aps,prd,superscriptaddress,nofootinbib,tighten,preprint]{revtex4}
\pdfoutput=1
\usepackage[utf8]{inputenc}
\usepackage[english]{babel}
\usepackage{amsmath}
\usepackage{hyperref}
\usepackage{slashed}
\usepackage{diagbox}
\usepackage{subfigure}
\usepackage{comment}
\usepackage{pstricks}
\allowdisplaybreaks
\usepackage{color}
\usepackage{amsfonts}
\usepackage{float}
\usepackage{amssymb}
\usepackage{graphicx}
\newcommand{\be}{\begin{equation}}
\newcommand{\ee}{\end{equation}}
\newcommand{\bea}{\begin{eqnarray}}
\newcommand{\eea}{\end{eqnarray}}
\newcommand{\noi}{\noindent}
\newcommand{\bse}{\begin{subequations}}
\newcommand{\ese}{\end{subequations}}
\newcommand{\smgauge}{${\rm SU}(3)_{c} \otimes {\rm SU}(2)_{L} \otimes {\rm U}(1)_{Y}$}
\newcommand{\zto}{\mathbb{Z}_2}
\newcommand{\nn}{\nonumber}
\newcommand{\x}{\chi}
\newcommand{\tf}{\tilde{\phi}}

\newcommand{\sigmaVindirect}{{\langle {\sigma {\rm v_{rel}}}\rangle}_{\bar{\x} \x \rightarrow\tf\tf}}

\catcode`@=12

\begin{document}
\title{Fermionic dark matter via UV and IR freeze-in and
its possible X-ray signature}

\author{Anirban Biswas}
\email{tpab3@iacs.res.in}
\affiliation{School of Physical Sciences, Indian Association for the Cultivation of Science,
2A $\&$ 2B Raja S.C. Mullick Road, Kolkata 700032, India}
\author{Sougata Ganguly}
\email{tpsg4@iacs.res.in}
\affiliation{School of Physical Sciences, Indian Association for the Cultivation of Science,
2A $\&$ 2B Raja S.C. Mullick Road, Kolkata 700032, India}
\author{Sourov Roy}
\email{tpsr@iacs.res.in}
\affiliation{School of Physical Sciences, Indian Association
for the Cultivation of Science, 2A $\&$ 2B Raja S.C. Mullick
Road, Kolkata 700032, India}

\begin{abstract} 
Non-observation of any dark matter signature at various direct detection
experiments over the last decade keeps indicating that immensely popular
WIMP paradigm may not be the actual theory of
particle dark matter. Non-thermal dark matter
produced through freeze-in is an attractive proposal, naturally
explaining null results by virtue of its feeble couplings with the
Standard Model (SM) particles. We consider 
a minimal extension of the SM by two gauge
singlet fields namely, a $\mathbb{Z}_2$-odd
fermion $\chi$ and a pseudo scalar $\tilde{\phi}$,
where the former has interactions with the SM
particles only at dimension five level and beyond. This
introduces natural suppression in the interactions of $\chi$ by a heavy
new physics scale $\Lambda$ and forces $\chi$ to be a non-thermal
dark matter candidate.\,\,We have studied the production of $\chi$
in detail taking into account both ultra-violate (UV), 
infra-red (IR) as well as mixed UV-IR freeze-in and found that for
$10^{10}\,{\rm GeV}\,\leq\,\Lambda\leq 10^{15}$ GeV, $\chi$ is
dominantly produced via UV and mixed UV-IR freeze-in when reheat temperature
$T_{\rm RH}\gtrsim 10^4$ GeV and below which the production
is dominated by IR and mixed freeze-in. Furthermore, we have considered 
the cascade annihilation$\chi \bar{\chi} \rightarrow \tilde{\phi}\tilde{\phi}\rightarrow 4\gamma$
to address the longstanding $\sim 3.5$ keV
X-ray line observed from various galaxies and galaxy clusters.
We have found that the long-lived intermediate state $\tilde{\phi}$
modifies dark matter density around the galactic centre to an
effective density $\rho_{\rm eff}$ which strongly depends on
the decay length of $\tilde{\phi}$. Finally, the allowed parameter
space in $\Lambda-g$ plane ($g$ is the coupling between $\chi\bar{\chi}$
and $\tilde{\phi}$) is obtained by comparing
our result with the XMM Newton observed X-ray flux from
the centre of Milky Way galaxy in $2\sigma$ range.
\end{abstract}
\maketitle
\section{Introduction}
\label{Intro}
The indirect evidence including flat galaxy rotation curve, bullet cluster
observation, gravitational lensing of distant objects etc.\,\,have firmly established
the fact that in addition to the visible baryonic matter, our Universe is made
of some mysterious non-baryonic, non-luminous matter which is commonly known
as the dark matter (DM). Besides these indirect evidences,
cosmic microwave background (CMB) anisotropy probing experiments like WMAP \cite{Hinshaw:2012aka} and
Planck \cite{Aghanim:2018eyx} have measured the amount of dark matter present in the Universe based
on $\Lambda$CDM model and the present value of dark matter relic density is
$\Omega_{\rm DM}\,h^2=0.120 \pm 0.001$. In spite of all these, the particle
nature of DM and its production mechanism are not known to us to date.
There exists a plethora of proposals in the literature for a viable dark matter candidate.
Among them thermally generated {\it cold} dark matter is the most studied scenario. This type
of dark matter candidates are classified as weakly interacting
massive particle (WIMP) \cite{Roszkowski:2017nbc}. 
The WIMP scenario naturally predicts a ``cold'' dark matter candidate
having mass in the GeV to TeV range with a weak scale total annihilation cross section
$\langle{\sigma {\rm v}\rangle} \simeq 3\times 10^{-26}$ cm$^3$/s. In addition, this
attractive scenario also predicts substantial scattering cross section with
first generation of quarks (unless they are prohibited by some exotic
symmetry of a particular model) and hence with nucleons as well, which can
easily be measured by the present dark matter direct detection
experiments. However, as of now no signature of dark matter has been
observed at direct detection experiments resulting in severe
exclusions of both spin independent ($\sigma_{\rm SI}$) and
spin dependent ($\sigma_{\rm SD}$) scattering cross sections.
In particular, $\sigma_{\rm SI}$ has the maximum exclusion
of $4.1 \times 10^{-47}$ cm$^2$ for $m_{DM}=30$ GeV by the XENON1T
experiment \cite{Aprile:2018dbl}. In future,
experiments like XENONnT \cite{Aprile:2015uzo},
LUX-ZEPLIN (LZ) \cite{Akerib:2018lyp} and
DARWIN \cite{Aalbers:2016jon} will be sensitive enough to explore all
the remaining parameter space above the {\it neutrino floor}, a
region dominated by coherent elastic neutrino-nucleus scattering
(CE${\nu}$NS), beyond which it is an extremely difficult task to
identify a dark matter signal from the neutrino background \cite{Boehm:2018sux}.

The null results of direct detection experiments raised a fundamental
question about the scale of interaction of dark matter with baryons. As a
result, there are many interesting proposals which predict diminutive interactions
between visible and dark sectors and at the same time attain correct relic
density as reported by the CMB experiment Planck. Non-thermal origin
of dark matter is one of such frameworks where dark matter possesses
extremely feeble interaction with other particles in the thermal
bath. In this framework, it is assumed that the initial abundance
of dark matter is almost negligible compared to other particles
maintaining thermal equilibrium among themselves. This may be
visualised in a situation where after the inflation, inflaton field
predominantly decays into the visible sector particles rather than
its dark sector counterparts. Here, dark
matter particles are produced gradually from the decay as well as
scattering of bath particles. This is known as the Freeze-in
mechanism \cite{Hall:2009bx}. Moreover, there are two types of freeze-in
depending on the time of maximum production of dark matter.
One of them is the ultra-violet (UV) freeze-in \cite{Hall:2009bx,
Elahi:2014fsa, McDonald:2015ljz, Chen:2017kvz} where dark
and visible sectors are connected by the higher dimensional
operators only. As a result, abundance of dark matter
becomes extremely sensitive to the initial history such
as the reheat temperature ($T_{\rm RH}$) of the Universe. In this
situation production of dark matter occurs only through
scatterings. On the other hand, renormalisable
interaction between dark and visible sectors leads
to another kind of dark matter production which
is dominated at around the temperature $T\sim$
mass of the initial state particles, when latter are
in thermal equilibrium. Beyond this, as the temperature of
the Universe drops below the mass of the mother particle, its number density
becomes Boltzmann suppressed and the corresponding production mode
of dark matter ceases. Unlike the previous case, this kind of
freeze-in is mostly effective at the lowest possible temperature
for a particular production process (i.e. either decay or scattering or both)
hence this is known as the infrared (IR) freeze-in \cite{Hall:2009bx,
Yaguna:2011qn, Blennow:2013jba, Biswas:2015sva, Co:2015pka, Shakya:2015xnx,
Biswas:2016bfo, Konig:2016dzg, Biswas:2016yjr,
Bernal:2017kxu, Biswas:2017tce, Pandey:2017quk, Biswas:2018aib, Borah:2018gjk}. 
Moreover, freeze-in can also be possible when initial state particles themselves
remain out of thermal equilibrium 
and in such cases one needs to calculate first the distribution
function of mother particle which later enters into the
Boltzmann equation of dark matter \cite{Biswas:2016iyh, Konig:2016dzg}
and in such cases the dark matter production era through
freeze-in depends on the nature of distribution function
of mother particle.   
In this work, we have studied both the UV freeze-in and IR freeze-in
in a single framework and we also discuss a possible signature of our
dark matter candidate via $\sim 3.5$ keV X-ray line. For that, we have extended
the SM by adding a gauge singlet and $\mathbb{Z}_2$-odd Dirac fermion ($\chi$)
and a gauge singlet pseudo scalar ($\tf$). The Dirac fermion $\x$ is absolutely
stable due to the unbroken $\mathbb{Z}_2$ symmetry and hence it is our dark
matter candidate. Both the SM gauge invariance and also the invariance
of $\mathbb{Z}_2$ symmetry dictate that $\x$ does not have any direct interactions
with the SM fields except that with the Higgs doublet $\Phi$ and that
too is possible only using the gauge invariant operators having
dimensions five (minimum) or more. As a result, the production of
dark matter via UV freeze-in is possible before the electroweak symmetry
breaking (EWSB) where pairs of $\x$ and $\bar{\x}$
are produced from scatterings of the components of $\Phi$ and also
from the electroweak gauge bosons and gluons as well as top quark
(involving a $\tf$ mediator). Moreover, the same operator
between $\x$ and $\Phi$ is also responsible for the late time
production of dark matter via IR freeze-in, as after
EWSB $\Phi$ gets a nonzero vacuum expectation value (VEV)
and the dimension five operator decomposes into a
four dimensional interaction term between the SM Higgs
boson $h$ and $\x\bar{\x}$. In the IR freeze-in regime,
in addition to the annihilations of $h\,h$, $g\,g$, $W^+W^-$, $Z\,Z$,
and $\gamma\,\gamma$, pair annihilations of the SM fermions, pseudo
scalar ($\tf$) and the decays of the
SM Higgs boson $h$ and $\tf$ are the possible sources
of $\x\bar{\x}$ production. However, we will see
that $\tf$ with a mass larger than 163 eV cannot be
produced thermally at the early Universe via
Primakoff processes as it will overclose the
Universe.\,\,This can be evaded if the freeze-out
temperature of $\tf$ production processes
is larger than the reheat temperature of
the Universe. In spite of this, in the present model
$\tf$ can also be produced non-thermally from processes involving
top quark in the initial state as well as from the
decay of the SM Higgs boson ($h$) if $m_{\tf}\leq m_{h}/2$. 
Non-thermal production of $\tf$ from UV processes and its subsequent decay 
into $\x \bar{\x}$ via a dimesion four operator
is dubbed as mixed UV-IR freeze-in scenario of $\x$.
In addition, depending on
its mass and lifetime (varies mainly with $\Lambda$, $m_{\tf}$
and $\tf\x\bar{\x}$ coupling $g$) the pseudo scalar $\tf$
may also act as a decaying dark matter component.
After solving the Boltzmann equation of $\x$ considering
all possible production processes in the collision term
we have shown the allowed parameter space in $\Lambda-T_{\rm RH}$ plane, 
which satisfies dark matter relic density in $1\sigma$ range, where
$\Lambda$ is the possible new physics scale. We have found that in the
lower end of $\Lambda-T_{\rm RH}$ plane, contributions to the relic density 
come mainly from IR as well as mixed freeze-in whereas  
UV and mixed freeze-in contribute to the 
relic density  for higher values of $T_{\rm RH}$ and $\Lambda$.

Finally, we have discussed an indirect signature of
our dark matter candidate in detail. For that we have
considered $\sim 3.5$ keV X-ray line emission from different galaxies
and galaxy clusters. The observation of $\sim 3.5$ keV X-ray line by
the XMM Newton X-ray observatory from various galaxy clusters
including Perseus, Coma,
Centaurus etc. was first reported in \cite{Bulbul:2014sua}.
Afterwards, there are studies by various groups claiming the
presence of this {\it line} in the X-ray spectrum from
the Andromeda galaxy \cite{Boyarsky:2014jta} and also from the
centre of our Milky Way galaxy \cite{Boyarsky:2014ska, Jeltema:2014qfa}.
Furthermore, the Suzaku X-ray observatory has searched for the signature
of the $3.5$ keV line in the four X-ray brightest galaxy clusters
namely Perseus, Virgo, Coma and Ophiuchus and they have detected signal in the
Perseus galaxy cluster only \cite{Urban:2014yda}. More recently, the evidence
of this X-ray line has also been found in the cosmic X-ray background
by the Chandra X-ray Observatory \cite{Cappelluti:2017ywp}.
There are plenty of studies focusing on the explanation of
this mysterious X-ray signal from dark matter decay
in a wide class of beyond Standard Model (BSM) scenarios \cite{1402.7335,
1403.0865, 1403.1536, 1403.1782,Kolda:2014ppa, 1403.6503, 1403.6621, 1404.2220,
1404.3676, 1405.6967, 1412.4253, Arcadi:2014dca, Biswas:2015sva, 1503.06130, Biswas:2015bca,
1604.01929, 1612.08621, Biswas:2017ait,Bae:2017dpt}\footnote{See Ref.\,\cite{Dessert:2018qih}
for a recent study on decaying dark matter interpretation of 3.5 keV
X-ray line.}. On the other hand, although in less numbers,
there exist proposals involving dark matter annihilation as well
\cite{Dudas:2014ixa, Baek:2014poa, Brdar:2017wgy}. Nevertheless,
the astrophysical explanation of this X-ray line
in the form of atomic transitions in helium-like potassium
and chlorine \cite{Phillips:2015wla, Iakubovskyi:2015kwa}, 
is also possible.\,\,In this work, our explanation using annihilating
dark matter is distinctly different from the earlier attempts.
Here, our dark matter candidate $\x$ undergoes a cascade annihilation in which
first a pair of $\x\bar{\x}$ annihilates into $\tf\tf$
and thereafter each $\tf$ decays into two $\gamma$s.
This type of dark matter annihilation produces a ``box''
shaped photon spectrum \cite{Ibarra:2012dw} which gets a {\it line}
shape as $m_{\x}\rightarrow m_{\tf}$. In this scenario, we
have derived necessary analytical expressions of the photon
flux and have compared our result with the observed X-ray flux
by XMM Newton from the centre of Milky Way galaxy
\cite{Boyarsky:2014ska} and finally have presented
the allowed parameter space in the $\Lambda-g$ plane.  

Rest of the article is organised as follows. In Section \ref{mod}
we describe our model briefly. Detailed analysis of dark matter
production via both UV and IR freeze-in has been presented
in Section \ref{uvir}. Possibility of $\tf$ as a decaying
dark matter candidate is discussed in Section \ref{phitDM}.
The Section \ref{indirect_sig} deals with a comprehensive study
of an indirect signature of our proposed dark matter candidate
$\x$ in the form of long-standing $\sim 3.5$ keV X-ray line.
Finally, we summarise in Section \ref{conclu}.   
 \section{Model}
\label{mod}
In this section we will describe our model briefly. As
we have mentioned in the previous section, we
consider a minimal extension of the SM, where one
can have both types of freeze-in (UV and IR) effects in the
relic density of dark matter. For that we have extended the
fermionic sector as well as the scalar sector of the SM by adding
a Dirac fermion $\x$ and a pseudo scalar $\tf$.
Both $\x$ and $\tf$ are singlet under the SM gauge group \smgauge.
Additionally, we have imposed a $\zto$ symmetry in the
Lagrangian and we demand that only $\x$ is odd under $\zto$.
Due to this, the Dirac fermion $\x$ cannot have any interaction with
the SM fields up to the level of dimension four. The minimal operator
describing interaction of $\x$ with SM Higgs doublet $\Phi$ is 
a five dimensional operator suppressed by a mass scale $\Lambda$.
However, $\x$ has renormalisable interaction with the remaining
non-standard particle $\tf$. On the other hand, being a pseudo
scalar, the CP invariance restricts interactions of $\tf$ as well.
Although unlike $\x$, $\tf$ has interaction with the
SM Higgs doublet at dimension four level, beyond that one can
have interactions between $\tf$ and SM gauge bosons which
have very rich phenomenology. Moreover, since $\tf$ does not
have any VEV, there is no spontaneous CP-violation as well after
symmetry breaking. In this scenario, Dirac fermion $\x$ is our dark matter
candidate which is absolutely stable due the unbroken $\mathbb{Z}_2$
symmetry. On the other hand, in some particular cases where the mass
of $\tf$ is less than $m_{\x}/2$, 
$\tf$ can be partially stable, contributing some fraction of total dark
matter relic density at the present epoch and thus can act as a decaying
dark matter candidate. Here the lifetime of $\tf$ is entirely controlled
by the cut-off scale $\Lambda$. \,The charges of all the fields under
\smgauge$\otimes\mathbb{Z}_2$\footnote{We have used the relation
$Q_{\rm EM} = T_3 + \dfrac{Y}{2}$ to determine the electromagnetic
charge of each field.}
symmetry are listed in Table \ref{tab1}.
\begin{center}
\begin{table}[hbt!]
\begin{tabular}{||c|c||}
\hline
Field content & Charge under \smgauge$\otimes\mathbb{Z}_2$ symmetry \\
\hline
\hline
$\ell_{L} = \begin{pmatrix}
\nu_e \\e
\end{pmatrix}_{L},\,
\begin{pmatrix}
\nu_\mu \\ \mu
\end{pmatrix}_{L},\,
\begin{pmatrix}
\nu_\tau \\ \tau
\end{pmatrix}_{L}$ & $(1,\,2,\,-1,\,+)$ \\
\hline
$\ell_{R}=e_R,\,\mu_R,\,\tau_R$ & $(1,\,1,\,-2,\,+)$\\
\hline
$Q_L$=$\begin{pmatrix}
u \\d
\end{pmatrix}_L,\,
\begin{pmatrix}
c \\ s
\end{pmatrix}_L,\,
\begin{pmatrix}
t \\ b
\end{pmatrix}_L$ & $(3,\,2,\,\frac{1}{3},\,+)$\\
\hline
$U_R=u_R,\,c_R,\,t_R$ & $(1,\,1,\,\frac{4}{3},\,+)$\\
\hline
$D_R=d_R,\,s_R,\,b_R$ & $(1,\,1,\,-\frac{2}{3},\,+)$\\
\hline
$\Phi = \begin{pmatrix}
\phi^+ \\ \phi^0
\end{pmatrix}$ & $(1,\,2,\,1,\,+)$\\
\hline
$\chi$ & $(1,\,1,\,0,\,-)$\\
\hline
\vspace{0.2 pt}
$\tilde{\phi}$ & $(1,\,1,\,0,\,+)$\\
\hline
\end{tabular}\\
\textit{}
\caption{Field content of our model and their charges under \smgauge$\otimes\mathbb{Z}_2$.}
\label{tab1}
\end{table}
\end{center}
The gauge invariant and CP conserving Lagrangian of our model is given by\footnote{Note that
for simplicity we have considered all the interactions of $\tf$ with gauge bosons and the interaction 
of $\x$ with Higgs boson to have the same coupling constant $\frac{1}{\Lambda}$. In general, different 
higher dimensional terms can have different coefficients and two different mass scales can also be involved
corresponding to the interactions of $\x$ and $\tf$.}
\be
\begin{aligned}
\mathcal{L}= & \mathcal{L}_{SM}+\overline{\chi}(i\,\slashed{\partial}-m_\chi)\chi
+\frac{1}{2}(\partial^\mu \tf)(\partial_\mu \tf)-\frac{1}{2} m_{\tf} \tf^2
-\frac{\tf\,B_{\mu \nu}{\tilde{B}}^{{\mu \nu}}}{2\,\Lambda}-
\frac{\tf\,W^a_{\mu \nu}{\tilde{W}_a}^{{\mu \nu}}}{2\,\Lambda}
-\frac{\tf\,G^b_{\mu \nu}{\tilde{G}_b}^{{\mu \nu}}}{2\,\Lambda}\\
&-\frac{y^{\ell}_{\alpha\,\beta}}{\Lambda}(i\,\overline{{\ell}_{L}}_{\alpha}
\,\Phi\,\gamma_5\,{\ell_{R}}_{\beta}\,
\tf +h.c)-\frac{y^d_{\alpha\,\beta}}{\Lambda}(i\,\overline{{Q_{L}}_\alpha}
\,\Phi\,\gamma_5\,{D_{R}}_{\beta}\tf +h.c) - 
\frac{y^u_{\alpha \,\beta}}{\Lambda}(i\,\overline{{Q_{L}}_{\alpha}}
\,\tilde{\Phi}\,\gamma_5\,{U_{R}}_{\beta} \tf +h.c)\\
&-\frac{\overline{\x}\x\,\Phi^\dagger \Phi}{\Lambda}-g\,\overline{\x}\gamma_5\x \tf
+\frac{\lambda}{2}\tilde{\phi}^2\Phi^\dagger \Phi \,,
\end{aligned}
\label{eq1}
\ee
where $\alpha,\,\beta$ are the generation indices,
$\Lambda$ is some mass scale which represents the
cut-off scale of our effective theory. $B_{\mu\nu}=\partial_\mu B_\nu-\partial_\nu B_\mu$
,  $W^a_{\mu\nu}=\partial_\mu W^a_\nu-\partial_\nu W^a_\mu +i\,g_2
\,\epsilon^{abc}\,W^b_\mu\,W^c_\nu$ ($a=1,\,2,\,3$) 
and $G^b_{\mu\nu}=\partial_\mu G^b_\nu-\partial_\nu G^b_\mu +i\,g_3
\,f^{bcd}\,G^c_\mu\,G^d_\nu$  ($b=1\,...\,8$)  are field strength
tensors for U(1)$_{Y}$, SU(2)$_{L}$ and SU(3)$_c$ gauge groups
while the corresponding gauge bosons are denoted by $B_{\mu}$,
$W^{a}_{\mu}$ and $G^{b}_{\mu}$ respectively. Moreover, $g_2\,(g_3)$
and $\epsilon^{abc}\,(f^{abc})$ are gauge couplings and structure
constants of SU(2)$_{L}$ (SU(3)$_c$) respectively. Further, in the above
$\tilde{X}_{\mu\nu}$ ($X = B_\mu$, $W^a_\mu$, $G^a_\mu$)
is the Hodge dual of field strength tensor and 
is defined as $\tilde{X}_{\mu\nu} = \frac{1}{2}\epsilon_{\mu\nu\rho\sigma} X^{\rho\sigma}$ 
with $\epsilon_{\mu\nu\rho\sigma}$ representing a four
dimensional Levi-Civita symbol. Now, the combination between field strength tensor $X_{\mu\nu}$
and its Hodge dual tensor $\tilde{X}_{\mu\nu}$ gives a
pseudo scalar which is invariant under the Lorentz transformation. Hence,
a combination of this product with $\tf$ remains CP invariant.

In the above Lagrangian, we have added an extra $i\,(\equiv\sqrt{-1})$
in each interaction term between SM fermions and $\tf$ so that
the fermionic bilinears and their hermitian conjugate form
a pseudo scalar. Here, $\tilde{\Phi}$ is an SU(2)$_L$ doublet
with hypercharge -1 and is defined as $\tilde{\Phi}=i\,\sigma_2 \Phi^*$,
where $\sigma_2$ is the second Pauli spin matrix and
all the fermionic fields have been defined in Table\,\ref{tab1}.
Further, we would like to note here
that the Yukawa couplings $y^{\ell}$, $y^u$ and $y^d$
are same as the Yukawa couplings associated with charged
leptons, up-type quarks and down-type quarks in the SM respectively. This is required
to avoid the flavour changing neutral current between
the SM fermions and $\tf$. All these interaction terms between $\tf$ and the SM
fields except Higgs boson are suppressed by the
new physics scale $\Lambda$. Furthermore, as in the SM, here also electroweak symmetry
is broken to residual U(1)$_{\rm EM}$
symmetry by the vacuum expectation value of the
neutral component of doublet $\Phi$. 
\section{Dark matter production via UV and IR freeze-in}
\label{uvir}
In this work, our principal goal is to study both types
of freeze-in mechanism in a single framework by minimally
extending the SM. As we have already seen in the previous
section, all the interactions of our dark matter candidate $\x$
with the SM particles are suppressed by a heavy new physics scale $\Lambda$.
This naturally ensures that our dark matter candidate $\x$ has
extremely feeble interactions with thermal bath containing SM
particles. Consequently, $\x$ always stays out of thermal equilibrium
and behaves as a non-thermal relic. The genesis of non-thermal
dark matter in the early Universe is known as the freeze-in
mechanism \cite{Hall:2009bx} and depending upon the nature of interaction
of non-thermal dark matter candidate with other bath particles,
there are two types of freeze-in namely UV freeze-in and IR freeze-in.
In the present case, both types of freeze-in mechanisms are
important for $\x$ production at two different epochs.
The UV freeze-in is possible due to 
the presence of higher dimensional
interactions between $\x$ and SM fermions, gauge bosons
and in this process maximum $\x$ production occurred
when the temperature of the Universe was equal to $T_{\rm RH}$,
the reheat temperature. On the other hand, after
electroweak symmetry breaking (EWSB) additional $\x$
particles are produced from the scatterings and decays
of the SM particles via IR freeze-in mechanism.
This is indeed possible because after SU(2)$_{L}\otimes{\rm U}(1)_{Y}$
breaking, one can construct dimension three (responsible
for both scattering and decay) as well as dimension four
(responsible for scattering only) interactions involving
$\chi$ and other SM particles from those higher dimensional
operators. We have calculated both the UV and IR contributions
to the relic density of our dark matter candidate $\x$. The required
interaction terms which are responsible for the UV contribution are
given by,
\bea
\mathcal{L} &\supset& -\frac{\overline{\x}\x \Phi^\dagger \Phi}{\Lambda}
-\frac{\epsilon^{\mu \nu\alpha\beta}\left(\partial_\mu B_{\nu}\right)
\left(\partial_\alpha B_\beta\right)\tf}{\Lambda}
-\frac{\epsilon^{\mu \nu\alpha\beta}\left(\partial_\mu W^a_{\nu}\right)
\left(\partial_\alpha W^a_\beta\right)\tf}{\Lambda}
-\frac{\epsilon^{\mu \nu\alpha\beta}\left(\partial_\mu G^b_{\nu}\right)
\left(\partial_\alpha G^b_\beta\right)\tf}{\Lambda}\, \nn \\
&&
-\dfrac{i}{\Lambda}\sum_{i=1}^3 \Bigg \{ \sum_{\alpha = u, d}
{y^\alpha_{ii}}\,\overline{q_i}_L\,\gamma_5\,{q_i}_R\,{\phi^0}^{*}\,
+{y^\ell_{ii}}\,\overline{\ell_i}_L\,\gamma_5\,{\ell_i}_R\,{\phi^0}^{*}\,
+ h.c.\Bigg \}\tilde{\phi} \, -g\,\bar{\chi}\gamma_5\chi\,\tf \,.
\label{L-UV}
\eea
Note that the Yukawa couplings will provide
significant contribution only for the top quark.
Since UV freeze-in occurs well above the electroweak symmetry breaking\footnote{We consider
reheat temperature $T_{\rm RH} >> T_{\rm EW}$, where $T_{\rm EW}$ being
the temperature of the Universe when electroweak symmetry breaking
happened.}, at that time there is no mixing between hypercharge gauge boson $B_{\mu}$
and $W^3_{\mu}$. Therefore, during UV freeze-in $B_{\mu}$, $W_{\mu}^a$ ($a=1,\,2,\,3$)  
and $G_{\mu}^b$ ($b=1\,....\,8$) are physical gauge bosons and $\x$ is produced from
their annihilations mediated by pseudo scalar $\tf$. Moreover, $\x$
can also be produced from the annihilations of 
$T_3 = \frac{1}{2}$ and $-\frac{1}{2}$ components of the Higgs doublet $\Phi$ respectively,
where the former one is no longer a Goldstone boson before the EWSB.
Furthermore, one can also have the
following scattering processes
$t\,\bar{t}\rightarrow\Phi\,\tf$, $t\,\Phi\rightarrow t\,\tf$ and 
$\bar{t}\,\Phi\rightarrow \bar{t}\,\tf$, which are dominant
in the UV regime. $\tf$ production involving these scattering processes with subsequent 
$\tf \rightarrow \bar{\x}\x$ is termed as mixed UV-IR freeze-in 
scenario as mentioned earlier. 

\noi
The Feynman diagrams for all these processes which are contributing
significantly towards $\x$ production via UV freeze-in are
shown in Fig.\,\ref{fig1} and gauge boson-pseudo scalar
vertices are given in Appendix \ref{app:vertices}.
\begin{figure}[h!]
\centering
\begin{tabular}{|c|c c c|}
\hline
Regime & \multicolumn{3}{|c|}{DM Production channels} \\ 
\hline 
&&& \\
Before EWSB  & \includegraphics[height=2.5cm,width=4.5cm]{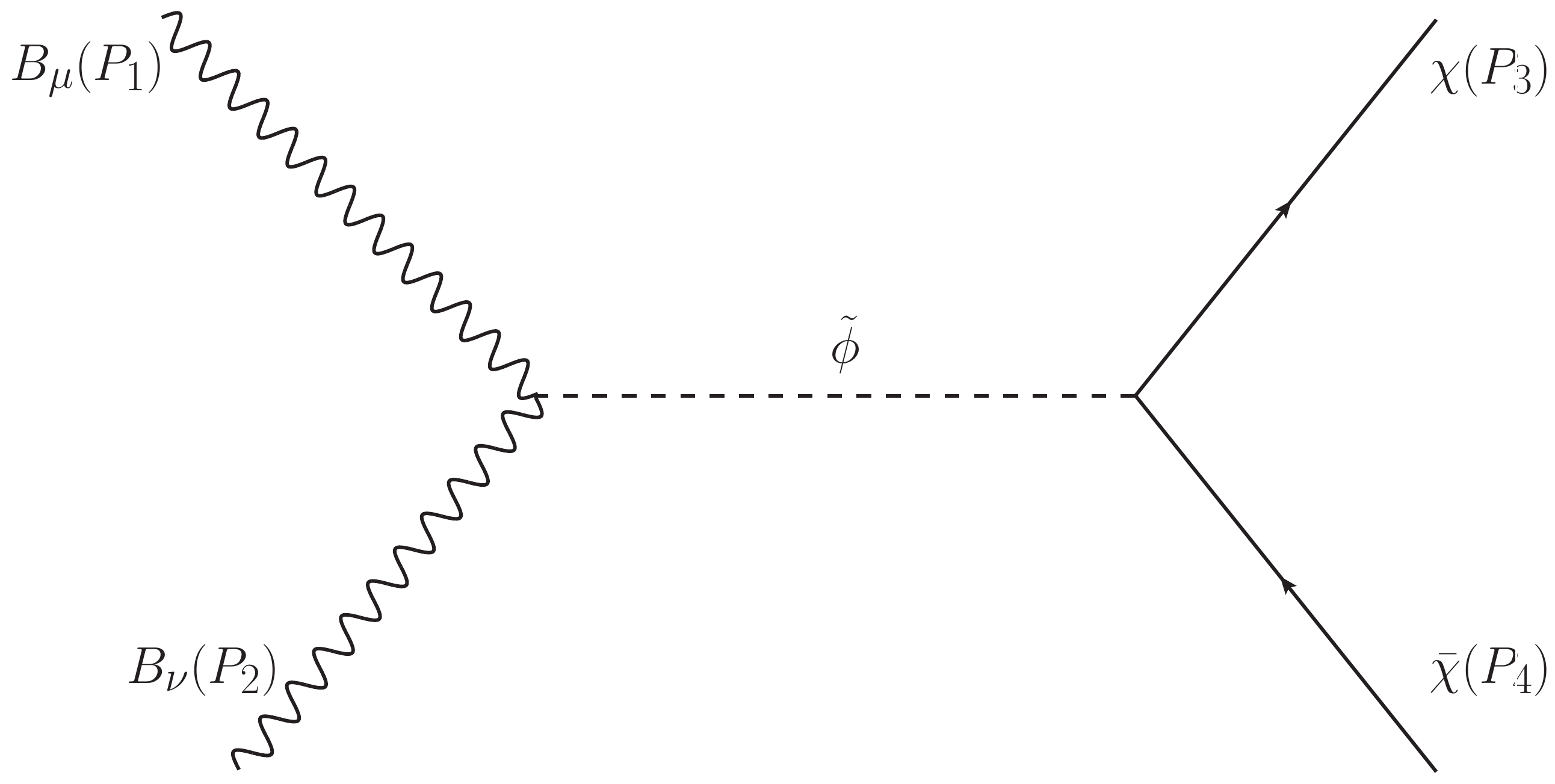} & 
\includegraphics[height=2.5cm,width=4.5cm]{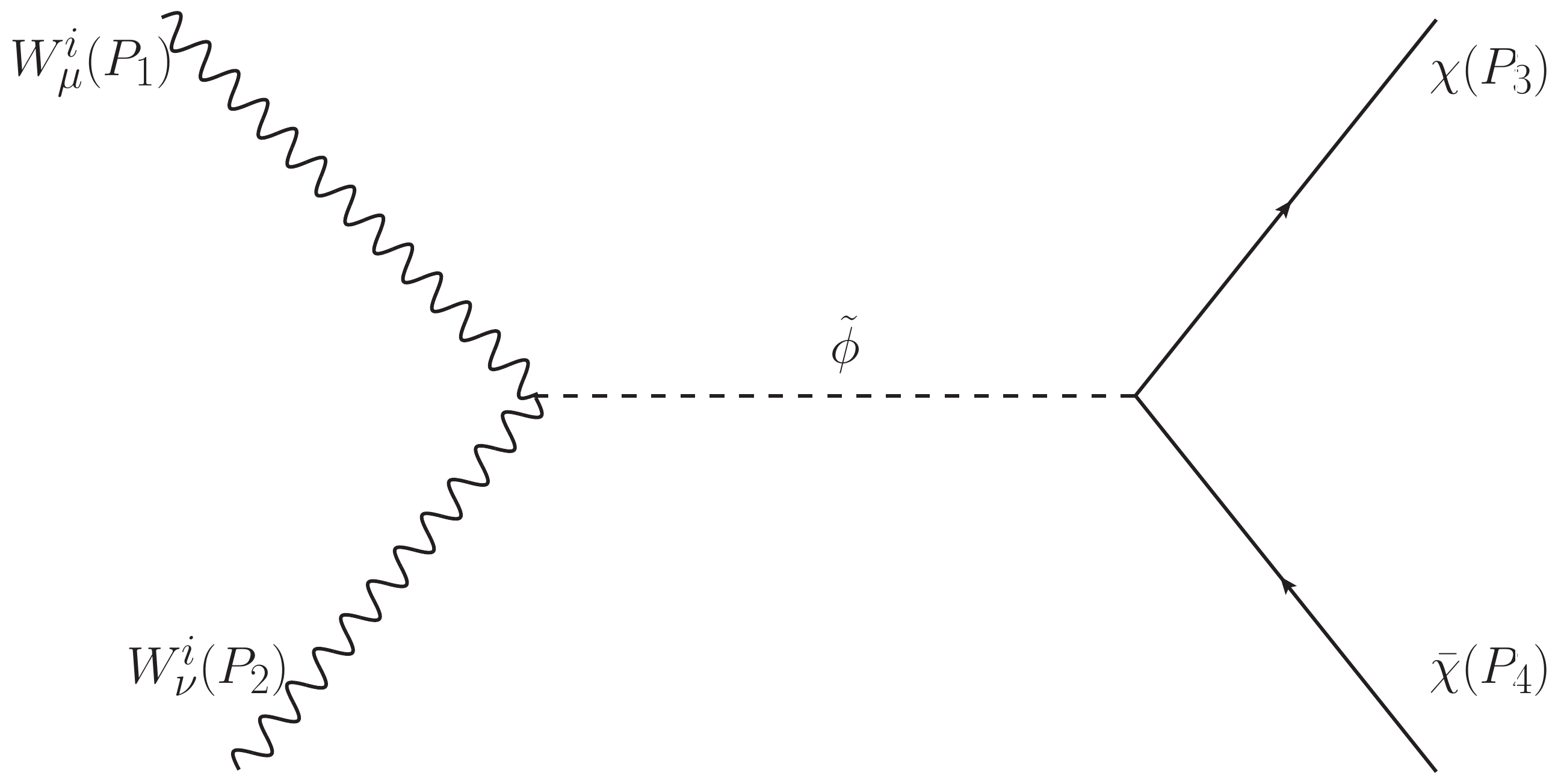} &
\includegraphics[height=2.5cm,width=4.5cm]{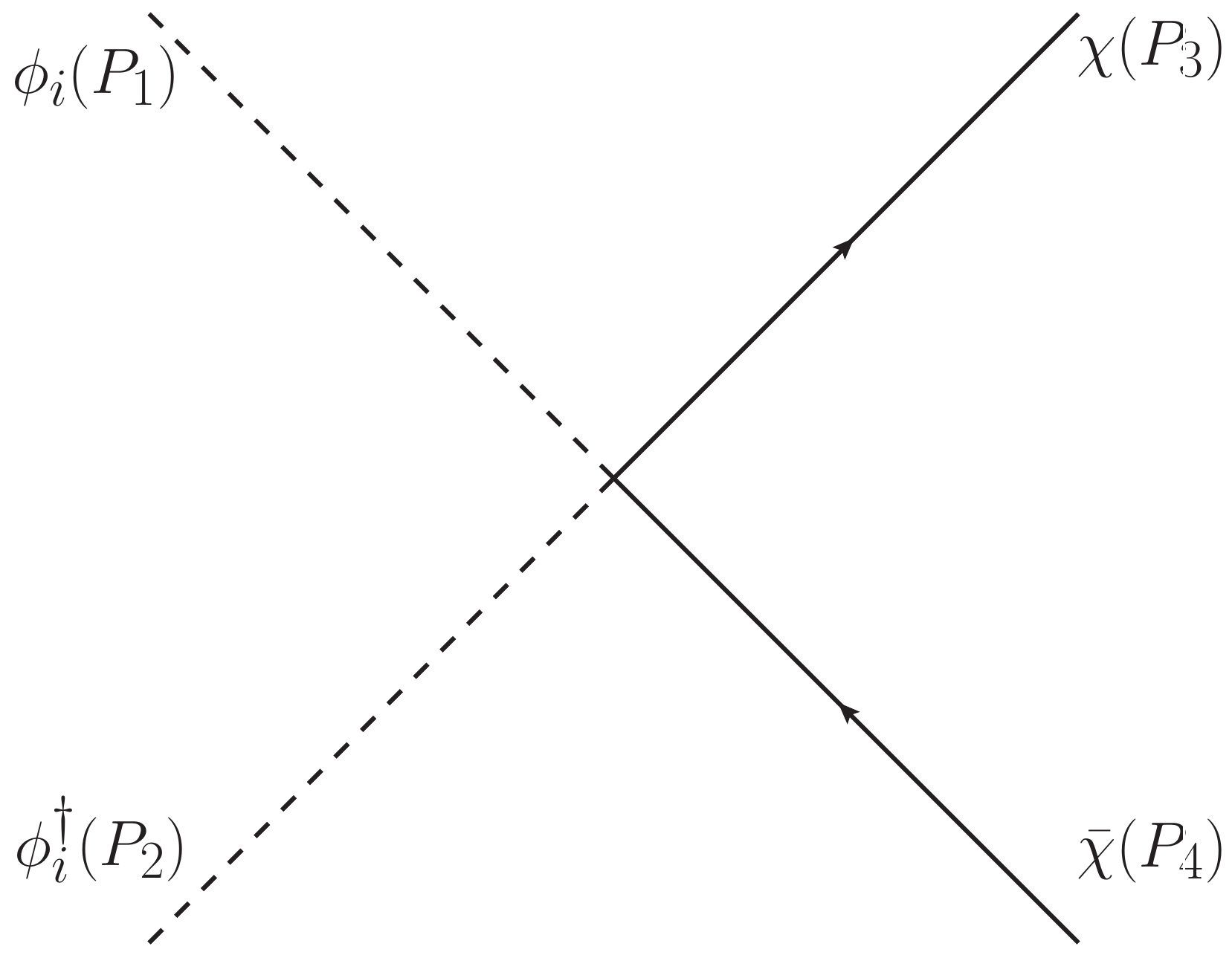}\\
UV freeze-in of $\x$ & & & \\
&\includegraphics[height=2.5cm,width=4.5cm]{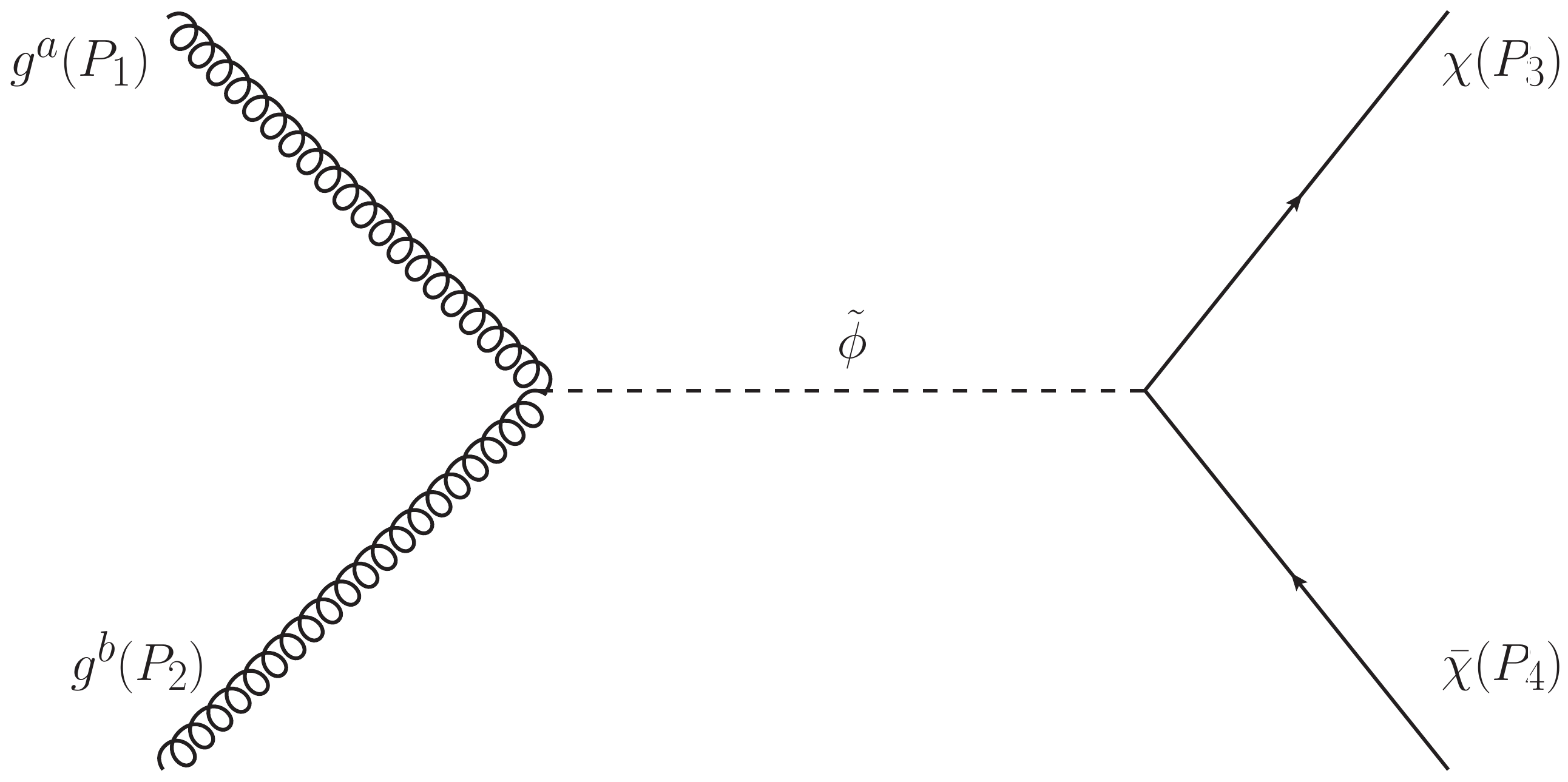}
&\includegraphics[height=2.5cm,width=4.5cm]{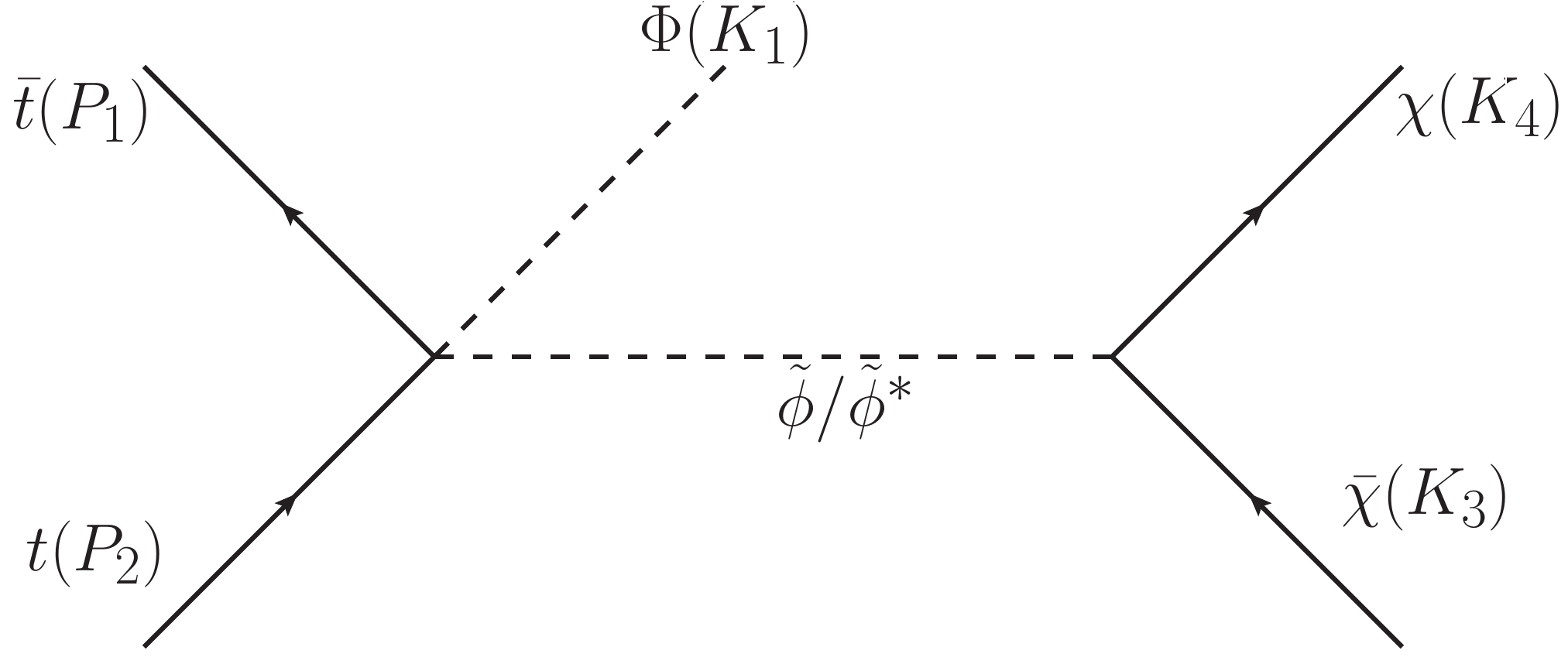}
&\includegraphics[height=2.5cm,width=4.5cm]{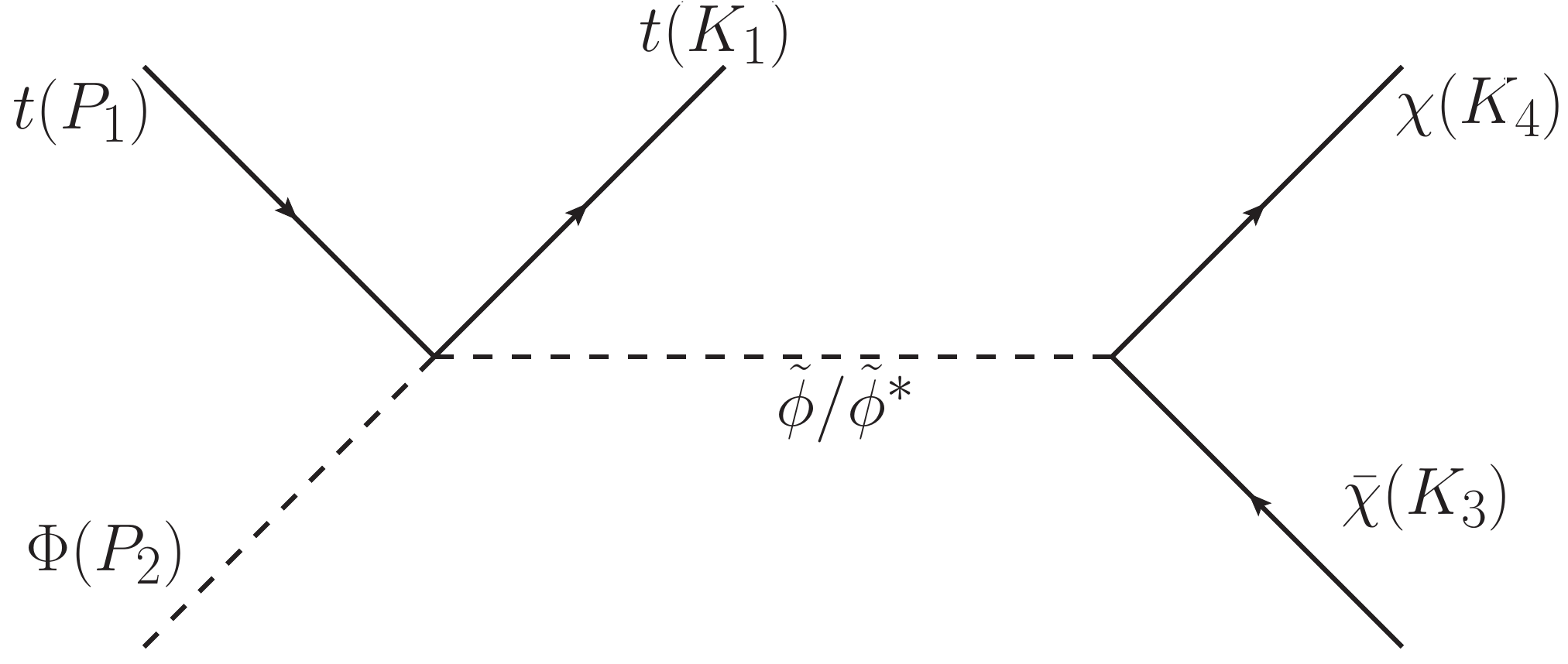}\\
&&\includegraphics[height=2.5cm,width=4.5cm]{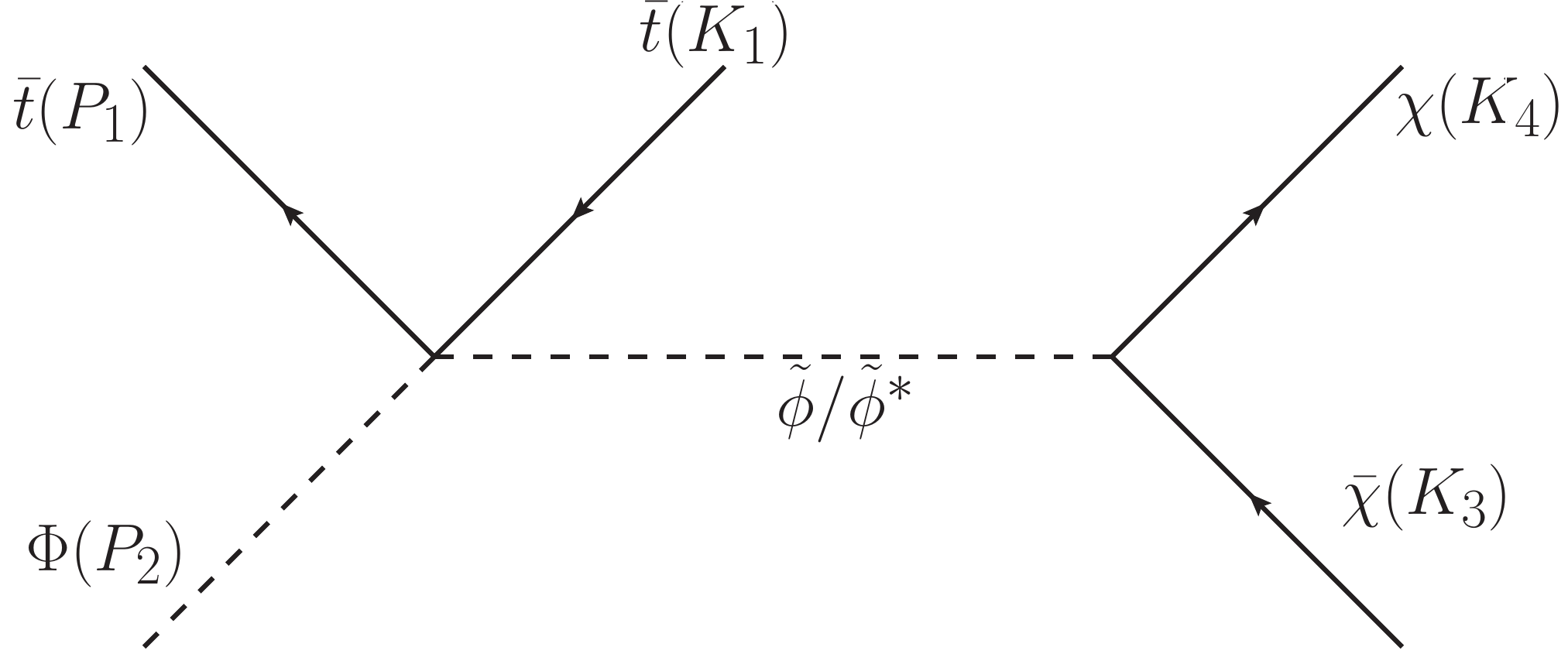}&\\
\hline
\end{tabular}
\caption{UV production channels of dark matter before EWSB.}
\label{fig1}
\end{figure}\\
On the other hand, IR freeze-in becomes effective only after the
EWSB where we have three massive gauge bosons $W^{\pm}_{\mu}$, $Z_{\mu}$ and
nine massless gauge boson $A_{\mu}$, $G_{\mu}^a$
as photon and gluons respectively in the particle
spectrum. Additionally, our dark matter candidate can now have interactions
with all the SM fermions and Higgs boson $h$. Depending
on the mass of $\x$, each of these particles have contributions to
$\x$ production in this regime. The interaction terms which govern
IR production of $\x$ are given by
\begin{align}
\mathcal{L} & \supset -\frac{\epsilon^{\mu\nu\alpha\beta}
\left(\partial_\mu W^-_\nu\right)\left(\partial_\alpha W^+_\beta\right)\tf}{\Lambda}-
\frac{\epsilon^{\mu\nu\alpha\beta}\left(\partial_\mu W^+_\nu\right)
\left(\partial_\alpha W^-_\beta\right)\tf}{\Lambda}
-\frac{\epsilon^{\mu\nu\alpha\beta}\left(\partial_\mu Z_\nu\right)\left(\partial_\alpha Z_\beta\right)
\tf}{\Lambda}\nn\\
&-\frac{\epsilon^{\mu\nu\alpha\beta}(\partial_\mu A_\nu)(\partial_\alpha A_\beta)\tf}{\Lambda}
-\frac{\epsilon^{\mu\nu\alpha\beta}\left(\partial_\mu G^b_\nu\right)\left(\partial_\alpha G^b_\beta\right)\tf}{\Lambda}
-\dfrac{i}{\Lambda}\sum_{j} m_{f_j}\overline{f_j}\gamma_5 f_j \tilde{\phi} \nn\\
&-\dfrac{\textit{v}}{\Lambda}\overline{\x}\x h -\frac{1}{2\Lambda}\overline{\x}\x\,h^2
-g\,\overline{\x}\gamma_5\x\,\tf\,,
\label{L-IR}
\end{align}
where $m_{f_j}$ is the mass of the SM fermion $f_j$ and the summation index $j$ is
taken over all the SM fermions. 
\newpage
\noi
Therefore, our dark matter candidate $\x$
has IR contributions to its relic density from the following scattering 
processes\footnote {$g\,g \rightarrow {h^*} \rightarrow \bar{\x} \x$
becomes important at low temperature where $g_3$ increases with
the decrease in energy scale. The $\tf\tf\rightarrow \overline{\x}\x$ process
will be suppressed due to the quadratic dependence of non-thermal
distribution function of $\tf$. $\bar{f}_jf_j\rightarrow \overline{\x}\x$
contribution comes essentially from top quark and that is
what we have considered in our numerical analysis.}:
$gg\rightarrow \overline{\x}\x$, 
$W^{+}W^{-}\rightarrow \overline{\x}\x$,
$ZZ\rightarrow \overline{\x}\x$,
$\bar{f}_jf_j\rightarrow \overline{\x}\x$,
$hh\rightarrow \overline{\x}\x$ and
$\tf\tf\rightarrow \overline{\x}\x$.
Apart from these scatterings, $\x$ can also be produced from the decays
of Higgs boson $h$ and pseudo scalar $\tf$ if such processes
are kinematically allowed. Feynman diagrams for
all these processes are shown in Fig.\,\ref{fig1a}.
Moreover, we would like to comment here that 
as in this work we are considering $m_{\tf} \leq 100$ GeV,
this choice does not allow $\tf$ to be generated thermally.
This will become clear in the next section where we have a detailed discussion
on this topic.
\begin{figure}[h!]
\centering
\begin{tabular}{|c|c c c|}
\hline
Regime & \multicolumn{3}{|c|}{DM Production channels} \\ 
\hline 
&&& \\
After EWSB&\includegraphics[height=2.5cm,width=4.5cm]{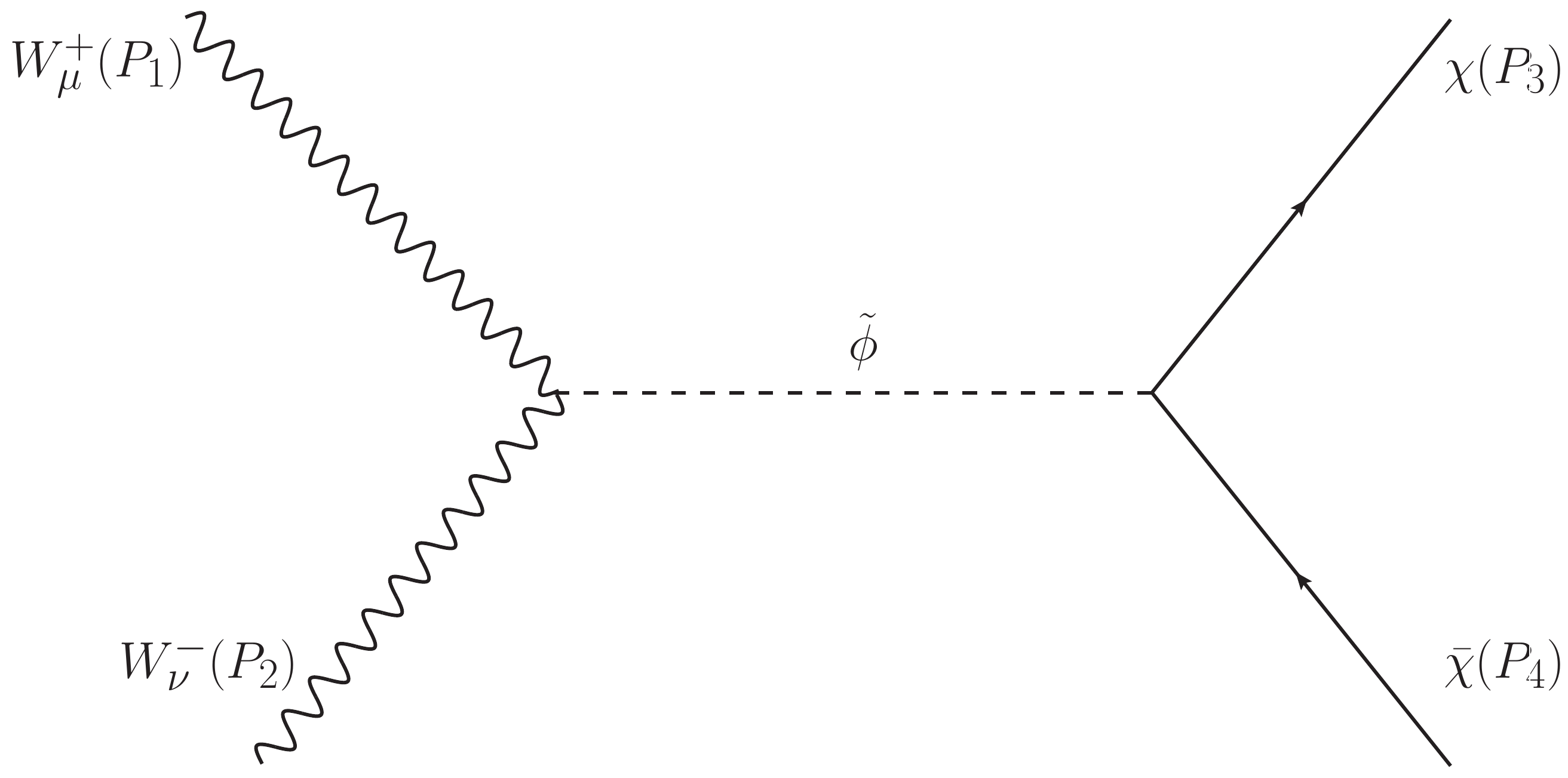} 
&\includegraphics[height=2.5cm,width=4.5cm]{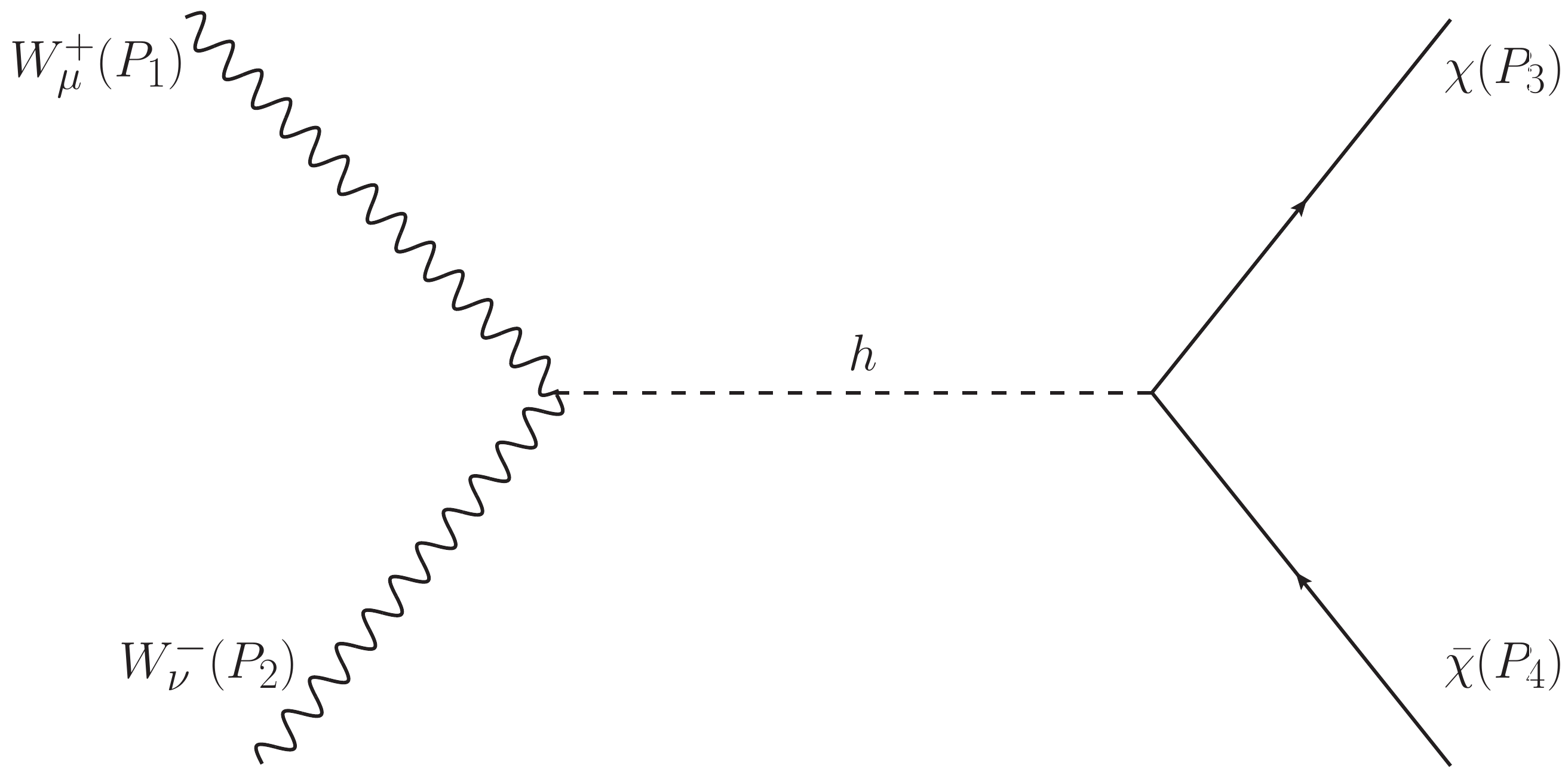}
&\includegraphics[height=2.5cm,width=4.5cm]{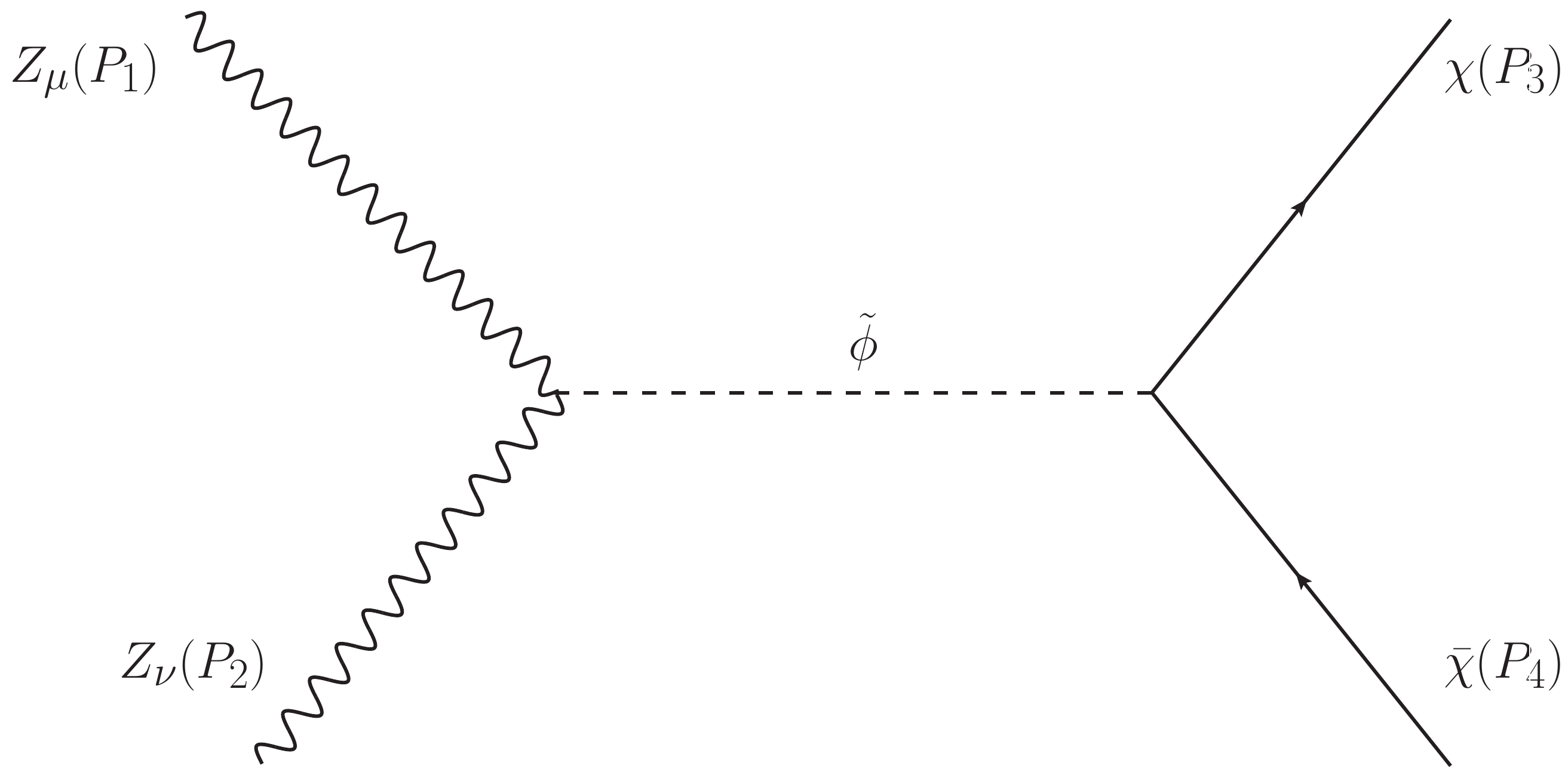}\\
IR production of $\x$&&&\\
&\includegraphics[height=2.5cm,width=4.5cm]{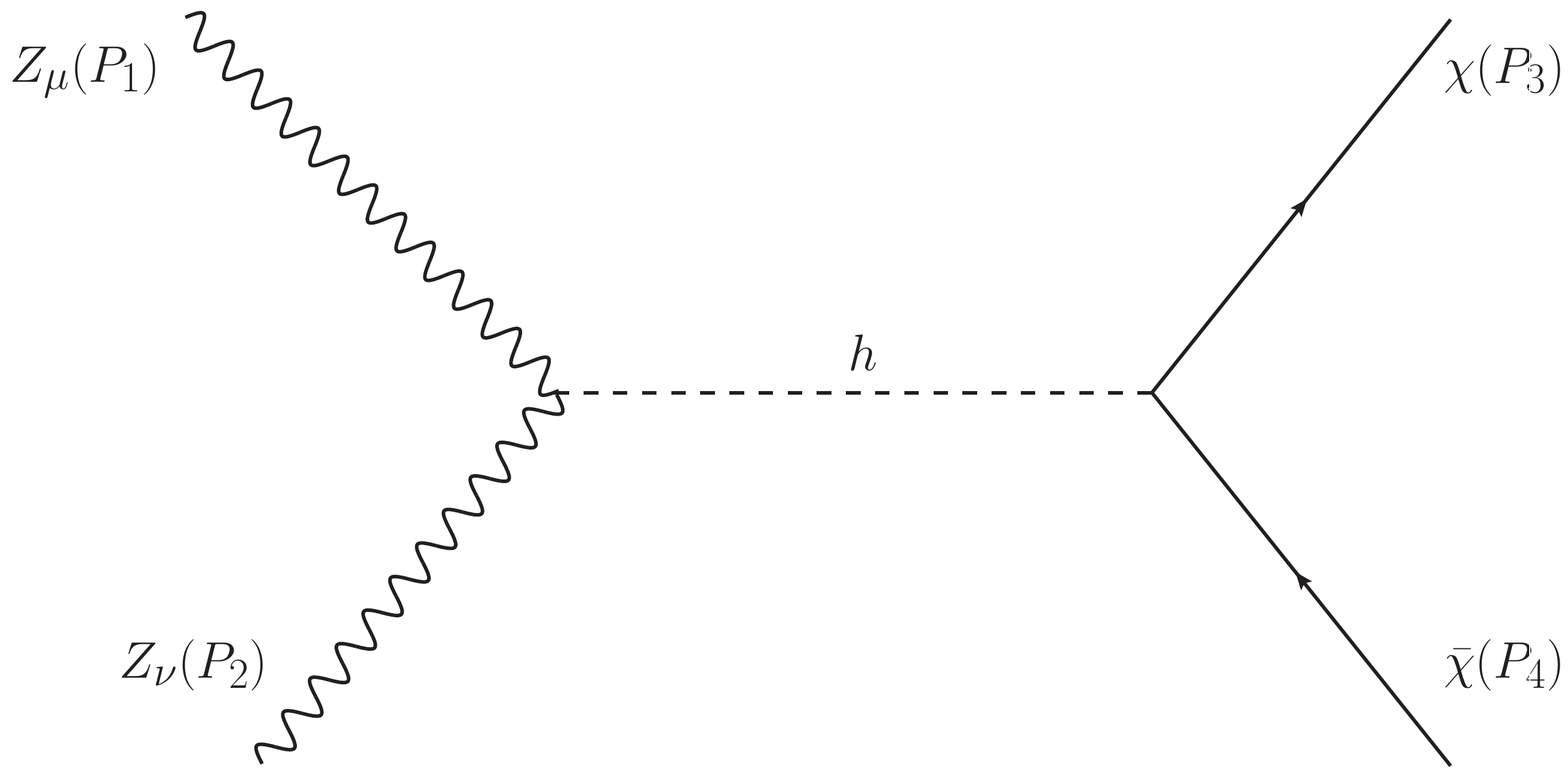} 
&\includegraphics[height=2.5cm,width=4.5cm]{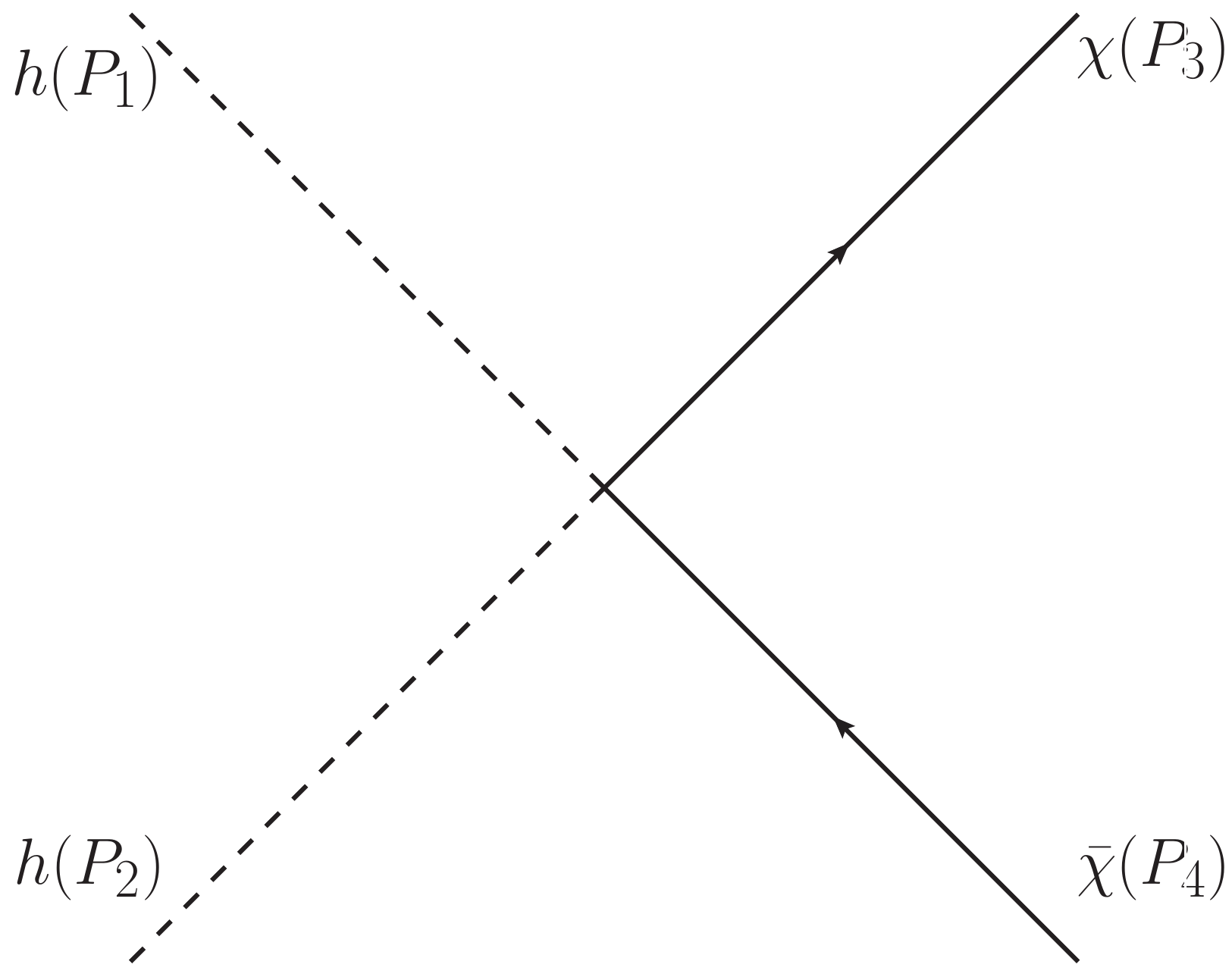}
&\includegraphics[height=2.5cm,width=4.5cm]{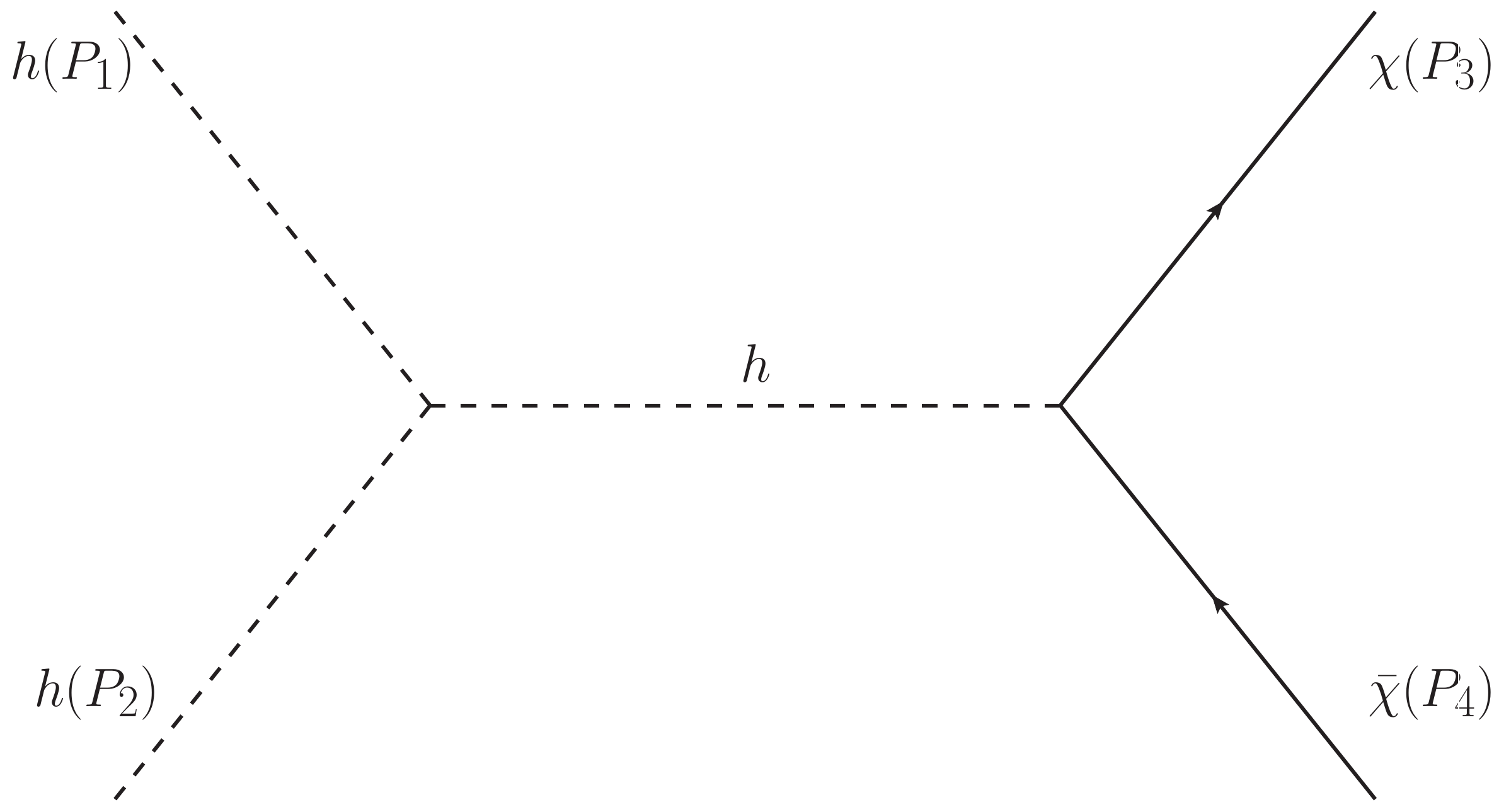}\\
&&&\\
&\includegraphics[height=2.5cm,width=4.5cm]{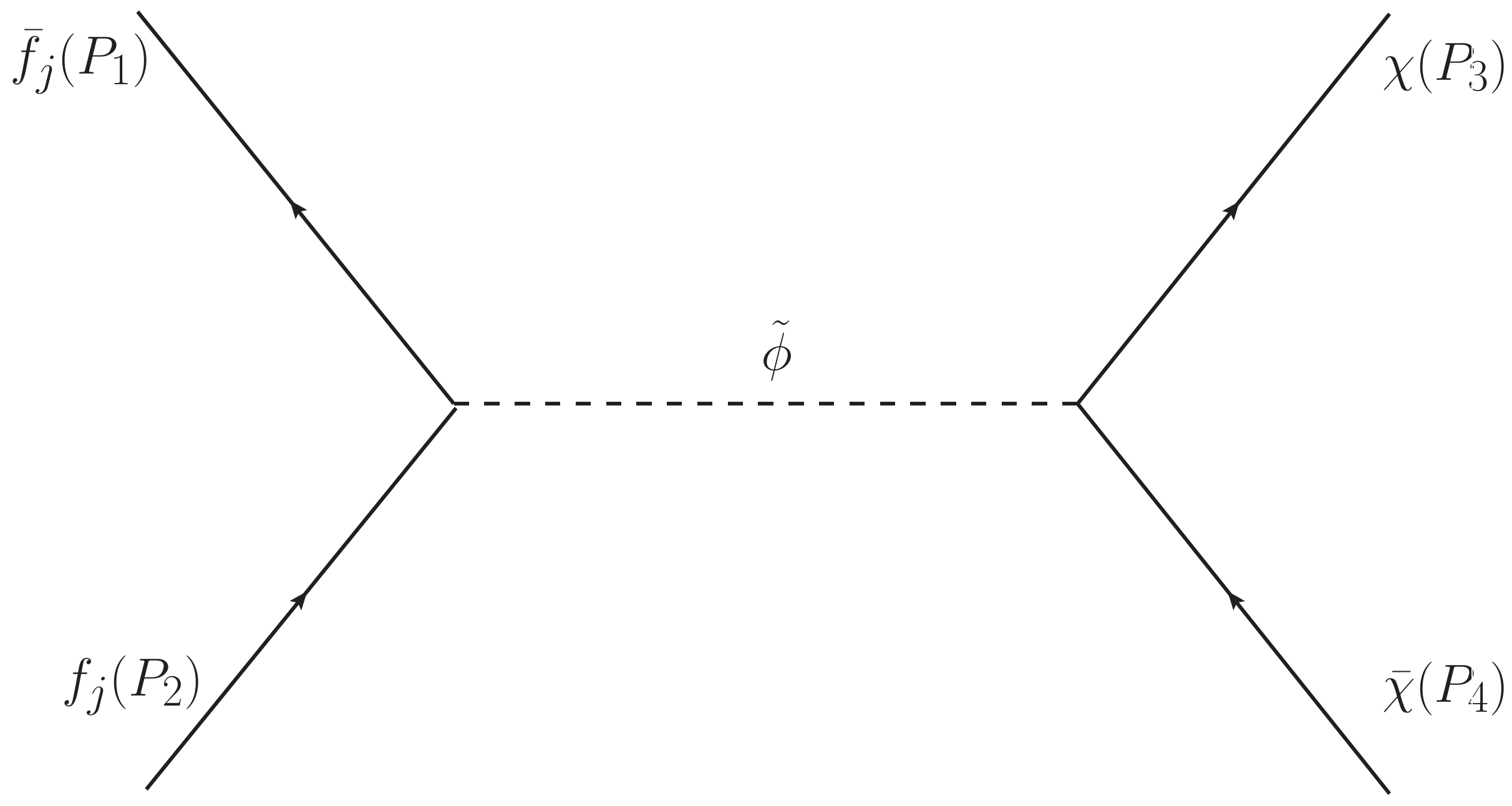}
&\includegraphics[height=2.5cm,width=4.5cm]{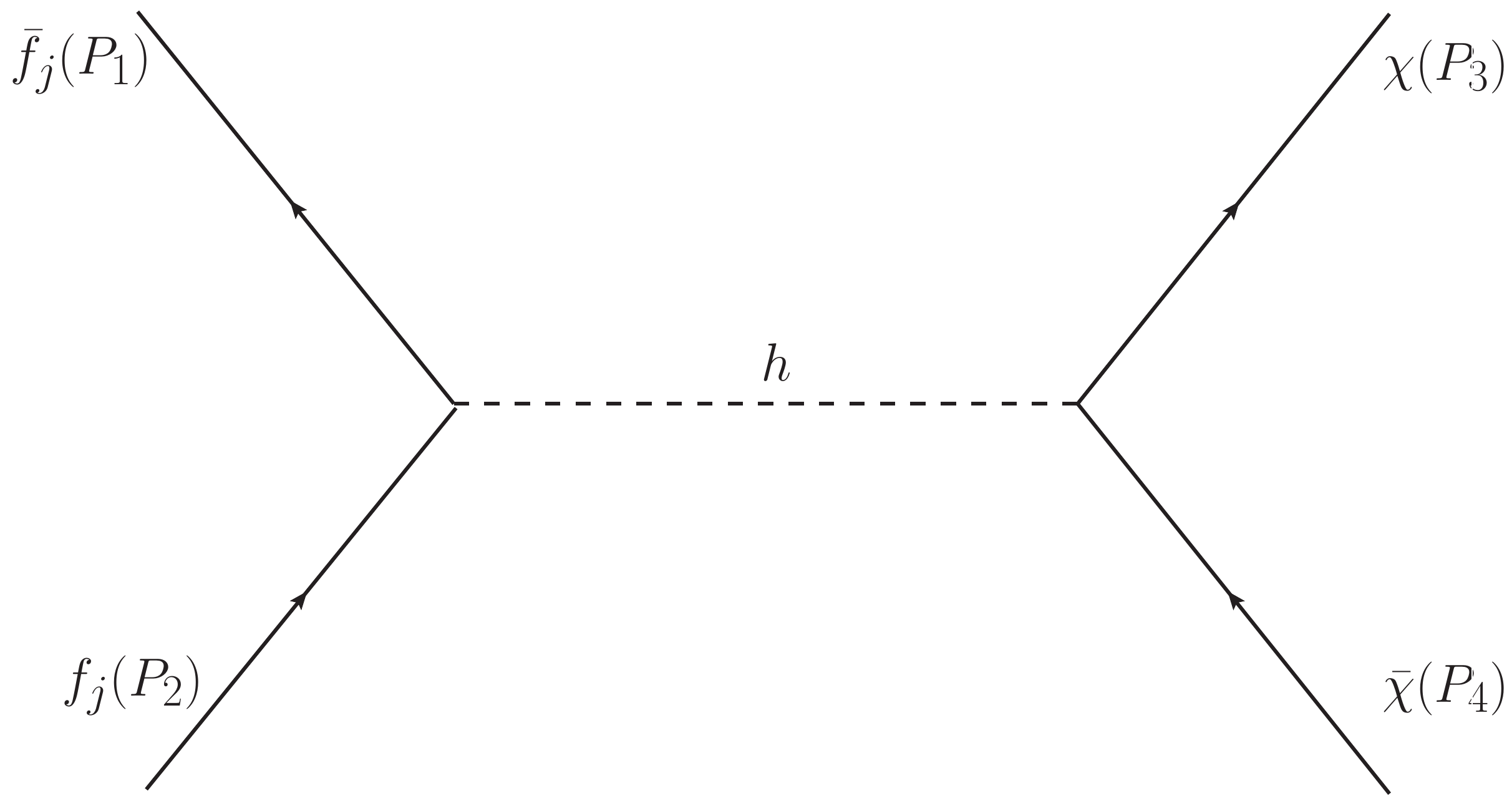}
&\includegraphics[height=2.5cm,width=4.5cm]{ggcc.pdf}\\
&\includegraphics[height=2.5cm,width=6.0cm]{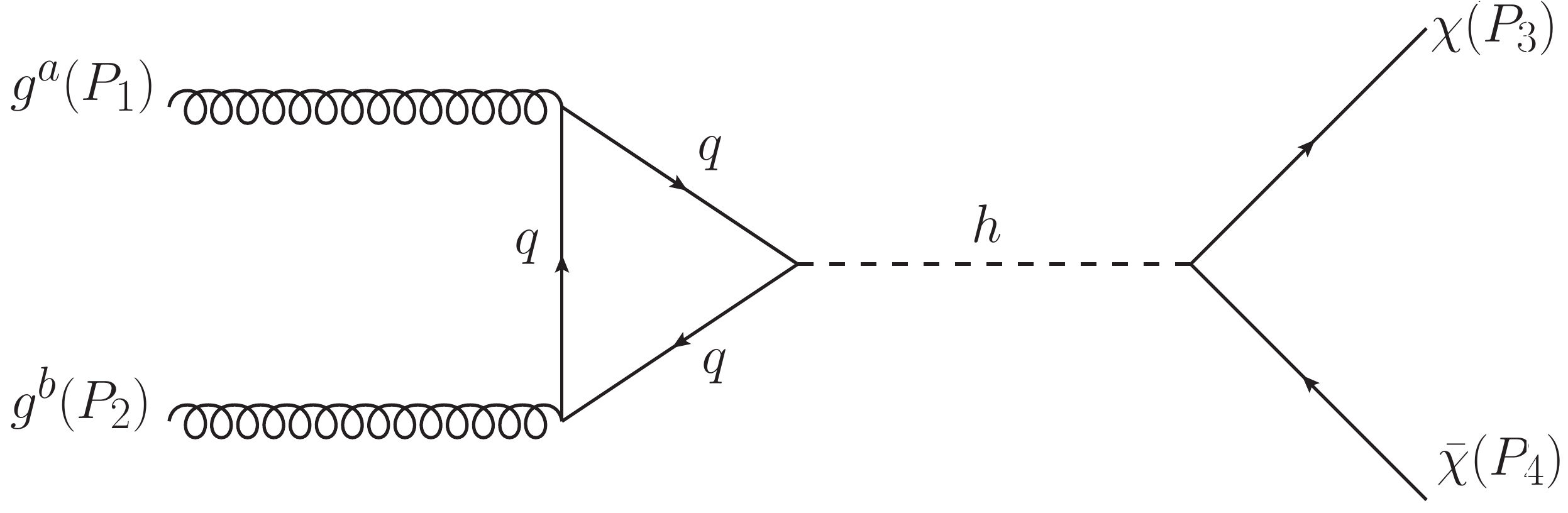}
&\includegraphics[height=2.5cm,width=4.5cm]{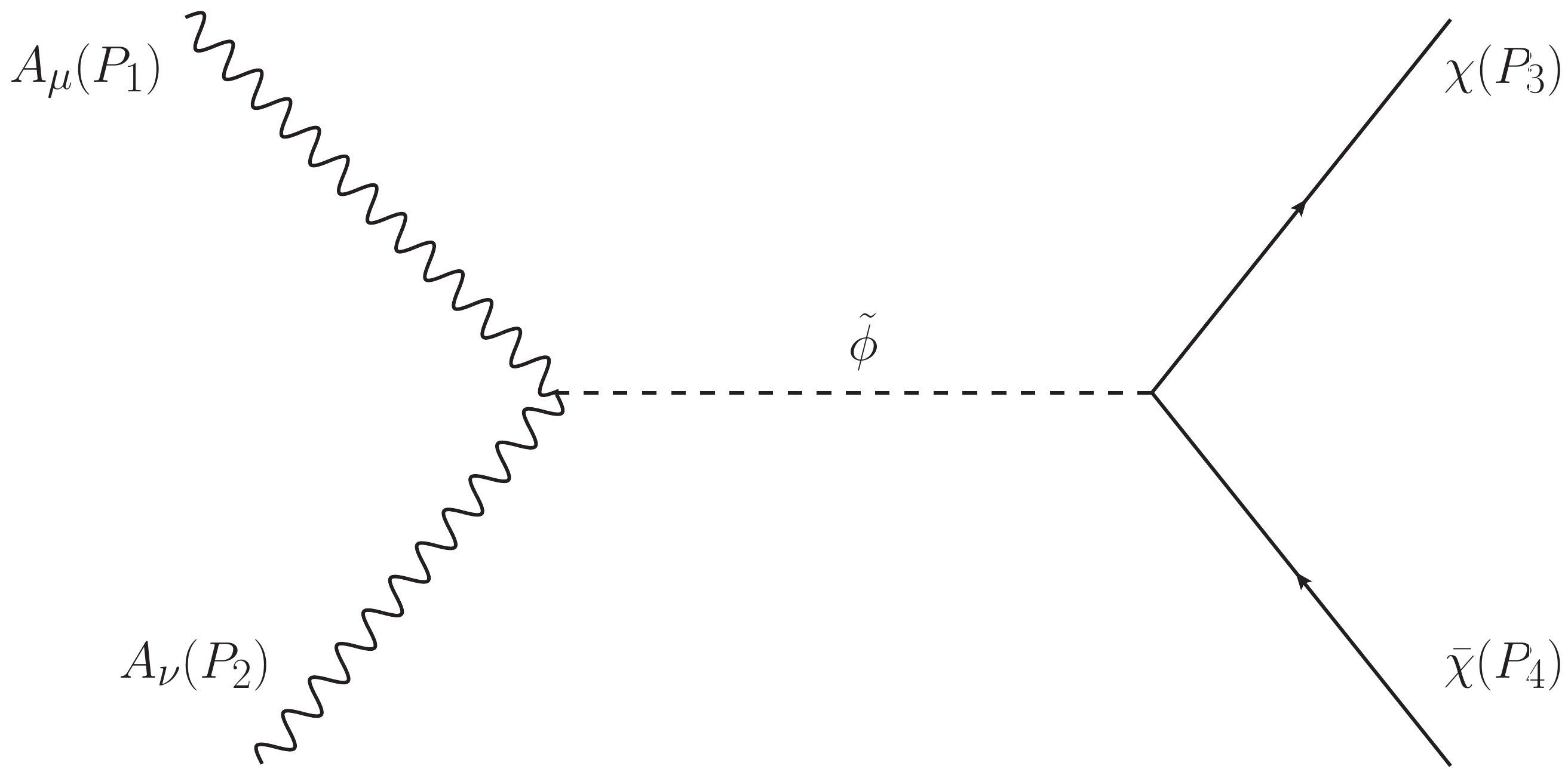}
&\includegraphics[height=2.5cm,width=4.5cm]{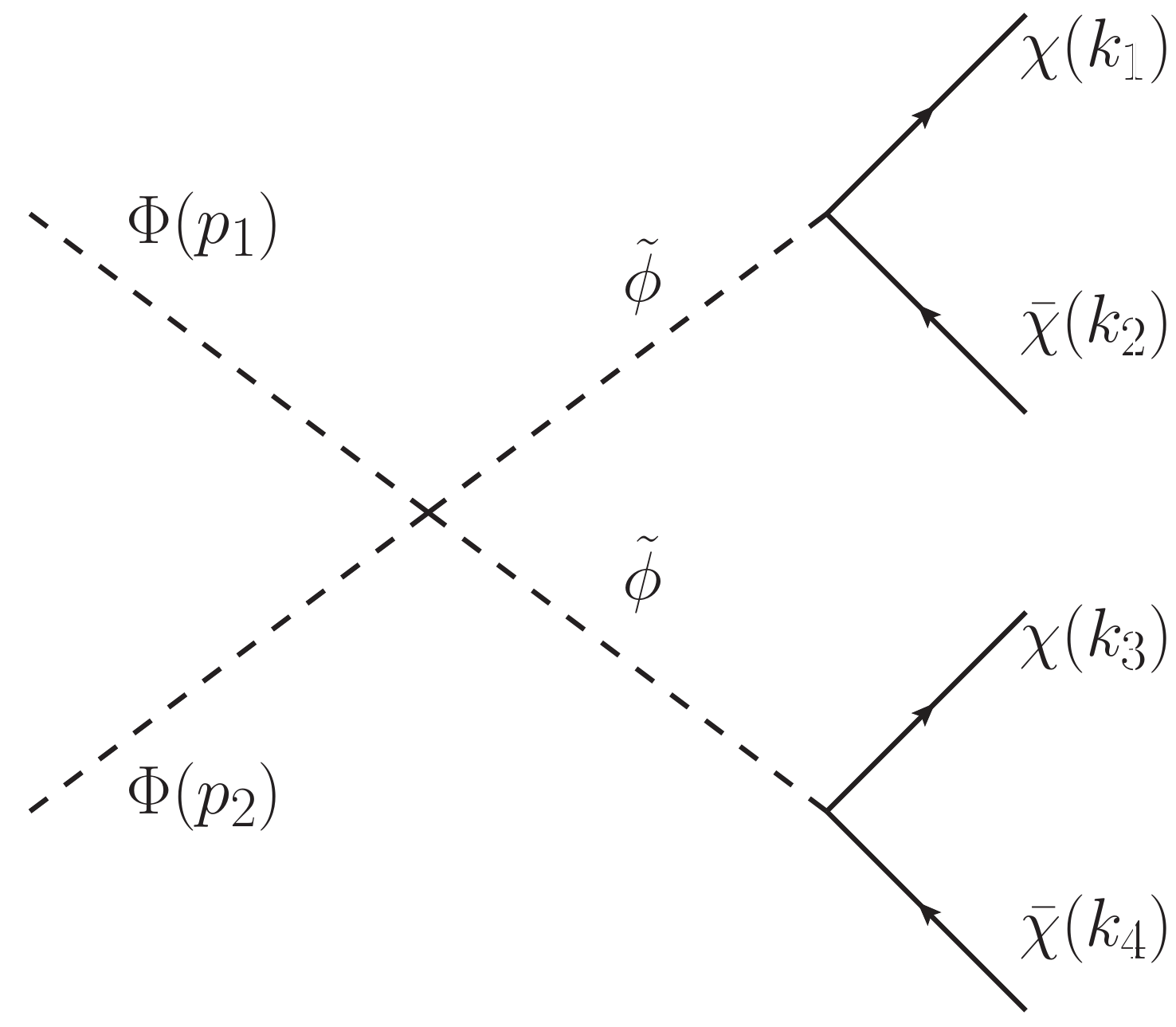}\\
&\includegraphics[height=2.5cm,width=4.5cm]{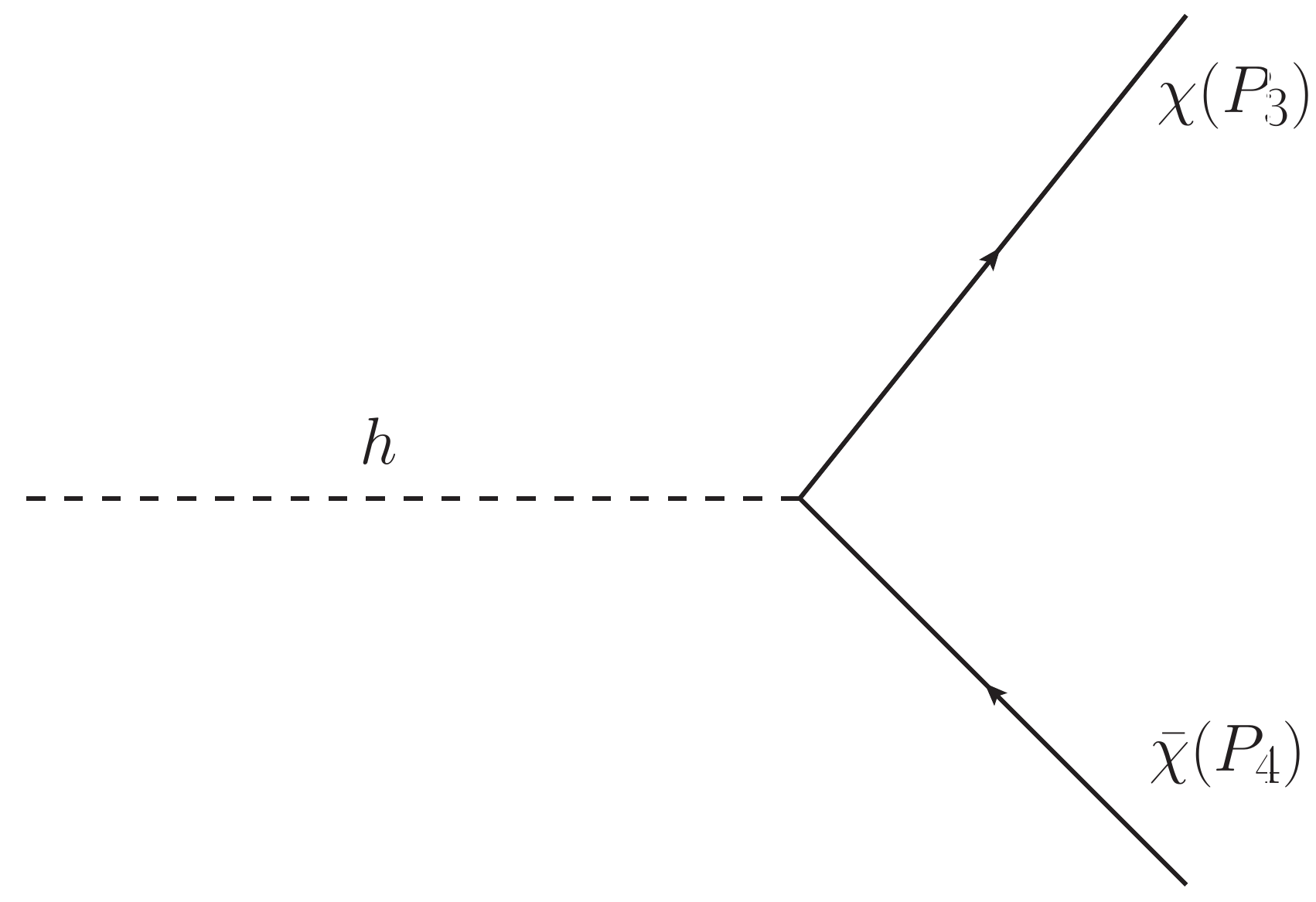}
&\includegraphics[height=2.5cm,width=4.5cm]{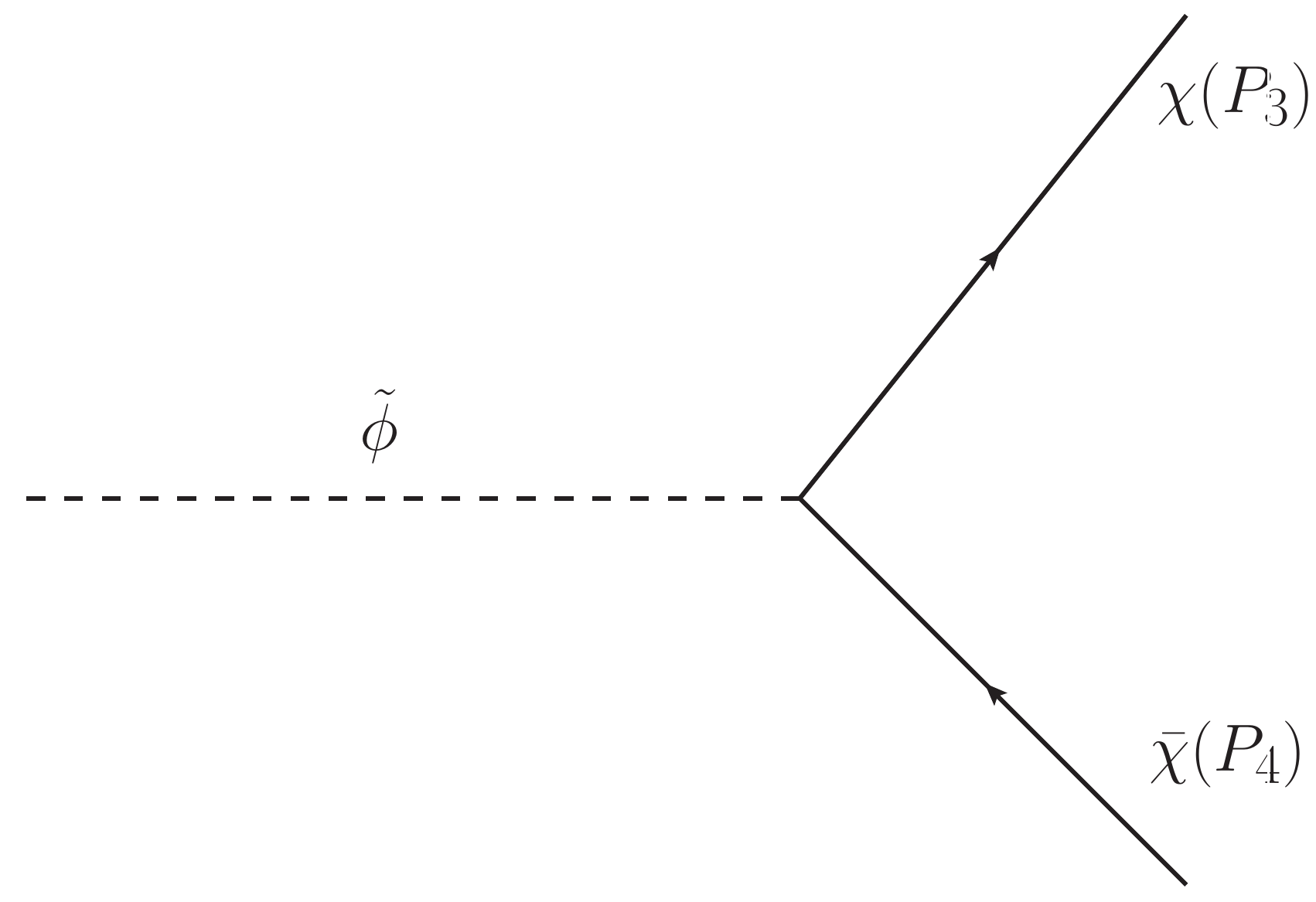}&\\

\hline
\end{tabular}
\caption{IR production channels of dark matter after EWSB.}
\label{fig1a}
\end{figure}
Now, we calculate
the relic abundance of $\x$ produced via freeze-in. For that one
needs to solve the Boltzmann equation of $\x$, taking into account
all possible interactions into the collision term. The Boltzmann
equation of $\x$ is given by
\begin{align}
\frac{dn_\x}{dt}+3Hn_{\x} &\simeq  \sum_{i} n^{\rm eq}_{i}\,n^{\rm eq}_{i}\,
\langle {\sigma\,{\rm v_{\rm rel}} \rangle_{i\,i \rightarrow \overline{\x}\x}}\nn
+n^{\rm eq}_{h}\,\langle {\Gamma_{h \rightarrow \overline{\x}\x}}\rangle\\
&+\dfrac{g_{\tf}m_{\tf} \Gamma_{\tf\rightarrow\x\bar{\x}}}{2\pi^2}\int_0^\infty f(p,T) 
\dfrac{p^2 dp}{\sqrt{p^2+m_{\tf}^2}}+
\mathcal{F}_{2\rightarrow3}(T)+\mathcal{F}_{2\rightarrow4}(T)\,,
\label{BE-main}
\end{align}
where, $n_{i}$ is the number density of species $i$ and
the corresponding equilibrium number density is denoted
by $n^{\rm eq}_i$. 
The second term in the left hand side of the Boltzmann
equation proportional to the Hubble parameter $H$
dilutes $n_{\x}$ due to the expansion
of the Universe. In the right hand side,
we have the usual collision term of the Boltzmann equation.
Since we are dealing with a non-thermal dark matter candidate
$\x$ with an insignificant initial number density, in the
collision term we have neglected backward reaction terms (inverse processes)
proportional to $n^2_{\x}$. Thus, in this case the collision term only
contains all types of production processes of $\x$ from annihilations
as well as decays of other particles. The first term in the right
hand side indicates the production of $\x$ from scattering of particles
$i$ which are in thermal equilibrium and their number densities
only depend on mass of the corresponding particle
and temperature of the Universe. The summation
over $i$ is there to include all possible scattering
processes.
In principle, there can also be two
different particles at the initial state of
scattering, however, we are not considering
such type of production processes of $\x$ because
those are not present in our present model.
The scattering term has both UV and
IR contributions. The epoch of UV production
is determined by the reheat temperature $T_{\rm RH}$
of the Universe. However, for IR processes
the production of dark matter from a particular
decay(annihilation) becomes significant at around
the temperature $T \sim m$, where $m$ being the mass of the
initial state particle(s). Particularly, this is due to the reason
that for temperature $T<<m$, the number density of
mother particles following Maxwell-Boltzmann
distribution gets exponentially suppressed. 

The second part of the collision term represents
the increase in number density of $\x$ from
decay of the SM Higgs ($h$).
A detail derivation of the collision term for both scattering
and decay processes are given in Appendices \ref{app:BE-scatt}
and \ref{app:BE-dec}. 

The third part of the collision term indicates the production of
$\x$ via mixed freeze-in. 
Since $\tf$ is not in thermal equilibrium with the SM bath,
the calculation of the abundance of $\x$ from the decay of $\tf$,
requires the knowledge of the momentum distribution function ($f(p,T)$) of $\tf$ . 
A detail derivation of momentum distribution function of $\tf$
is given in Appendix \ref{dist_function}.
The fourth and fifth terms in Eq. (\ref{BE-main}) indicate the production of $\x$ from
$2\rightarrow3$ and $2\rightarrow4$ processes via off-shell\footnote{$2 \rightarrow 3$
processes contribute maximally at $T \sim T_{\rm RH}$
and hence we have neglected these processes in the IR regime.} $\tf$ 
and the detail calculation of these collision terms
are given in Appendix \ref{n_body_scat}.

To solve the above equation,
it is useful to consider a dimensionless variable
namely the comoving number density of $\x$,
$Y_{\x} = \dfrac{n_{\x}}{s}$, which absorbs the effect
of expansion of the Universe and indicates only the
change in $n_{\x}$ due to number changing processes
involving dark matter candidate $\x$ and other bath
particles. Here the quantity $s$ is the entropy density of the Universe.
Following the procedure as discussed in Appendices \ref{app:BE-scatt}
and \ref{app:BE-dec} for solution of the Boltzmann equation
for both UV as well as IR freeze-in case, we now present the
solution of the Boltzmann equation (value of $Y_{\x}$ at the
present epoch) as given below,
\bea
Y_{\x}(T_0) &\simeq & 2\times\frac{1}{16\pi^4} \Bigg[\int_{T_{\rm EW}}^{T_{\rm RH}}
\frac{dT}{sH}\,
\int^{\infty}_{4\,m^2_{i}} d\hat{s}\,\sum_{i}\,\left(g_{i}^2\,\sigma_{ii\rightarrow
\overline{\x}\x}\right)\,F \sqrt{\hat{s}-4\,m^2_{i}}\,{\rm K_1}
\left(\frac{\sqrt{\hat{s}}}{T}\right)  \nn \\ &+&
16\pi^4
\left(\int_{T_{\rm EW}}^{T_{\rm RH}}\dfrac{dT}{sHT} \mathcal{F}_{2\rightarrow3}(T)
+
\dfrac{g_{\tf}m_{\tf} \Gamma_{\tf\rightarrow\x\bar{\x}}}{2\pi^2}
\int_{T_{0}}^{T_{\rm RH}}\dfrac{dT}{sHT}
\int_0^\infty f(p,T) \dfrac{p^2 dp}{\sqrt{p^2+m_{\tf}^2}}\right)\nn\\
&+& 16\pi^4\int_{T_0}^{T_{\rm EW}}  \dfrac{dT}{sHT}\,\mathcal{F}_{2\rightarrow 4}(T)+
\int_{T_0}^{T_{\rm EW}} 
\frac{dT}{sH}\,
\int^{\infty}_{4\,m^2_{i}} d\hat{s}\,\sum_{i}\,\left(g_{i}^2\,
\sigma_{ii\rightarrow
\overline{\x}\x}\right)\,F \sqrt{\hat{s}-4\,m^2_{i}}\,{\rm K_1}
\left(\frac{\sqrt{\hat{s}}}{T}\right) \nn \\
&+& 8\pi^2 g_h\,m^2_{h}\,
\Gamma_{h \rightarrow \overline{\x}\x}\int_{T_0}^{T_{\rm EW}}
\dfrac{dT}{sH} {\rm K_1}\left(\frac{m_h}{T}\right) \Bigg]\,.
\label{ychi}
\eea
Here, the first two terms are the UV contribution to $Y_{\x}$
and it is clearly visible that these parts depend on the
reheat temperature $T_{\rm RH}$. The third term is the
production of $\x$ via mixed freeze-in 
which is present before 
and after EWSB. On the other hand, the
next part is coming from the IR contributions to $Y_{\x}$
and it becomes effective only after the EWSB. This is because 
after the EWSB, Higgs doublet gets a nonzero VEV and there are
renormalisable interactions (up to a level of dimension four)
between dark matter and Higgs.
The last term is another IR contribution to $Y_{\x}$ originating
from the decay of $h$. In the above, $m_i$, $g_i$ are mass and internal
degrees of freedom of particle $i$ while $T_{\rm EW}$ and
$T_0$ are EWSB temperature and temperature at the present epoch. 
Moreover, as discussed in the Appendices \ref{app:BE-scatt} and
\ref{app:BE-dec}, one can further simplify UV contribution and
IR contribution from $h$ decay. On the other hand, the decay 
$\tf\rightarrow\bar{\x}\x$,  UV freeze-in terms originating 
from $2\rightarrow3$ processes and IR freeze-in 
term originating from $2\rightarrow4$ process
require knowledge of $f(p,T)$, $\mathcal{F}_{2\rightarrow3} $, 
$\mathcal{F}_{2\rightarrow4}$ respectively, which are 
discussed in detail in Appendices \ref{dist_function} and \ref{n_body_scat}. Simplification of
IR freeze-in term (fifth term in the right hand side of Eq.\,\,(\ref{ychi}))
originating from the scatterings requires knowledge about cross sections of
all the production processes. Therefore, the solution of the Boltzmann
equation after those simplifications can be written in a more compact
form as follows
\bea
Y_{\x}(T_0) &\simeq & 
2\times\dfrac{M_{pl}}{1.66\,(2\pi)^3}
\Bigg[\dfrac{180\,\left(16 g^2 + 32g^2 + 4\right)}
{(2\pi)^4\,\sqrt{g_{\rho}(T_{\rm RH})}\,g_s(T_{\rm RH})}
\dfrac{T_{\rm RH}-T_{\rm EW}}{\Lambda^2} + 135 \left(
\dfrac{g_{h}\,
\Gamma_{h \rightarrow \overline{\x}\x}}{m^2_{h}\,
\sqrt{g_{\rho}(m_h)}\,g_s(m_h)}\right)\Bigg] \nn \\
&+&2
\left(\int_{T_{\rm EW}}^{T_{\rm RH}}\dfrac{dT}{sHT} \mathcal{F}_{2\rightarrow3}(T)
+\dfrac{g_{\tf}m_{\tf} \Gamma_{\tf\rightarrow\x\bar{\x}}}{2\pi^2}
\int_{T_{0}}^{T_{\rm RH}}\dfrac{dT}{sHT}
\int_0^\infty f(p,T) \dfrac{p^2 dp}{\sqrt{p^2+m_{\tf}^2}}\right) \nn  \\ 
&+& 2\int_{T_{0}}^{T_{\rm EW}}\dfrac{dT}{sHT} \mathcal{F}_{2\rightarrow4}(T)
+\frac{1}{8\pi^4}
\int_{T_0}^{T_{\rm EW}} 
\frac{dT}{sH}\,
\int^{\infty}_{4\,m^2_{i}} d\hat{s}\,\sum_{i}\,\left(g_{i}^2\,
\sigma_{ii\rightarrow
\overline{\x}\x}\right)\,F \sqrt{\hat{s}-4\,m^2_{i}}\,{\rm K_1}
\left(\frac{\sqrt{\hat{s}}}{T}\right)\,.\nn \\
\label{ychiT0}
\eea
Here, the factor $\left(16 g^2 + 32g^2 + 4 \right)$ in the first term
comes from fourteen scattering diagrams including four electroweak gauge boson
annihilations, eight gluon annihilations and two scalar annihilations into
$\overline{\x}\x$ pairs (see Feynman diagrams in Fig.\,\ref{fig1}).
 Below we have listed all the relevant scattering cross sections
and decay widths which are required to find $Y_{\x}$
at $T_0$ using Eq.\,(\ref{ychiT0}) and the corresponding
Feynman diagrams are illustrated in Fig.\,\ref{fig1a}.\\
\begin{subequations}\label{eq5}
\begin{align}
\sigma_{Z Z \rightarrow \bar{\chi} \chi}=&
\frac{1}{16\pi \hat{s}}\sqrt{\frac{\hat{s}-4m_{\chi}^2}{\hat{s}-4m_{Z}^2}}
\Bigg[\frac{8m_Z^4}{9\Lambda^2}\frac{(\hat{s}-4m_{\chi}^2)(3+\frac{\hat{s}^2}{4m_{Z}^4}
-\frac{\hat{s}}{m_{Z}^2})}{(\hat{s}-m_{h}^2)^2} 
+\frac{4g^2}{9\Lambda^2}\frac{\hat{s}^2(\hat{s}-4m_{Z}^2)}
{(\hat{s}-m_{\tilde{\phi}}^2)^2}\Bigg]\,,
\label{eq5a} \\
\nn\\
\sigma_{W^+ W^- \rightarrow \bar{\chi} \chi}=&
 \frac{1}{16\pi \hat{s}}\sqrt{\frac{\hat{s}-4m_{\chi}^2}{\hat{s}-4m_{W}^2}}
 \Bigg[\frac{8m_W^4}{9\Lambda^2}\frac{(\hat{s}-4m_{\chi}^2)(3+\frac{\hat{s}^2}{4m_{W}^4}
-\frac{\hat{s}}{m_{W}^2})}{(\hat{s}-m_{h}^2)^2} 
+\frac{4g^2}{9\Lambda^2}\frac{\hat{s}^2(\hat{s}-4m_{W}^2)}
{(\hat{s}-m_{\tilde{\phi}}^2)^2}\Bigg]\,,
\label{eq5b}\\
\nn \\
\sigma_{\gamma \gamma  \rightarrow \bar{\chi} \chi}=&
 \frac{g^2\hat{s}^{\frac{3}{2}}}{16\pi \Lambda^2}
\frac{\sqrt{\hat{s}-4m_{\chi}^2}}{(\hat{s}-m_{\tilde{\phi}}^2)^2} \,,
\label{eq5c}\\
\nn\\
\sigma_{\bar{f_i}f_i\rightarrow \bar{\chi}\chi}=&
\frac{1}{16\pi \hat{s}}\sqrt{\frac{\hat{s}-4m_{\chi}^2}{\hat{s}-4m_{f_i}^2}}
\Bigg[\frac{m_{f_i}^2(\hat{s}-4m_{f_i}^2)(\hat{s}-4m_{\chi}^2)}{\Lambda^2(\hat{s}-m_h^2)^2}
+\frac{g^2 \hat{s}^2\,m_{f_i}^2}{\Lambda^2(\hat{s}-m_{\tilde{\phi}^2})^2}\Bigg] \,,
\label{eq5d}\\
\nn \\
\sigma_{h h \rightarrow \bar{\chi}\chi}=&
\frac{1}{8\pi \hat{s} \Lambda^2}\Bigg[1+\frac{9m_h^4}{(\hat{s}-m_h^2)^2}\Bigg]
\frac{(\hat{s}-4m_{\chi}^2)^{\frac{3}{2}}}{\sqrt{\hat{s}-4m_h^2}}\,,
\label{eq5e} \\
\nn \\
\sigma_{gg\rightarrow \bar{\x} \x}=&
\frac{1}{16\pi\Lambda^2}
\left[\frac{g^2 \hat{s}^{\frac{3}{2}}\sqrt{\hat{s}-4 m_{\x}^2}}{8 (\hat{s}-m_{\tf}^2)^2}\,
+ \frac{|c_{ggh}|^2 v^2 \sqrt{\hat{s}(\hat{s}-4m_{\x}^2)^3}}{8 (\hat{s}-m_h^2)^2}\right]\,,
\label{eq5f}\\
\Gamma_{h\rightarrow \bar{\chi}\chi}=&
\frac{\textit{v}^2}{\Lambda^2}\frac{m_h}{8\pi}
\Bigg(1-\frac{4m_{\chi}^2}{m_h^2}\Bigg)^{\frac{3}{2}} \,,
\label{eq5g}\\
\nn\\
\Gamma_{\tf \rightarrow \bar{\x}\x}=& \dfrac{g^2 m_{\tf}}{8\pi}\sqrt{1-\dfrac{4 m_{\x}^2}{m_{\tf}^2}}\,.
\label{eq5h}
\end{align}
\end{subequations}
In Eq.\,(\ref{eq5f}), $c_{ggh}$ is the loop factor of the process
$gg \rightarrow \bar{\x} \x$ which is given by \cite{Barger:1987nn}
$$c_{ggh}(\hat{s}) = - (\sqrt{2}G_F)^{\frac{1}{2}} \frac{\alpha_s(\hat{s})}{12\pi} \sum_q I_q\,,$$
and $$I_q = 3 \int_0^1 dx \int_0^{1-x} dy \, \dfrac{1-4 x y}{1-\dfrac{x y}{\lambda_q}}\,,$$
where $\lambda_q = \frac{m_q^2}{\hat{s}}$. The loop integral $I_q$ has the following form
$$I_q = 3\left[2 \lambda_q + \lambda_q \left(4\lambda_q - 1\right)f(\lambda_q)\right]\,.$$
The function $f(\lambda_q)$ is given by
\bea
f(\lambda_q) &= & -2\left(\arcsin\dfrac{1}{2\sqrt{\lambda_q}}\right)^2\,,\,\,\,\,\,\,\, 
\,\,\,\,\,\,\,\,\,\,\,\,\,\,\,\,\,\,\,\,\,\,\,\,\text{for} \,\lambda_q >\frac{1}{4}\,,\nn\\
& = & \frac{1}{2}\left(\log \frac{\eta^+}{\eta^-} \right)^2-\frac{\pi^2}{2} - 
i\pi \log \frac{\eta^+}{\eta^-}\,,\,\,\, \,\,\,\,\,\text{for} \, \lambda_q < \frac{1}{4}\,,\nn
\eea
with $\eta^{\pm} =  \dfrac{1}{2} \pm \sqrt{\dfrac{1}{4}-\lambda_q}$\,\,\,.

\vspace{0.2cm}

Finally, the relic density ($\Omega_{\x} h^2$) of $\x$ is defined as the ratio of
dark matter mass density to the critical density ($\rho_{\rm crit}$) of the Universe
and it is related to comoving number density ($Y_{\x}(T_0)$) by the following
relation \cite{Gondolo:1990dk, Edsjo:1997bg},
\begin{equation}
\Omega_{\x} h^2= 2.755 \times10^8 \dfrac{m_{\x}}{\rm GeV}\, Y_{\x}(T_0)\,.
\label{omega}
\end{equation}

Let us note in passing that the dominant production channels of $\x$
are $\phi_i^\dagger \phi_i \rightarrow \bar{\x} \x$ in the UV regime and 
$ W^+W^- \rightarrow \bar{\x} \x$, $ Z Z \rightarrow \bar{\x} \x$, 
$ h h \rightarrow \bar{\x} \x$, $ h \rightarrow \bar{\x} \x$ in IR regime.
Mixed freeze-in can contribute to the relic density of $\x$ in both
UV and IR regime.

\subsection{Numerical results of the Boltzmann Equation}
\label{BE_numeric}
\begin{figure}[h!]
\includegraphics[height=7cm,width=8.55cm,angle=0]{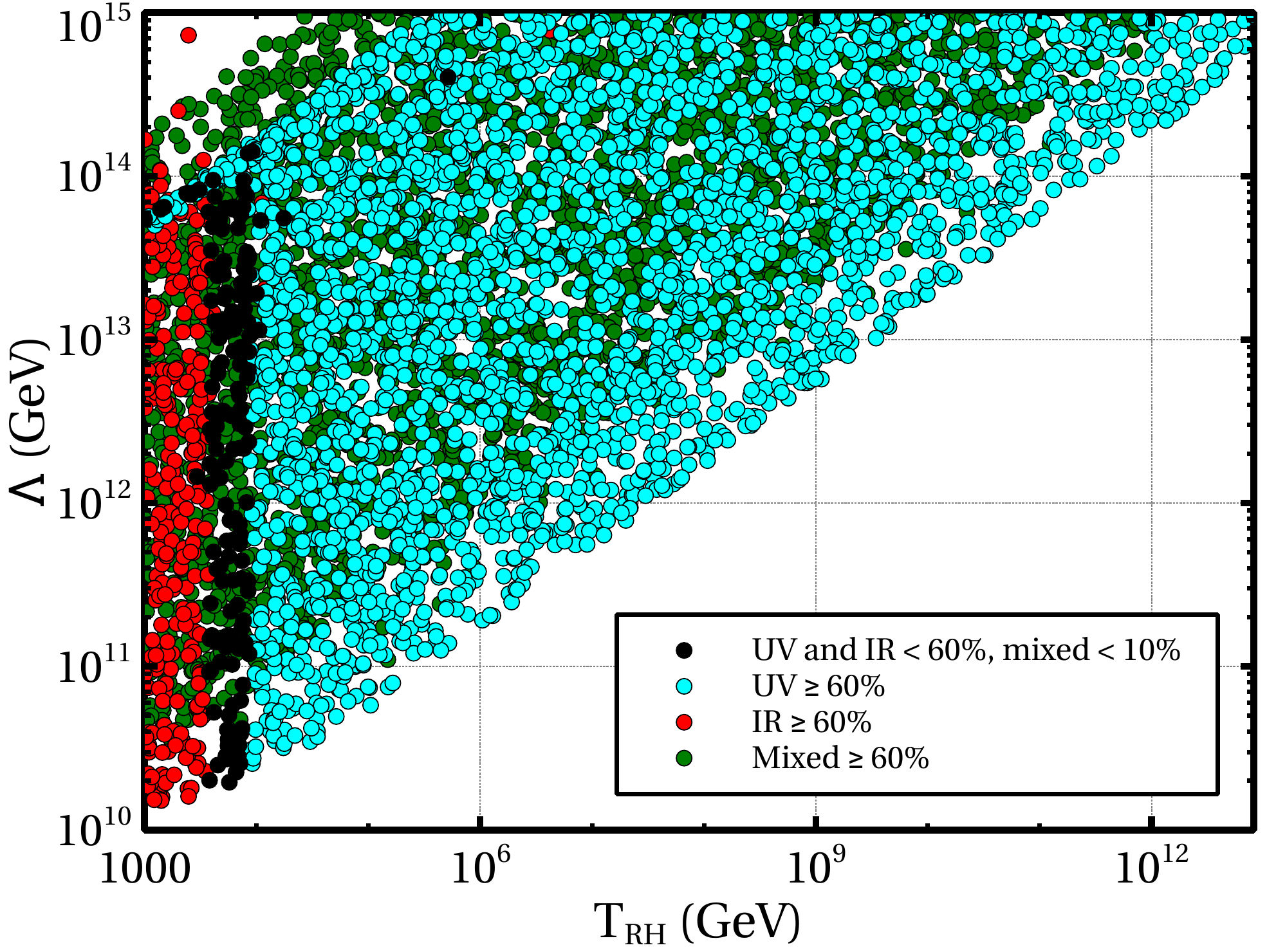}
\includegraphics[height=7cm,width=8.55cm,angle=0]{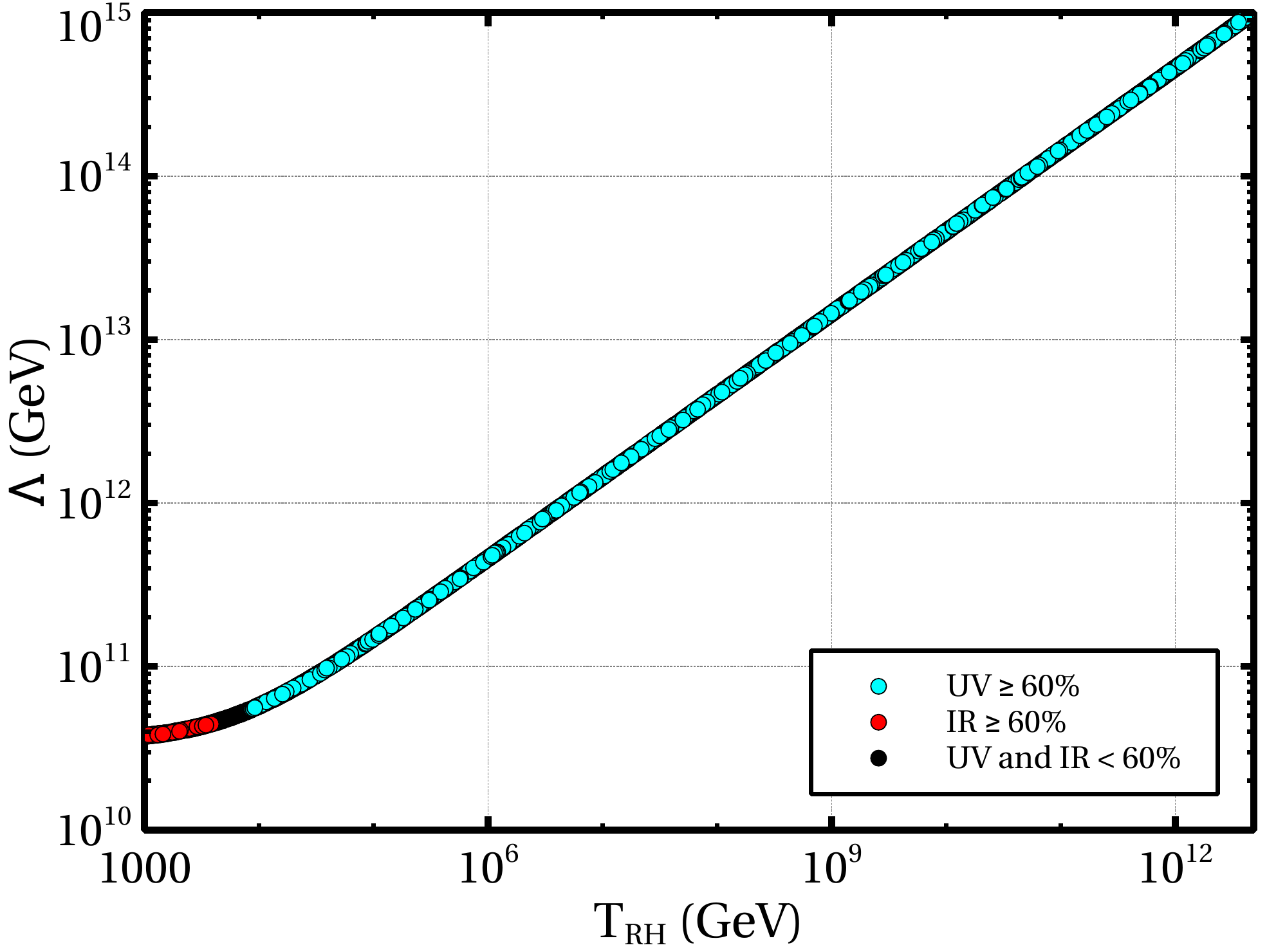}
\caption{Left panel: The $T_{\rm RH}-\Lambda$ parameter space reproducing
correct dark matter relic density via freeze-in mechanism for $m_{\tf}=100$ GeV
and $1\,\text{keV} \leq m_{\x} \leq 100$ GeV. Here cyan, red, green  points indiacate 
fractional contribution of UV, IR and mixed freeze-in greater than 60 $\%$  respectively. 
The black points indicate that the fractional contributions both of UV and IR freeze-in less than 60$\%$.
Right panel: Corresponding values of $T_{\rm RH}$ and $\Lambda$ for
$m_{\tf}\simeq m_{\x}=7.06$ keV. Colour codes are same as those in
the left panel.} 
\label{Fig:TRH-vs-Lambda}
\end{figure}

In this section, we present the allowed parameter space that we
have obtained by computing dark matter relic density using
Eqs.\,(\ref{ychiT0}-\ref{omega}) and comparing this with
the reported range of $\Omega_{\rm DM}h^2=0.120\pm0.001$ by the
Planck experiment. In order to do this, we have varied the
independent parameters $\Lambda$, $T_{\rm RH}$, $m_{\x}$,
$g$ and $\lambda$ within the following ranges. 
\begin{eqnarray}
\begin{array}{cccccc}
10^{10}\,\text{GeV}  \le & \Lambda & \le  10^{15}\,\text{GeV}\,\,,\\ 
10^{3}\,\text{GeV} \le & T_{\rm RH}  &\le  10^{13}\,\text{GeV} \,\, ,\\
10^{-6}\,\text{GeV}  \le & m_{\x} & \le  10^{2}\,\text{GeV}\,\, ,\\
10^{-8}\le &g& \le 10^{-2}\,\, ,\\
10^{-12}\le & \lambda &\le  10^{-8}\,\, ,
\label{para-ranges}
\end{array}
\end{eqnarray}
The allowed parameter space in the $T_{\rm RH}-\Lambda$
plane by the relic density constraint is shown in the
left panel of Fig.\,\ref{Fig:TRH-vs-Lambda}.  
In the present work, as mentioned earlier, different types of
freeze-in mechanisms are contributing to the relic density
of our dark matter candidate $\x$. 
From this figure it 
is clearly visible that UV freeze-in contributes maximally for 
large reheat temperature i.e. $T_{\rm RH} > 10^4$ GeV. This can be
understood easily from the first term of Eq.\,(\ref{ychiT0}), where we
see that the UV effect on $Y_{\x}(T_0)$ is proportional to $T_{\rm RH}$.
Moreover, the contribution of UV freeze-in to $Y_{\x}(T_0)$ has a $1/\Lambda^2$
suppression, as a result one needs larger $\Lambda$ for higher $T_{\rm RH}$
such that $\Omega_{\x} h^2$ does not exceed $1\sigma$ allowed range of
$\Omega_{\rm DM} h^2$.
We already know that
IR freeze-in can occur from both decay and scattering of the parent particles. 
For low reheat temperature ($T_{\rm RH}<10^{4}$ GeV), scatterings
of $W$, $Z$, $h$ and decay of the Higgs boson are the dominant sources of
IR freeze-in. An additional source of significant production of $\x$ is via 
mixed freeze-in.
As a result, $\tf$ production
from the scatterings involving top quark and subsequent decay of $\tf$ into 
$\bar{\x} \x$ pair gives substantial contribution to
$\Omega_{\x} h^2$. 
Furthermore, the variation of $\Lambda$ for a particular value of $T_{\rm RH}$
is mostly due to $g$ and $m_{\x}$ where the latter is varying
between 1 keV to 100 GeV. The same parameter
space for $m_{\tf}\simeq m_{\x}=7.06$ keV is shown\footnote{In Section \ref{indirect_sig},
it will be clear that in order to address the 3.53 keV X-ray line
from the centre of our Milky Way galaxy we need $m_{\tf} \simeq
m_{\x}=7.06$ keV.} in the right panel of Fig.\,\ref{Fig:TRH-vs-Lambda}. 
Here we get UV freeze-in dominance for higher values of $T_{\rm RH}$
and $\Lambda$ whereas for low reheating temperature ($T_{\rm RH}<10^4$ GeV)
IR freeze-in becomes superior. The absence of mixed freeze-in
of $\x$ is due to the fact that in this case the decay 
of $\tf$ to ${\x}\bar{\x}$ pair is kinematically
forbidden. On top of that another noticeable difference is that instead
of getting an allowed region in the $T_{\rm RH}-\Lambda$
plane here we get a line and this is mainly due to the reason that
in the present plot we have kept $m_{\x}$ fixed at $7.06$ keV.
\section{Possibility of $\tf$ as one of the dark matter components}
\label{phitDM}
In this section, we discuss about the possibility of having $\tf$
as another dark matter component besides $\x$. Since $\tf$ has interaction
with photons, it can be produced thermally from the inelastic scattering
between any charged fermions ($f$) and $\gamma$ at the early Universe. Such type of scatterings
$f+\gamma\rightarrow f + \tf$ are known as the Primakoff process, where
at the final state from each scattering a $\tf$ and a charged fermion
are produced. Moreover, as we have interactions like $\bar{f}\gamma_5 f \tf$
with $f$ being any SM fermion, the Primakoff scattering can also be
possible here through an $s$-channel mediator $f$ after EWSB.
However, extremely suppressed coupling (a ratio between the SM Yukawa coupling and
the new physics scale $\Lambda$) of $\bar{f}f$ with
$\tf$ makes such processes insignificant\footnote{Although top
quark has Yukawa coupling order unity, its number density
after EWSB is Boltzmann suppressed making top quark
contribution also insignificant.}. On the other hand, in the
present scenario, $\tf$ can also be produced from the Primakoff
like processes where photons are replaced by other SM gauge
bosons $Z$ and $W^{\pm}$\footnote{For $W^{\pm}$ one has to change the final
state fermion accordingly to maintain electromagnetic charge
conservation.}. Therefore, following the procedure given in
\cite{Cadamuro:2011fd}, we have found that the freeze-out
temperature $T_f$ of $\tf$ always remains larger than 100 GeV as long
as $\Lambda\gtrsim10^9$ GeV. This implies $\tf$ (of mass
$m_{\tf}\lesssim100$ GeV) freezes-out relativistically when
$\Lambda\gtrsim10^9$ GeV, which is the range of $\Lambda$
we are considering in this work and also is allowed from various
astroparticle physics experiments \cite{Cadamuro:2011fd}.  
As a result of this relativistic freeze-out, 
once we know the freeze-out temperature $T_f$, the
number density of $\tf$ at the present epoch becomes
fixed from the following relation as 
\bea
n_{\tf} (T_0) = \dfrac{n_{\gamma}(T_0)}{2}\dfrac{g_s(T_0)}{g_s(T_f)} \,,
\eea 
where $n_{\gamma}(T_0)$ is the number density of $\gamma$ at
the present epoch, $T_0=2.73$ K being the present average temperature
of the Universe and $g_s(T)$ is the number of relativistic degrees of freedom
present at temperature $T$ which are contributing to the entropy density of the
Universe. Using $n_{\tf}$, now one can calculate the relic density of $\tf$
easily which is given by \cite{Kolb:1990vq}
\bea
\Omega_{\tf} h^2 =0.12\times \left(\dfrac{m_{\tf}}{163\,{\rm eV}}\right)
\times \left(\dfrac{106.75}{g_s(T_f)}\right)\,\,.
\eea
Therefore, $\tf$ with mass larger than 163\,\,eV will overclose the Universe\footnote{
{However, in some non standard scenarios, the present density $\Omega_{\tf} h^2$
will be diluted by a factor $\kappa=\frac{s(T)}{s(T_f)}$
if there is a significant entropy production per comoving volume of
the Universe after decoupling of $\tf$ i.e. for $T<T_f$.}}. 
In order to avoid this unpleasant situation, one needs the freeze-out temperature $T_f$ of $\tf$
to be larger than the reheat temperature ($T_{\rm RH}$) of the Universe such that $\tf$ will
never be produced thermally from the Primakoff process. This condition
is indeed satisfied in the present work since the allowed range of $T_{\rm RH}$
for a particular value of new physics scale $\Lambda$ (see Fig.\,\ref{Fig:TRH-vs-Lambda}
in Section \ref{BE_numeric}) always lies below the freeze-out temperature
$T_f$ of the Primakoff process, which has the following approximate
dependence on $\Lambda$ as $T_f \simeq 1.259\times10^{-16}\,\Lambda^2$ \cite{Cadamuro:2011fd}.

As discussed in Section \ref{uvir}, in the present model,
there are some additional sources of $\tf$ production via UV and IR freeze-in.
However, $\tf$ will be produced dominantly via UV-freeze-in from the processes like
$t\,\bar{t}\rightarrow\Phi\,\tf$, $t\,\Phi\rightarrow t\,\tf$
and $\bar{t}\,\Phi\rightarrow\bar{t}\,\tf$ where the abundance 
of $\tf$ depends on $T_{RH}$ and $\Lambda$. Assuming the mass
of $\tf$ to be $\sim 7$ keV and $m_{\x}\gtrsim m_{\tf}$
(the reason for such a choice will be discussed in the next section)
$\tf$ has only $\gamma\,\gamma$ decay mode available. 
Therefore,the way to make $\tf$ partially stable is by increasing $\Lambda$. 
One can easily check from Eq.\,(\ref{GFF}) that for $\Lambda \geq 10^{12}$ GeV,
the lifetime of $\tf$ becomes larger than the present age of the Universe,
which is $\sim 10^{17}$ s. In that case the contribution of $\tf$ to the total dark matter 
relic density will be comparable to that coming from $\x$.
In such a scenario the value of $\Lambda$ in the range between $10^{12}$ GeV and $10^{17}$ GeV 
is strongly disfavored from\footnote{When
$\tf$ contributes to the entire dark matter relic density, the
allowed value is  $\Lambda \geq 10^{17}$ GeV and the
bound on $\Lambda$ relaxes with $\sqrt{x_{\tf}}$, where $x_{\tf}$
is the factional contribution of $\tf$ to the total
dark matter relic density.}
extragalactic background light (EBL) and X-ray observations \cite{1402.7335}.
Nevertheless, the new physics scale
$\Lambda > 10^{17}$ GeV is allowed for a keV scale $\tf$. However,
for such a large value of $\Lambda$, the abundance of $\tf$ is inadequate
to contribute significantly to the overall dark matter relic density
as the UV production processes of $\tf$ are suppressed by $\Lambda^{-2}$.
It may appear that such a large value of $\Lambda$ will
also affect the abundance of our principal dark matter candidate $\chi$.
One can easily avoid this by assigning two different scales of interactions
associated with $\tilde{\phi}$ and $\chi$ respectively. In this case, the UV production of $\chi$
from the scatterings of the components of $\Phi$ will be sufficient enough
to reproduce the observed dark matter relic density.
\section{Indirect signature of $\x$ via ${\sim\mathbf{3.5}}$ keV X-ray line} 
\label{indirect_sig}  
In the present model, our dark matter candidate $\x$ can annihilate
into a pair of $\tf$, which further decays into $\gamma\gamma$ final state\footnote{
We have checked the s-channel prompt annihilation process 
$\bar{\x} \x \rightarrow \gamma \gamma$ and $\sigma \rm v_{\rm rel}$ 
for this process is very small due to large value of $\Lambda$ 
(assuming the cross-section is not resonantly enhanced).} i.e.
$\bar{\x}{\x}\rightarrow\tf\tf \rightarrow 4 \gamma$. Such cascade annihilation
of $\x$ results in a box shaped diffuse $\gamma$-ray spectrum \cite{Ibarra:2012dw},
with each emitted photon has energy $m_{\tf}/2$ in the rest frame of $\tf$.
The photon energy $E_{\gamma}$ in the laboratory frame where dark matter particle 
$\x$ is assumed to be non-relativistic (i.e.\,\,each intermediate scalar
($\tf$) has energy equal to $m_{\x}$) is 
\bea
E_{\gamma} = \dfrac{m^2_{\tf}}{2\,m_{\x}}\left(1-
\cos \theta\,\sqrt{1-\dfrac{m^2_{\tf}}{m^2_{\x}}}\right)^{-1}\,,
\eea
where $\theta$ is the angle between $\tf$ and $\gamma$ in the laboratory
frame. The maximum and the minimum energies $E^{max}_{\gamma}$
and $E^{min}_{\gamma}$ of $\gamma$ are obtained by putting $\theta = 0$
and $\theta=\pi$ in the above expression. Moreover, as the intermediate
state is a scalar, we will always have an isotropic photon emission
in the rest frame of the scalar and also with respect to the laboratory
frame if the intermediate scalar is non-relativistic.
Thus, the resulting spectrum remains constant 
in energy with two sharp cut-off at
energies $E^{max}_{\gamma}$ and $E^{min}_{\gamma}$
respectively and mimics a ``box-shaped'' spectrum.
The width of the spectrum is given by,
\bea
\Delta{E}&=&E^{max}_{\gamma}-E^{min}_{\gamma}\,,\nn \\
&=& \sqrt{m^2_{\x}-m^2_{\tf}}\,, 
\eea  
which depends on the mass splitting between dark matter $\x$
and pseudo scalar $\tf$. Therefore, the box-shaped photon spectrum
becomes line like if $\x$ and $\tf$ are exactly
degenerate in mass. However, in this case we will have four photons
per $\x$ annihilation with each having energy $m_{\x}/2$,  which is not
the situation when a $\gamma$-ray line spectrum is obtained from a prompt
annihilation of $\x$ and $\bar{\x}$. In the later case, one gets two
photons each having energy $m_{\x}$ from a single $\x\bar{\x}$ annihilation.  

Now, we want to demonstrate that the cascade annihilation of $\x$ into
4\,$\gamma$ final state can explain the long-standing $\sim 3.5$ keV X-ray
line initially observed by the XMM Newton observatory from various
galaxy clusters including Perseus, Centaurus, Coma etc.
and also from the centre of our Milky Way galaxy.
The excess X-ray flux observed from centre of the Milky Way galaxy
within an angle of $14^\prime$ is $\left(29 \pm 5\right)\times 10^{-6}$
cts/sec/cm$^2$ at an energy $E_{\gamma} = 3.539\pm0.011$ keV \cite{Boyarsky:2014ska}.
Therefore, for the rest of this section we have considered $m_{\x}=7.06$ keV
and $\delta = (m_{\x}-m_{\tf})/m_{\x} \sim 10^{-5}$ to match with the
observed line like X-ray spectrum. However, there are strong
bounds on the coupling of a keV scale pseudo scalar $\tf$ with
photons from various astrophysical and cosmological phenomena.
These include bounds \cite{Cadamuro:2011fd, Bauer:2017ris} from
the detection of EBL,
distortion in the cosmic microwave background (CMB) spectrum,
measurement of effective number of relativistic degrees of freedom ($N_{eff}$)
at the time of CMB formation, alternation in the successful prediction
of Helium and Deuterium abundances from Big-Bang-Nucleosynthesis (BBN),
length measurement of the neutrino burst from the SN1987a Supernova,
the energy loss of stars through radiation (Horizontal Branch stars)
etc. The present upper bound on $\tf \gamma\gamma$ coupling,
which is inversely proportional to $\Lambda$ in our case, from all the
above mentioned observations is $C_{\tf \gamma\gamma}\lesssim 10^{-17}$ GeV
for a keV scale pseudo scalar $\tf$ \cite{1402.7335,Cadamuro:2011fd}. 

However, this bound has been derived by considering $\tf$
contributing $100 \%$ to the dark matter relic density. Here as
discussed in the previous section, a keV scale $\tf$ depending upon
its two photon coupling $C_{\tf \gamma\gamma}$ can be stable over the
cosmological time scale and its abundance at the present epoch will be
determined by an interplay between the rate of production of $\tf$
and the rate of decay into $\gamma\gamma$. In our case, $\tf$ is
mostly produced via UV-freeze in\footnote{Thermal production
of $\tf$ via Primakoff process has to be forbidden, otherwise
the thermal abundance of a keV scale $\tf$ would overclose the Universe.} 
from annihilations and scatterings of top quarks and for $\Lambda\gtrsim 10^{12}$ GeV, 
the contributions of $\tf$ and $\x$ are comparable in the relic density.
In this scenario with $\Lambda$ ranging between $10^{12}$ GeV to $10^{17}$ GeV
is disfavoured from EBL and X-ray observations. 
Thus, for $m_{\tf}\simeq m_{\x}$ and $m_{\tf}\sim$ 7 keV, we have
considered the mass scale $10^{10}$ GeV$\leq$ $\Lambda$ $\leq 10^{12}$ GeV.
In this range of $\Lambda$, $\tf$ is not stable over the
cosmological time scale and $\x$ is the only dark matter candidate.

Therefore, we are in a situation where
our dark matter candidate $\x$ can annihilate to produce a pair of
intermediate long lived scalars ($\tf$) and each $\tf$ later on
decays into a pair of $\gamma$ after travelling a certain distance
in the galaxy. To compute X-ray flux in this case, one
has to take into account both annihilation
cross section of $\bar{\x}\x\rightarrow\tf \tf$ as well as
decay width of $\tf \rightarrow\gamma\gamma$ properly.
The differential photon flux from the cascade
annihilation of dark matter is given by, 
\bea
\frac{d\Phi_{\gamma}}{dE_{\gamma}}
&=&2\times \dfrac{1}{4}\,\dfrac{r_{\odot}}{4\pi}
\left(\dfrac{\rho_{\odot}}{m_{\x}}\right)^2
\sigmaVindirect\,\dfrac{dN_{\gamma}}{dE_{\gamma}}\,
J_{\rm eff}\Delta{\Omega}\,,
\label{photon-flux} 
\eea
where, $\rho_{\odot} = 0.3$ GeV/cm$^3$ is the
dark matter density at the solar neighbourhood,
$r_\odot=8.5$ kpc is the distance of the solar location
from the galactic centre and $\Delta{\Omega}$
is the solid angle corresponding to an angle $14^\prime(\sim 0.25^0)$
around the galactic centre. $\sigmaVindirect$ is the
thermally averaged annihilation cross section
for $\bar{\x} \x \rightarrow \tf \tf$. A detailed
derivation of the differential photon flux for
the case of dark matter annihilation into a pair of
long lived intermediate particles has been performed
in Appendix \ref{X-ray_line}. Moreover, as we
are getting a monochromatic photon spectrum,
in the above equation $\dfrac{dN_{\gamma}}{dE_{\gamma}}$
is just a Dirac delta function. The differential flux
formula in this case is similar to the 
photon flux from direct annihilation of dark matter   
except a few noticeable changes which are as follows.
The extra 2 factor in front of the right hand side of Eq.\,(\ref{photon-flux})
is due to the fact that instead of two here we are
getting four photons per annihilation of $\bar{\x} \x$.
However, the most important thing lies within the $J$-factor
where the effect of large lifetime of $\tf$ modifies
the $J$-factor, which is a measure of amount of dark matter
present in the region of interest, into an effective
one that has the following expression
\bea
J_{\rm eff} = \dfrac{1}{\Delta{\Omega}}
\int_{\Delta{\Omega}} d{\Omega} \int_{l.o.s}
\dfrac{dr}{r_{\odot}}
\dfrac{\rho^2_{\rm eff}(x)}{\rho^2_{\odot}}\,,
\label{jeff}
\eea
and the expression of effective density $\rho^2_{\rm eff}$
at point B (see Fig. \ref{Fig:flux_dia} in Appendix \ref{X-ray_line}), 
where $\tf$ decays into two photons, is given in Eq.\,(\ref{rhoeff}). 
We would like to note here that while calculating the $l.o.s$ integration 
in Eq.\,(\ref{jeff}), one has to  use the transformation
$\vec{x}=\vec{r}+\vec{r}_\odot$, where both $\vec{x}$
and $\vec{r}$ are clearly depicted in Fig.\,\ref{Fig:flux_dia}.
From Eq.\,(\ref{photon-flux}), it is clearly seen that the photon
flux depends on the annihilation cross section
$\sigmaVindirect$ and $J_{\rm eff}$. Furthermore, as we can see from 
Eq.(\ref{jeff}) that $J_{\rm eff}$ depends on the effective dark
matter density $\rho_{\rm eff}$ which eventually is
a function of decay length $\lambda_{\tf}$ of $\tf$.
Thus the combined effect of both $\sigmaVindirect$ and $\lambda_{\tf}$
will determine the final photon flux.

\begin{figure}[h!]
\includegraphics[height=6cm,width=11cm,angle=0]{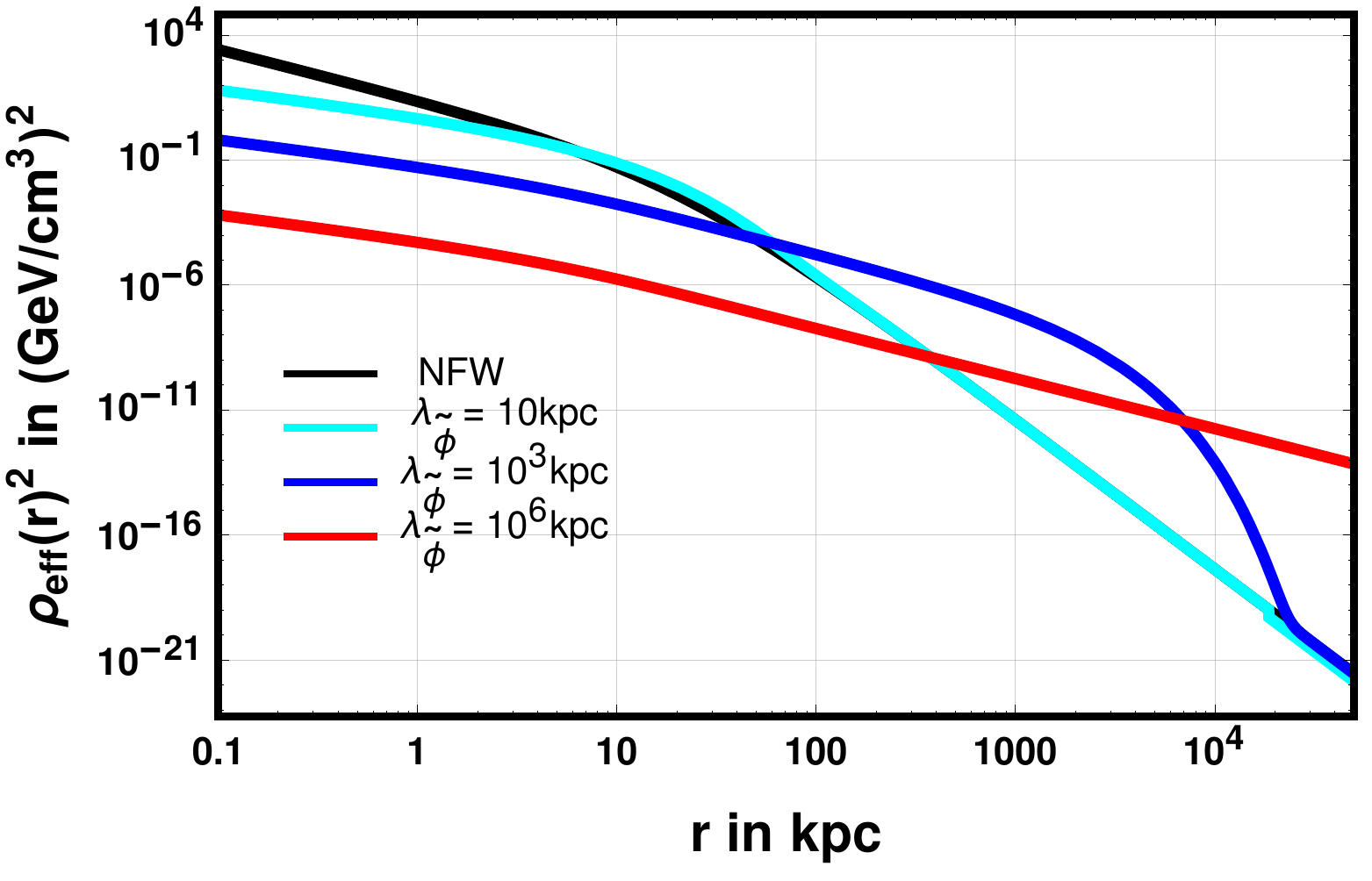}
\caption{Variation of $\rho^2_{\rm eff}$ with respect to the
distance $r$ measured from the centre of Milky Way galaxy for three different
values of decay length $\lambda_{\tf}$. Moreover, for comparison we have
also superimposed the variation of density square for the
Standard NFW profile.}
\label{Fig:rho_eff}
\end{figure}

In Fig.\,\ref{Fig:rho_eff}, we show
the variation of $\rho^2_{\rm eff}$
as a function of distance $r$ from the galactic
centre for three different values of decay length
namely $\lambda_{\tf} = 10$ kpc, $10^3$ kpc and $10^6$
kpc respectively. We have considered the Standard NFW
halo profile \cite{Navarro:1996gj} to describe dark matter distribution
around the galactic centre in Eq.\,(\ref{rhoeff}).
Additionally, to compare the variation
of $\rho_{\rm eff}$ with distance from
the standard case of prompt dark matter annihilation  
we have also plotted the NFW profile
in the same figure using a black solid line. 
From this plot it is seen
that for small decay length i.e. $\lambda_{\tf}=$ 10
kpc (indicated by cyan coloured solid line), the effective
dark matter density closely follows
the NFW profile above a threshold value of $r$ which
depends on decay length $\lambda_{\tf}$ of the intermediate
particle $\tf$. Below the threshold of $r$, the deviation 
of $\rho^2_{\rm eff}$ from the NFW profile increases
as we move towards the galactic centre ($r\rightarrow 0$).
This is due to the fact that for $r < \lambda_{\tf}$, $\tf$
does not have sufficient time to decay into $\gamma\gamma$
final states. Moreover, as $r$ becomes slightly 
larger than the threshold distance the effective
dark matter density overshoots the NFW profile 
and its magnitude increases with $\lambda_{\tf}$. This
is clearly visible by comparing $\rho^2_{\rm eff}$
for $\lambda_{\tf} = 10$ kpc (cyan solid line) and $\lambda_{\tf}=10^3$ kpc
(blue solid line) respectively.   
Finally, for $r>>\lambda_{\tf}$ when almost all $\tf$s
are converted into $\gamma$s, the effective density
reduces to the density given by the Standard NFW
profile.  
\begin{figure}[h!]
\includegraphics[height=8cm,width=12cm,angle=0]{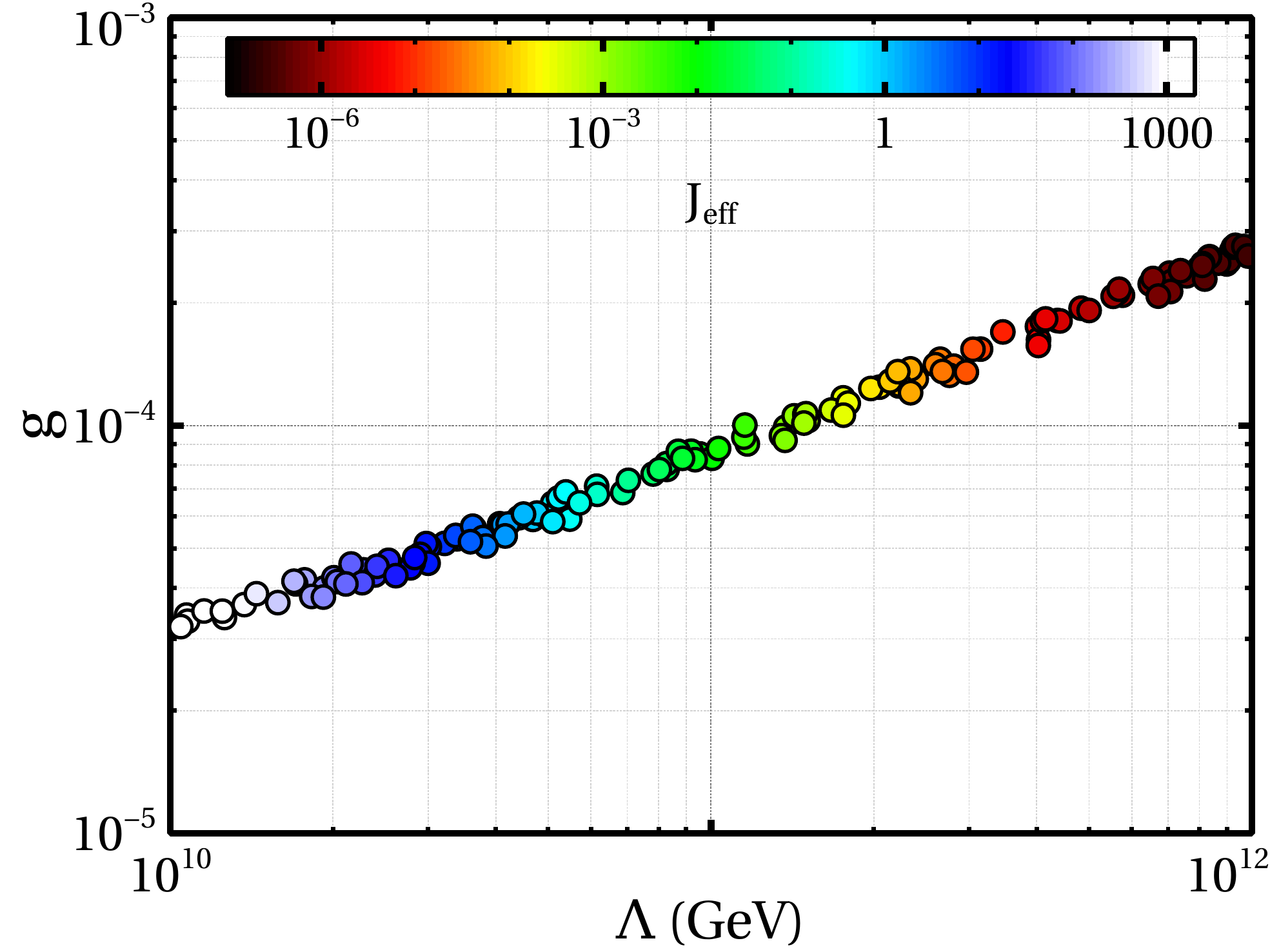}
\caption{Allowed values of $\Lambda$ and $g$ which
reproduce observed X-ray flux ($19\times10^{-6}
\text{cts}/\text{cm}^2/s \leq \Phi_{\gamma}\leq 39
\times 10^{-6}\text{cts}/\text{cm}^2/s$ in 2$\sigma$ range)
from the centre of Milky Way galaxy. The corresponding
values of $J_{\rm eff}$ is illustrated by using
the colour bar.}
\label{Fig:lambda-vs-g}
\end{figure}

Finally, using Eqs.\,(\ref{photon-flux}), (\ref{jeff}) and (\ref{rhoeff}) we have
computed the X-ray flux from cascade annihilation of $\bar{\x} \x$.
As we want to explain the 3.53 keV X-ray line which has been observed
by XMM Newton from the galactic centre \cite{Boyarsky:2014ska}, we consider\footnote{
For prompt annihilation process $\bar{\x} \x \rightarrow \gamma \gamma$, $\sigma \rm v_{\rm rel}\approx 
\mathcal{O}(10^{-47})\text{ cm}^3 \rm s^{-1}$ for $\Lambda=10^{11}\text{ GeV}$ 
and $\delta=5\times10^{-6}$, which is too small to reproduce the observed 
$\sim 3.5 \text{ keV}$ photon flux  \cite{Dudas:2014ixa}.}
$m_{\x}=7.06$ keV and $\delta = 5 \times 10^{-6}$. As a result,
the computed X-ray flux will depend on two unknown parameters namely
the coupling $g$ which enters into the flux formula through
$\sigmaVindirect$ (see Eq.\,(\ref{sigmaVind})) and
the other one is the new physics scale $\Lambda$ which
has a deep impact on $J_{\rm eff}$ through $\rho_{\rm eff}$
(see Eqs.\,(\ref{decay_length}),\,(\ref{rhoeff}) and (\ref{jeff})). The
dependence of $\rho_{\rm eff}$ on $\Lambda$
in Fig.\,\ref{Fig:rho_eff} can be understood from
Eqs.\,(\ref{decay_length}) and (\ref{rhoeff}). Now, in order to reproduce the
observed X-ray flux in $2\sigma$ range i.e. $19 \times 10^{-6}\,
{\rm cts}/{\rm cm}^2/{\rm s} \leq \Phi_{\gamma} \leq 39
\times 10^{-6}\,{\rm cts}/{\rm cm}^2/{\rm s}$ we have varied
both $g$ and $\Lambda$ and the allowed parameter space
in the $\Lambda-g$ plane is shown in Fig.\,\ref{Fig:lambda-vs-g}.
The corresponding values of $J_{\rm eff}$ are depicted
by the colour bar. While calculating $J_{\rm eff}$
we have considered a solid angle $\Delta{\Omega}$
corresponding to an angular aperture of $14^\prime(\sim 0.25^0)$
around the galactic centre, which is equal to half of the
field of view (FOV) of XMM in the energy range 0.15 keV
to 15 keV \cite{Turner:2000jy}. From this plot one can see that
as $\Lambda$ increases we have to increase $g$ also
so that X-ray flux lies within the $2\sigma$ band.
This can be understood in the following way. Any
increment in $\Lambda$ increases the decay length $\lambda_{\tf}$ (or lifetime)
of the intermediate particle $\tf$, which results in a reduction
in $\rho_{\rm eff}$ and hence also in $J_{\rm eff}$.
Consequently, one requires an adequate enhancement in
the annihilation cross section to compensate this
suppression in $J_{\rm eff}$. As both $m_{\tf}$
and $m_{\x}$ are fixed, this can only be
possible by increasing the associated coupling $g$.
Physically this implies that the probability of getting
photons from $\tf$ decays keeps decreasing as $\Lambda$
continues to increase from a lower to higher value. Therefore,
in order to get the same amount of photon flux, the
number density of $\tf$ must increase and for the
present situation that is possible only by increasing
the $\bar{\x}\x\tf$ coupling $g$. 
Finally, for a specific benchmark point one can consider
$\Lambda \simeq 10^{11}$ GeV which corresponds
to $J_{\rm eff} \sim 10^{-2}$. This requires $g \lesssim 10^{-4}$
to get the X-ray flux of energy 3.53 keV within $2\sigma$
error bar.
\section{Conclusion}
\label{conclu}
In this work, we have considered a minimal extension of the Standard Model
by a gauge singlet $\mathbb{Z}_2$-odd fermion $\x$ which couples to
the SM Higgs boson by a dimension five effective operator. As a result,
all the interactions of $\x$ with the SM fields are suppressed by a
large new physics scale $\Lambda$, which naturally ensures $\x$ to be a
non-thermal dark matter candidate. The production of $\x$ at the early
Universe occurs through freeze-in mechanism and depending on the
time of maximum production of dark matter there are two types of
freeze-in namely ultra-violet freeze-in and infra-red freeze-in.
The former case is characterised by the presence of a higher
dimensional interaction and the dark matter comoving number
density generated via UV freeze-in becomes directly proportional
to the reheat temperature ($T_{\rm RH}$), making the dark
sector extremely sensitive to the early history of the Universe.
On the other hand, IR freeze-in does not require any involvement of higher
dimensional operator and occurs mostly around a temperature $T\sim$
mass of the mother particle, when latter is in thermal equilibrium.
Here, $\x\bar{\x}$ pairs before EWSB are produced only from the scatterings
of scalars (both charged and neutral components of the Higgs doublet $\Phi$), 
gauge bosons $W^{a}_{\mu}\,($a=1,\,2,\,3$)$, $B_{\mu}$ and
gluons $G^b_\mu \, (b=1\,....\,8)$ via
UV freeze-in. After EWSB, $\bar{\x} \x$ pairs can also be produced
from IR freeze-in, where scatterings of electroweak gauge bosons ($W^{\pm}_{\mu}$,
$Z_{\mu}$, $G^a_\mu$), SM fermions and both decay as well as scattering of SM
Higgs boson are dominant sources of dark matter production. In addition,
the decay of $\tf$ to $\bar{\x} \x$ is also a dominant
source of ${\x}$ in both before and after EWSB, where
the parent particle $\tf$s are produced copiously from the
scattering involving top quarks in UV regime, which is 
the mixed freeze-in of $\x$.
In the
present scenario, we have solved the Boltzmann equation of $\x$
to compute its relic density considering all possible dark matter
production processes from the thermal plasma. We have presented our
result in $\Lambda-T_{\rm RH}$ plane, which is allowed by the
relic density constraint of dark matter. We have found that for the
lower end of this plane where $T_{\rm RH}\lesssim10^4$ GeV, the maximum fraction
of dark matter is produced by IR and mixed freeze-in while
UV and mixed freeze-in are dominant for $T_{\rm RH}$ beyond $10^{4}$ GeV.  
   
Moreover, we have also presented a detailed discussion on
the indirect signature of our non-thermal dark matter
$\x$ in the light of the unexplained $\sim 3.5$ keV X-ray line from
various galaxies including our own Milky Way galaxy and also from
galaxy clusters. In order to do this, we have further introduced a SM
gauge singlet pseudo scalar $\tf$ in the particle spectrum. This $\tf$
becomes a long lived particle in keV scale as in this mass range
$\tf$ has only $\gamma\gamma$ decay mode which is extremely suppressed
by the large $\Lambda^2$. Furthermore,
this keV scale $\tf$ with a new physics scale $\Lambda\gtrsim10^{12}$ GeV
will be partially stable over the cosmological scale since it has lifetime
greater than $10^{17}$ s, the present age of the Universe. 
However, in our model $\Lambda$ lying between $10^{12}$ GeV to $10^{17}$ GeV 
is disfavored from EBL and X-ray observations.
Therefore, in the present model for $\Lambda \leq 10^{12}$ GeV we have 
only one dark matter candidate and our dark matter $\x$ of mass 
$\sim7$ keV pair annihilates into a pair
of long lived $\tf$ and each $\tf$ thereafter
decays into two photons. Generally, this type
of cascade annihilation results in a box shaped
spectrum which turns into a line like X-ray spectrum
when $m_{\x}$ and $m_{\tf}$ are almost degenerate.
The calculation of photon flux in this situation is distinctly
different from the standard case of prompt dark matter
annihilation, where we have large astrophysical $J$-factor
for a particular solid angle $\Delta\Omega$ around
the region of interest (here galactic centre) and
hence smaller annihilation cross section for the
annihilation channel $\x\bar{\x}\rightarrow \gamma\gamma$
are required.
On the other hand, for this cascade annihilation of
dark matter via an intermediate long lived state,
we have derived necessary analytical expressions
which are required to compute the photon flux. We have
found that due to the presence of long lived intermediate states
all the dark matter particles present around the galactic centre
are not able to produce photons. It depends on decay length
$\lambda_{\tf}$ of the intermediate scalar $\tf$. This is represented
by an effective dark matter density profile $\rho_{\rm eff} (r)$
which depends on the decay length (or lifetime) of $\tf$.
Furthermore, we have noticed that $\rho_{\rm eff}(r)$ (and hence $J_{\rm eff}$)
is suppressed compared to the standard NFW profile for $r\lesssim \lambda_{\tf}$
as we increase the new physics scale $\Lambda$ which is controlling
the lifetime of $\tf$. In this circumstances, we need enhanced
$\x\bar{\x}\rightarrow \tf \tf$ annihilation cross section (or larger
coupling $g$) to produce the observed X-ray from the galactic centre. Finally, we
have identified the allowed parameter space in $\Lambda-g$ plane
which reproduces the XMM Newton observed X-ray flux from
the galactic centre of Milk way in $2\sigma$ range.
\section{Acknowledgements} 
Authors would like to thank Eung Jin Chun for a
very useful discussion on the indirect detection during his
visit at IACS, Kolkata. One of the authors SG would like to
acknowledge University Grants Commission (UGC) for financial
support as a junior research fellow. AB and SG acknowledge the
cluster computing facility at IACS (pheno-server).
\appendix
\section{Pseudo scalar gauge boson vertices after EWSB:}
\begin{figure}[hbt!]
\centering
\subfigure[$A_\mu\, A_\nu\, \tf$ vertex~~~~~~~~~~]{\includegraphics[width=0.3\textwidth]{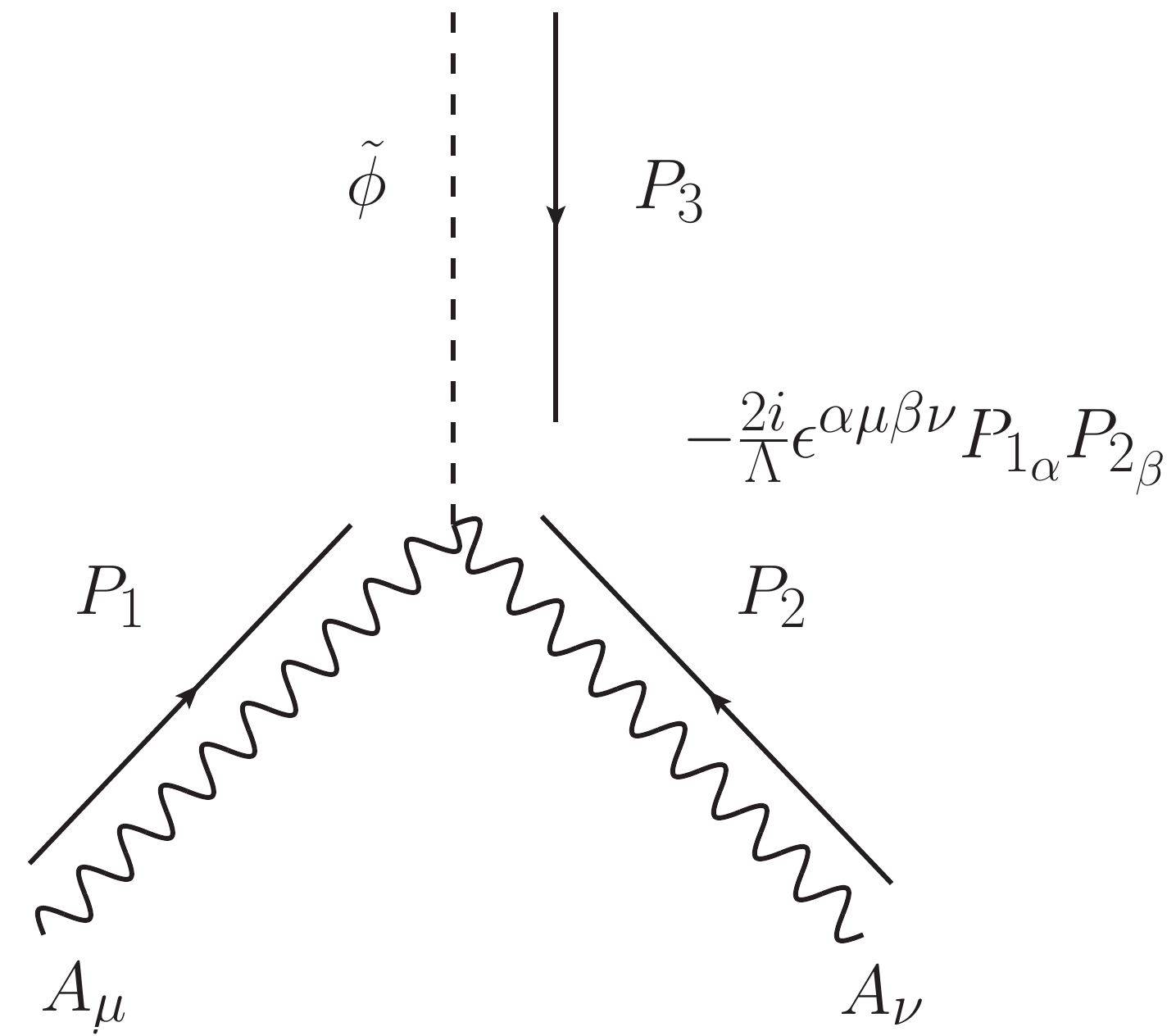}}\label{fig:GBPa}
\subfigure[$W_\mu^+\, W_\nu^-\,\tf$ vertex~~~~~~~~~~] {\includegraphics[width=0.3\textwidth]{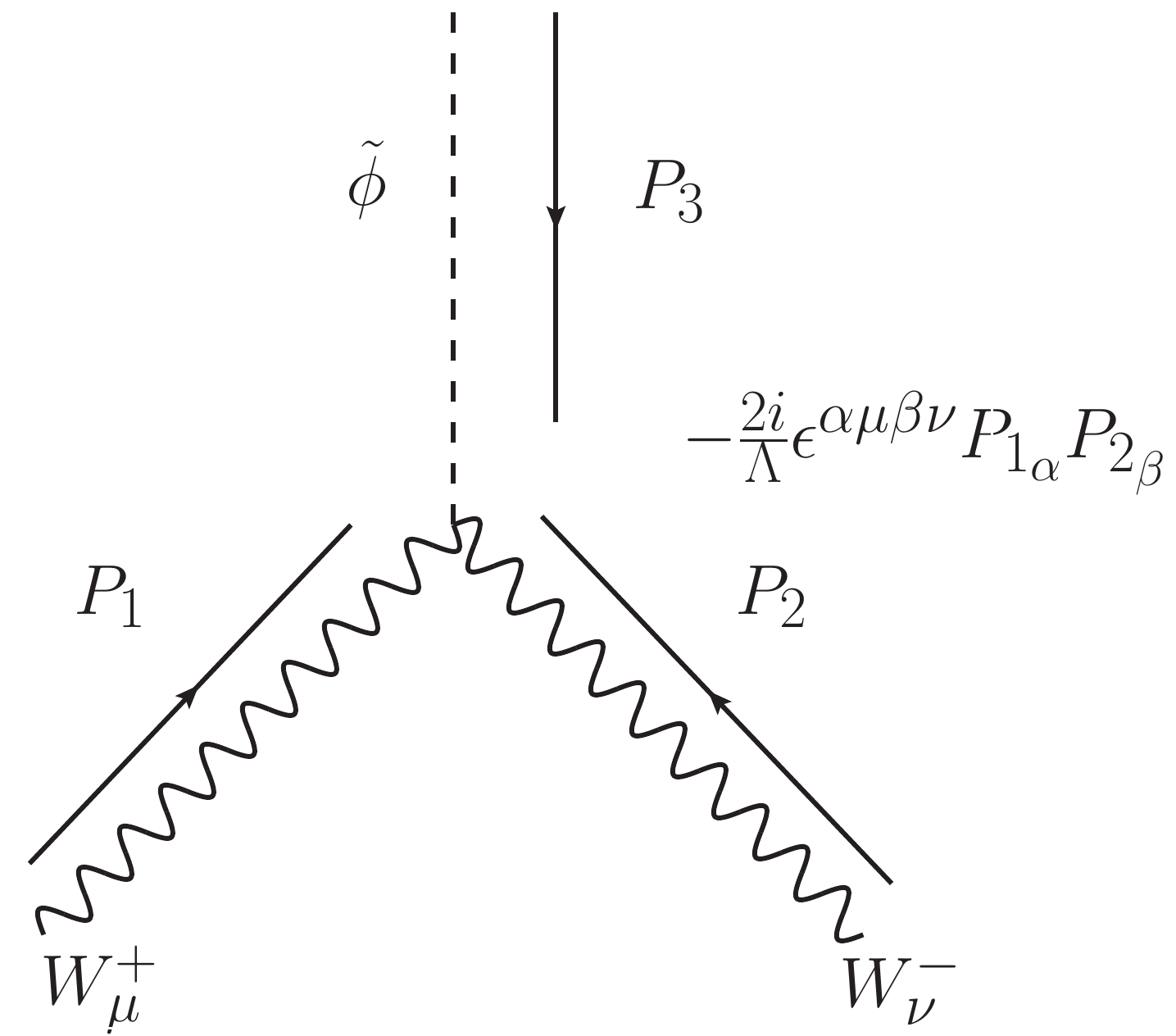}}\label{fig:GBPb}
\subfigure[$Z_\mu\, Z_\nu\, \tf$ vertex~~~~~~~~~~]{\includegraphics[width=0.3\textwidth]{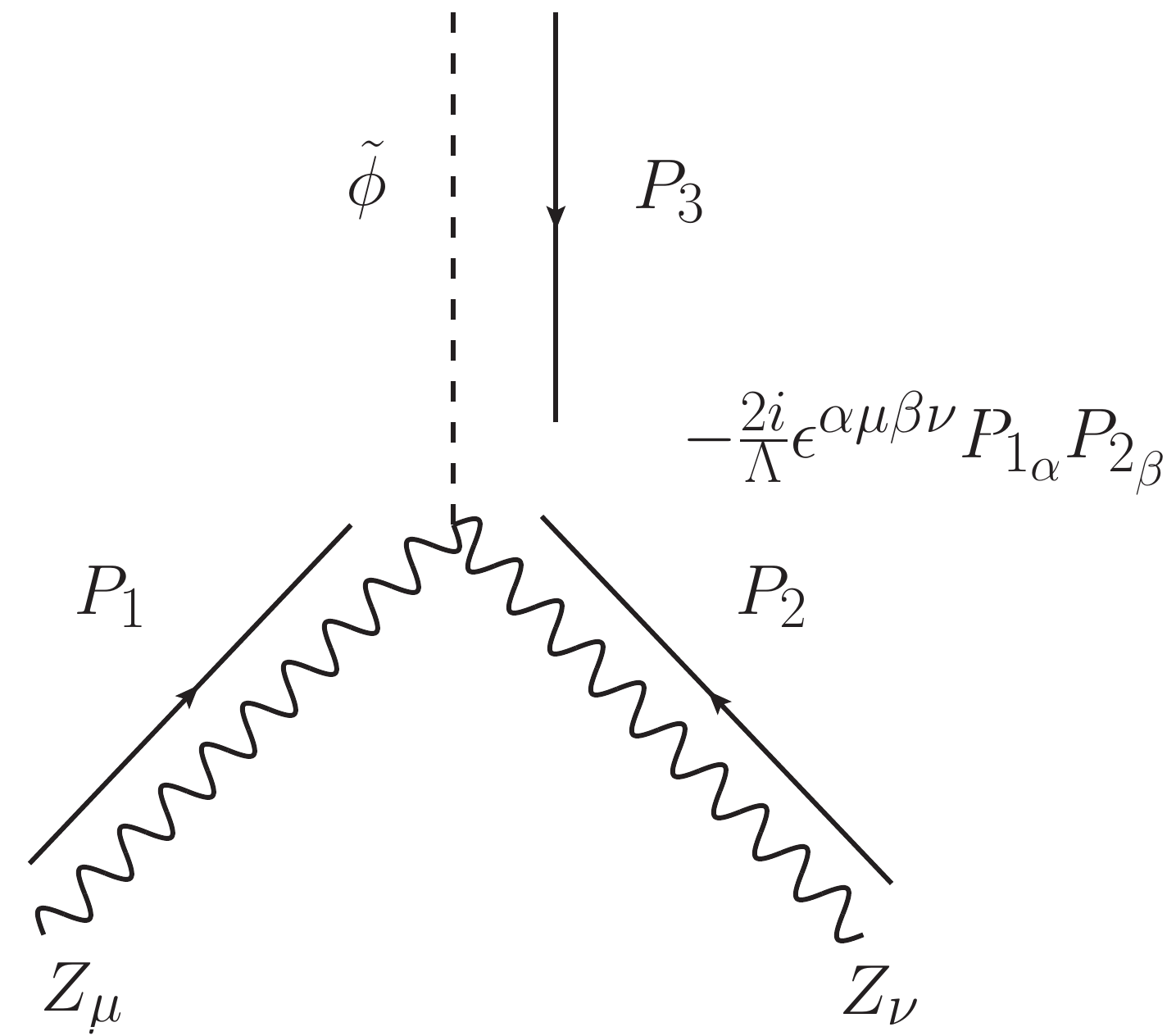}}\label{fig:GBPc}
\caption{Gauge boson pseudo scalar vertices after EWSB} \label{GBP}
\end{figure}

The gauge bosons-pseudo scalar vertices which are genrated after 
electroweak symmetry breaking are shown in Fig.\,\ref{GBP}.
\label{app:vertices}
\section{Calculation of momentum distribution function of $\tilde{\phi}$}
\label{dist_function}
In this section, we have briefly discussed the calculation of momentum distribution function 
$f(p,T)$ of  $\tf$ which is produced from the annihilation of SM particles and $\tf$ can also decay \
to the DM particles. 
The Boltzmann equation at the level of momentum distribution function can be written as
\be
\hat{L}f(p,T)=\hat{\mathcal{C}}[f(p,T)]\,,
\label{BE-dist}
\ee
where $\hat{L}=\dfrac{\partial}{\partial t}-Hp\dfrac{\partial}{\partial p}$,
is the Liouville operator, 
$\hat{\mathcal{C}}[f(p,T)]$ is the collision term and $p\equiv |\vec{p}|$.\\
To simplify the form of Liouville operator, we have made the following
variable transformations as in Ref.\,\cite{Biswas:2016iyh,Konig:2016dzg}
\begin{align}
r &= \dfrac{M_0}{T} \,.\\
\xi & =\left(\dfrac{g_{s}(T_0)}{g_{s}(T)}\right)^{1/3}\dfrac{p}{T}\,.
\label{variable trans}
\end{align} 
Where $M_0$ and $T_0$ are some reference mass and temperature scale respectively.
Using the above transformations, Liouville operator takes the following form
\be
\hat{L}=rH\left(1+\dfrac{1}{3}\dfrac{d\,(\ln \,g_{s}(T))}
{d \,(\ln T)}\right)^{-1}\dfrac{d}{d r}\,.
\label{modified L}
\ee
Now substituting this expression of $\hat{L}$ in Eq.(\ref{BE-dist}), we can write
\be
rH\left(1+\dfrac{1}{3}\dfrac{d\,(\ln \,g_{s}(T))}{ d \,(\ln T)}\right)^{-1}\dfrac{d f(\xi,r)}{d r}\,
=\, \hat{\mathcal{C}}[f(\xi,r)]_{\rm prod.}\,+\,\hat{\mathcal{C}}[f(\xi,r)]_{\rm decay}\,,
\label{final BE}
\ee
where $\hat{\mathcal{C}}[f(\xi,r)]_{\rm prod.}$ and $\hat{\mathcal{C}}[f(\xi,r)]_{\rm decay}$ are the collision terms
for the production of $\tf$ (from annihilations and scatterings of top quark) and decay of $\tf$ respectively and the analytical forms are given below
\begin{align}
\hat{\mathcal{C}}[f(\xi\,r)]_{\rm prod.}\,=&\, \dfrac{3}{8\pi^3\Lambda^2}\left(\dfrac{m_t}{v}\right)^2
\left(\dfrac{M_0}{r}\right)^3\exp\left[-\xi \,\beta(r)\right]\,\,\nn,\\
\hat{\mathcal{C}}[f(\xi,r)]_{\rm decay}\,=&\, -\dfrac{m_{\tf}\,r}{M_0\sqrt{\xi^2 \beta(r)^2+
\left(\dfrac{m_{\tf}\,r}{M_0}\right)^2}} f(\xi,r) \Gamma_{\tf\rightarrow \bar{\x} \x}\,\,,
\label{collision term}
\end{align}
where $\beta(r)= \left(\dfrac{g_{s}(M_0/r)}{g_{s}(M_0/r_{in})}\right)^{1/3}$. 
Using Eq.(\ref{collision term}), we have solved Eq.(\ref{final BE}) numerically
to calculate the momentum distribution function of $\tf$.

\section{Calculation of the collision term for $2\rightarrow3$ and $2\rightarrow4$ processes }
\label{n_body_scat}
In this section,we have derived the collision term for $2\rightarrow3$ and $2\rightarrow4$ processes which 
are denoted by $\mathcal{F}_{2\rightarrow 3}(T)$ and $\mathcal{F}_{2\rightarrow 4}(T) $ respectively in Eq.(\ref{BE-main}).
The forms of $\mathcal{F}_{2\rightarrow 3}(T)$ and $\mathcal{F}_{2\rightarrow 4}(T)$ are given by 
\begin{align}
\mathcal{F}_{2\rightarrow3}(T)=&\int \prod_{p_i=1}^2 d\Pi_{p_i} \prod_{k_i=1}^3 d\Pi_{k_i}
\left(2\pi\right)^4
\delta^4 \left(\sum_{i=1}^2 p_i-\sum_{i=1}^3 k_i\right)\overline{|\mathcal{M}|}^2_{2\rightarrow3}
\exp\left(-\dfrac{E_1+E_2}{T}\right)
\label{F1}
\end{align}
\begin{align}
\mathcal{F}_{2\rightarrow4}(T)=&\int \prod_{p_i=1}^2 d\Pi_{p_i} \prod_{k_i=1}^4 d\Pi_{k_i} \left(2\pi\right)^4
\delta^4 \left(\sum_{i=1}^2 p_i-\sum_{i=1}^4 k_i\right)\overline{|\mathcal{M}|}^2_{2\rightarrow4}
\exp\left(-\dfrac{E_1+E_2}{T}\right)
\label{F2}
\end{align}
and $$\int d\Pi_i=g_i\int \dfrac{d^3 \vec{p_i}}{(2\pi)^3 2E_i}.$$
To simplify Eq.(\ref{F1}), we have decomposed the 3 body phase space into two 2 body
phase space and the phase space integral is
\be
\Phi_3(s)=\int_{(m_1+m_2)^2} ^{(\sqrt{s}-m_3)^2} \dfrac{dX^2}{(2\pi)}\,\Phi_2(X^2,\,m_1^2,\,m_2^2)\,\Phi_2(s,\,m_3^2,\,X^2)
\label{dlips3}
\ee
and 
\be
\Phi_2(s,\,m_i^2,\,m_j^2)= \dfrac{1}{8\pi}\sqrt{1+\dfrac{m_i^4}{s^2}+\dfrac{m_j^4}{s^2}-\dfrac{2m_i^2}{s}
-\dfrac{2m_j^2}{s}-\dfrac{2 m_j^2m_i^2}{s^2}}\int\dfrac{d \Omega_{ij}}{4\pi}\nn
\label{dlips2}
\ee
Using Eq.(\ref{dlips3}), Eq.(\ref{F1}) for $2\,\rightarrow\,3$ processes 
($t\,\bar{t}\rightarrow \x\,\bar{\x}\,\Phi$, $t\,\Phi\rightarrow \x\,\bar{\x}\,t$, 
$\bar{t}\,\Phi\rightarrow \x\,\bar{\x}\,\bar{t}$) can be written as
\bea
\mathcal{F}_{2\rightarrow 3}(T)=\dfrac{\alpha\, T}{(2\pi)^4}
\int_0^\infty \dfrac{\sqrt{s}}{4} \left[\mathcal{A}^{\rm s-channel}_{2\rightarrow 3}(s)
+ 2\,\mathcal{A}^{\rm t-channel}_{2\rightarrow 3}(s)\right]
{\rm K}_1\left(\dfrac{\sqrt{s}}{T}\right)ds\,,
\label{berhs23}
\eea
where $\mathcal{A}^{\rm s-channel}_{2\rightarrow 3}(s)$
and $\mathcal{A}^{\rm t-channel}_{2\rightarrow 3}(s)$ have the following
expressions,
\bea
\mathcal{A}^{\rm s-channel}_{2\rightarrow 3}(s) &=& 
{s \,\ln}\,\left(\dfrac{s}{4 m_{\x}^2}\right)- s\,, \nn \\
\mathcal{A}^{\rm t-channel}_{2\rightarrow 3}(s) &=&
\dfrac{1}{4} \left({s \,\ln}\,\left(\dfrac{s}{4 m_{\x}^2}\right)-\dfrac{3}{2}s \right)\,.
\eea
Here $\alpha=\dfrac{1}{32\,\pi^3\,N_c}\,\left(\dfrac{g}{\Lambda} \dfrac{m_t}{v}\right)^2$
and $N_c=3$ is the color factor for top quark.\\
\vspace{0.1cm}\\
Similarly, for $2\,\rightarrow \,4$ process
($\Phi\,\Phi\,\rightarrow\,\x\,\bar{\x}\,\x\,\bar{\x}$)
\be
\mathcal{F}_{2\rightarrow 4}(T)=\dfrac{\pi^2 T}{(2\pi)^6}\int_0^\infty\sqrt{s}\,
{\rm K}_1\left(\dfrac{\sqrt{s}}{T}\right)\mathcal{A}(s) ds\,\,\,,
\label{berhs24}
\ee
where,
\be
\mathcal{A}(s)=\dfrac{\lambda^2g^4}{8(2\pi)^3}\int_0^s \dfrac{dX^2}{2\pi}\dfrac{1}{X^2}
\int_0^{(\sqrt{s}-X)^2} \dfrac{dY^2}{2\pi}\dfrac{1}{Y^2}\sqrt{1+\dfrac{X^4}{s^2}+\dfrac{Y^4}{s^2}-\dfrac{2X^2}{s}
-\dfrac{2Y^2}{s}-\dfrac{2 X^2Y^2}{s^2}}\,\,\,.\nn
\ee
Using Eqs.(\ref{berhs23}) and (\ref{berhs24}), one can solve Eq.(\ref{BE-main}) to calculate the 
co-moving number density of $\x$ from these $2\rightarrow3$ and $2\rightarrow4$ processes.
\section{Boltzmann equation for freeze-in via scattering}
\label{app:BE-scatt}
In this section, first we have derived a general solution of the Boltzmann equation
of a dark matter candidate $\x$ produced at the early Universe from scattering
of bath particles. Such production process via freeze-in mechanism
has both UV and IR contributions and particularly it depends
on the type of interactions that our dark matter candidate
$\x$ has with other particles present in the Universe.  
After deriving a general expression of comoving number density ($Y_{\x}$)
of $\x$ for freeze-in, we have given a more simplified
expression of $Y_{\x}$ for the case of UV freeze-in. 
As we have already known that for UV freeze-in one needs higher dimensional
effective interactions between $\x$ and other bath particles hence, we will consider
a dimension five interaction $\dfrac{\Phi^{\dagger}\Phi\,\overline{\x}{\x}}{\Lambda}$
between $\x$ and Higgs doublet $\Phi$ (the first term of Eq.\,(\ref{L-UV}))
as an example. Once we derive an expression of $Y_{\x}$ for this
dimension five operator, it can easily be generalised for any higher
dimensional operator. Let us assume that our dark matter candidate $\x$ is produced
at the early Universe from annihilations of $\Phi$
and $\Phi^\dagger$ (depicted in Fig.\,\ref{fig1}).\,\,To calculate the
net number density (hence relic density) of $\x$ generated from
$\Phi^{\dagger} \Phi \rightarrow \overline{\x}\x$\footnote{Before EWSB, both
components of the doublet $\Phi$ are physical fields and can
be represented by two complex scalar fields $\phi^\pm$ and $\phi^0 \equiv\dfrac{h+i\,a}{\sqrt{2}}$
. However, after EWSB only the real part of $\phi^0$ ($h$) remains
as a physical scalar and other two ($\phi^{\pm}$, $a$) become massless
Goldstone bosons. In generic notation, we denote everything by
$\Phi$ only.},
one needs to solve the Boltzmann equation which is given by,
\begin{equation}
\frac{dn_\x}{dt}+3Hn_\x\simeq\int d\Pi_{\Phi} d\Pi_{\Phi^{\dagger}} d\Pi_{\x} d\Pi_{\overline{\x}}
\Bigg[f^{\rm eq}_{\Phi}(\vec{p}_1,T)f^{\rm eq}_{\Phi^\dagger}(\vec{p}_2,T)
\overline{\left|\mathcal{M}\right|}^2_{\Phi^{\dagger} \Phi \rightarrow \overline{\x}\x}
(2\pi)^4 \delta^4(P_1+P_2-P_3-P_4)\Bigg] \,,
\label{eq7}
\end{equation}
where $P_1$, $P_2$, $P_3$ and $P_4$ are the four
momenta of $\Phi$, $\Phi^\dagger$, $\x$ and $\overline{\x}$
respectively while the corresponding three momenta are denoted
by $\vec{p_i}$s. Besides, $d\Pi_i \equiv g_i\frac{d^3\vec{p_i}}{(2\pi)^3 2E_i}$
is invariant under the Lorentz transformation and $g_i$ is the number of internal
degrees of freedom of the particle having three momentum $\vec{p_i}$
while $\overline{\left|\mathcal{M}\right|}^2_
{\Phi^{\dagger} \Phi \rightarrow \overline{\x}\x}$
is the Lorentz invariant matrix amplitude square for the process
${\Phi^{\dagger} \Phi \rightarrow \overline{\x}\x}$ and
it is averaged over spins of initial and final state particles.   
Here we have assumed that all the particles
except $\x$ are in thermal equilibrium and they obey Maxwell
Boltzmann distribution. Moreover, as the initial
number densities of $\x$ and $\overline{\x}$ are negligible compared to those
of $\Phi$ and $\Phi^\dagger$, we have neglected the back reaction
term in the right hand side of the Boltzmann equation. To proceed
further, let us define a dimensionless quantity $Y_{\x}$ which is
known as the co-moving number density $\x$ and it is defined as
$Y_{\x}(T)=\frac{n_\x(T)}{s(T)}$, where $s(T)$ is the entropy
density of the Universe at temperature $T$. In terms of $Y_{\x}(T)$
the left hand side of Eq.\,(\ref{eq7}) takes the following form
\be
s \frac{dY_{\x}}{dt}\simeq
\int d\Pi_{\Phi} d\Pi_{\Phi^{\dagger}} d\Pi_{\x} d\Pi_{\overline{\x}}
\Bigg[f^{\rm eq}_{\Phi}(\vec{p}_1,T)f^{\rm eq}_{\Phi^\dagger}(\vec{p}_2,T)
\overline{\left|\mathcal{M}\right|}^2_{\Phi^{\dagger} \Phi \rightarrow \overline{\x}\x}
(2\pi)^4 \delta^4(P_1+P_2-P_3-P_4)\Bigg] \,.
\label{eq8}
\ee
Now, using the time-temperature relation for
the radiation dominated era $\frac{dT}{dt}\simeq-HT$, the above
equation can be expressed as  
\be
\frac{dY_{\x}}{dT}\simeq-\frac{1}{sHT}
\int d\Pi_{\Phi} d\Pi_{\Phi^{\dagger}} d\Pi_{\x} d\Pi_{\overline{\x}}
\Bigg[f^{\rm eq}_{\Phi}(\vec{p}_1,T)f^{\rm eq}_{\Phi^\dagger}(\vec{p}_2,T)
\overline{\left|\mathcal{M}\right|}^2_{\Phi^{\dagger} \Phi \rightarrow \overline{\x}\x}
(2\pi)^4 \delta^4(P_1+P_2-P_3-P_4)\Bigg]\,,
\label{eq9}
\ee
where $H$ being the Hubble parameter. Now, using the definition of
cross section $\sigma$ for ${\Phi^{\dagger} \Phi \rightarrow \overline{\x}\x}$
the collision term of the Boltzmann equation further simplifies to
\be
\frac{dY_{\x}}{dT}\simeq-\frac{1}{sHT}
\int d\Pi_{\Phi} d\Pi_{\Phi^{\dagger}} 
4\,E_{\Phi}\,E_{\Phi^\dagger}
\sigma\,{\rm v_{\rm rel}}\,f^{\rm eq}_{\Phi}(\vec{p}_1,T)
\,f^{\rm eq}_{\Phi^\dagger}(\vec{p}_2,T)
\,.
\label{eq10}
\ee
The integration in the right hand side is nothing but a product of two
well known quantities $\langle \sigma\,{\rm v_{\rm rel}}
\rangle_{\Phi^{\dagger} \Phi \rightarrow \overline{\x}\x}$ and
$n^{\rm eq}_{\Phi} n^{\rm eq}_{\Phi^\dagger}$, where thermal averaged
cross section and equilibrium number densities are defined as
\be
\langle \sigma\,{\rm v_{\rm rel}} \rangle_{\Phi^{\dagger}
\Phi \rightarrow \overline{\x}\x} = \dfrac{
\int d\Pi_{\Phi} d\Pi_{\Phi^{\dagger}} 
E_{\Phi}\,E_{\Phi^\dagger}
\sigma\,{\rm v_{\rm rel}}\,f^{\rm eq}_{\Phi}(\vec{p}_1,T)
\,f^{\rm eq}_{\Phi^\dagger}(\vec{p}_2,T)}{\int d\Pi_{\Phi} d\Pi_{\Phi^{\dagger}} 
E_{\Phi}\,E_{\Phi^\dagger}
f^{\rm eq}_{\Phi}(\vec{p}_1,T)
\,f^{\rm eq}_{\Phi^\dagger}(\vec{p}_2,T)}\,,
\label{sigmaVav}
\ee
and 
\be
n^{\rm eq}_{\Phi(\Phi^\dagger)} = 2\int d\Pi_{\Phi(\Phi^\dagger)}
E_{{\Phi}(\Phi^\dagger)}\,f^{\rm eq}_{\Phi(\Phi^\dagger)}(\vec{p}_{1(2)},T)\,.
\label{neq}
\ee
Therefore, using Eqs.\,(\ref{sigmaVav}, \ref{neq}), the Boltzmann equation
can be written in a more compact notation as
\be
\frac{dY_{\x}}{dT}\simeq-\frac{1}{sHT}\,n^{\rm eq}_{\Phi}\,n^{\rm eq}_{\Phi^{\dagger}}
\,\langle \sigma\,{\rm v_{\rm rel}}
\rangle_{\Phi^{\dagger} \Phi \rightarrow \overline{\x}\x}\,.
\label{BEfrinsc}
\ee  
If we assume there is no asymmetry in the number densities of $\Phi$
and $\Phi^\dagger$  then $n^{\rm eq}_{\Phi}=n^{\rm eq}_{\Phi^\dagger}$.
Also, we have neglected the effect of change in number density of $\x$
from the inverse process i.e. $\overline{\x}\x\rightarrow \Phi\Phi^\dagger$
due to insufficient number densities of $\x$ and $\overline{\x}$. Otherwise,
we will have a term with a +ve sign and proportional to $n_{\x}\,n_{\overline{\x}}$
in the right hand side of the above equation. If one incorporates these two
things then the above equation will reduce to the more familiar form
of the Boltzmann equation. Now, following the mathematical steps
given in \cite{Gondolo:1990dk} for the calculation of thermal averaged
cross section, one can further express the right hand side as
\be
\frac{dY_{\x}}{dT}\simeq-\frac{1}{sHT}\,\,\dfrac{g_{\Phi}^2\,T}
{16\,\pi^4} \int^{\infty}_{4\,m^2_{\Phi}} d\hat{s}\,\sigma\,F \sqrt{\hat{s}-4\,m^2_{\Phi}}\,{\rm K_1}
\left(\frac{\sqrt{\hat{s}}}{T}\right)\,.
\label{BERHSsigma}
\ee 
Where $g_{\Phi}$ and $m_{\Phi}$ are the internal degrees of
freedom and mass of $\Phi$ and ${\rm K_1}
\left(\frac{\sqrt{s}}{T}\right)$ is the first order Modified
Bessel function of second kind. The quantity $F$
is related to the initial state momentum ($\vec{\mathfrak{p}}_i$) in 
centre of momentum frame as $F= |\vec{\mathfrak{p}}_{i}|\sqrt{\hat{s}}$
with $\sqrt{\hat{s}}$ being the total energy of scattering
in centre of momentum frame. Moreover,
by using the standard expressions of differential cross section $\frac{d\sigma}{d\Omega}$
in centre of momentum frame, the above equation reduces to
another well known form as given in \cite{Hall:2009bx, Elahi:2014fsa},
\be
\frac{dY_{\x}}{dT}\simeq-\frac{1}{sHT}\,\,\dfrac{g_{\Phi}^2\,T}
{512\,\pi^6} \int^{\infty}_{4\,m^2_{\Phi}} d\hat{s} \int d\Omega\,|\vec{\mathfrak{p}}_f|\,|\vec{\mathfrak{p}}_i|
\dfrac{\overline{\left|\mathcal{M}\right|}^2_{\Phi^{\dagger}
\Phi \rightarrow \overline{\x}\x}}{\sqrt{\hat{s}}}\,{\rm K_1}
\left(\frac{\sqrt{\hat{s}}}{T}\right)\,.
\label{BERHSm2}
\ee  
Here, $\vec{\mathfrak{p}}_f$ is the final state momentum in centre of momentum frame.
So far, we have not used the information about the nature of interaction
between dark matter and mother particles into the derivation of Boltzmann equation. 
Thus Eqs.\,(\ref{BEfrinsc}-\ref{BERHSm2}) are equally applicable for both
types of freeze-in where dark matter candidate $\x$ is produced from
annihilation of $\Phi$ and $\Phi^{\dagger}$. However, for the case of UV
freeze-in one can further simplify the collision term of the Boltzmann
equation. As mentioned in the beginning, here we will consider a
dimension five operator $\dfrac{\Phi^{\dagger}\Phi\,\overline{\x}{\x}}{\Lambda}$.
The matrix amplitude square for dark matter production process
$\Phi^{\dagger} \Phi \rightarrow \overline{\x} \x$ is given by
\begin{eqnarray}
\overline{\left|\mathcal{M}\right|}^2_{\Phi^{\dagger}
\Phi \rightarrow \overline{\x}\x} = \,\dfrac{2}{\Lambda^2}
\left(\hat{s}-4\,m^2_{\x}\right)\,.
\end{eqnarray}
As UV freeze-in occurs at very early Universe when temperature
$T$ is much larger compared to the masses of associated
particles, in the following we consider both dark matter
as well as initial state particles to be massless. Now, using the expressions
of $\overline{\left|\mathcal{M}\right|}^2_{\Phi^{\dagger}
\Phi \rightarrow \overline{\x}\x}$, $|\vec{\mathfrak{p}}_i|$ and
$|\vec{\mathfrak{p}}_f|$ (in massless limit) in Eq.\,(\ref{BERHSm2})
we get
\bea
\frac{dY_{\x}}{dT}&\simeq&-\frac{1}{sHT}\,\dfrac{T}{256\pi^5\,\Lambda^2}
\int_{0}^{\infty} d\hat{s}\,\,\hat{s}^{3/2}\,{\rm K_1}
\left(\frac{\sqrt{\hat{s}}}{T}\right)\,,\nn \\
&\simeq& -\dfrac{1}{s\,H} \dfrac{T^5}{8\pi^5\Lambda^2}\,, \nn \\
&\simeq& - \dfrac{45 M_{pl}}{\,1.66\times 16\,\pi^7\sqrt{g_{\rho}}
\,g_s\,\Lambda^2}\,.
\label{YUV_general}
\eea 
In the last line we have used the expressions of Hubble parameter
$H=\dfrac{1.66\sqrt{g_{\rho}}}{M_{pl}}T^2$ and entropy density
$s=\dfrac{2\pi^2}{45}\,g_s T^3$ for radiation dominated era. Furthermore,
we assume that the production of $\x$ via UV freeze-in is effective
in a period when degrees of freedoms $g_{\rho}$ and $g_s$ remain
constant with $T$. This assumption is indeed true for our
scenario where UV freeze-in occurs much before EWSB. Consequently,
$\frac{dY_{\x}}{dT}$ becomes independent of temperature $T$ and
$Y_{\x}$ depends linearly on $T$ with maximum contribution
to $Y_{\x}$ coming at the maximum possible temperature
after reheating, which can be taken to be equal to
the reheat temperature $T_{\rm RH}$. Therefore,  
\bea
Y_{\x} \simeq \dfrac{360\,M_{pl}}{\,1.66\times (2\pi)^7\sqrt{g_{\rho}(T_{\rm RH})}
\,g_s(T_{\rm RH})\,\Lambda^2} T_{\rm RH} \,,
\label{YUV_simplified}
\eea 
with $M_{pl}=1.22\times 10^{19}$ GeV, the Planck mass.
\section{Boltzmann equation for Freeze-in via decay}
\label{app:BE-dec}
In this section, we have briefly discussed the Boltzmann
equation of a dark matter candidate $\x$ produced
from the decay of a mother particle $\phi$ ($\phi\rightarrow \overline{\x}{\x}$)
which is in thermal equilibrium and following Maxwell-Boltzmann distribution.
The Boltzmann equation for $\x$ in this case is given by,
\bea
\frac{dn_\x}{dt}+3Hn_\x\simeq\int d\Pi_{\phi} d\Pi_{\x} d\Pi_{\overline{\x}}
\,\overline{\left|\mathcal{M}\right|}^2_{\phi \rightarrow \overline{\x}\x}
(2\pi)^4 \delta^4(P_1-P_2-P_3) f^{\rm eq}_{\phi}(\vec{p}_1,T)\,,
\eea
where, similarly to the previous case, here also we denote
four momenta by $P_i$ and the corresponding three momenta
by $\vec{p}_i$. Moreover, we also neglect the effect of
inverse decay on $n_{\x}$ due to non-thermal nature of $\x$.
Using the standard definition of decay width for a two
body process like $\phi \rightarrow \overline{\x}\x$ and
also the definition of $n^{\rm eq}_{\phi}$ (Eq\,(\ref{neq})),
the right hand side of the Boltzmann equation can be expressed
as
\bea
\frac{dn_\x}{dt}+3Hn_\x &\simeq& 2\,m_{\phi}\,
\Gamma_{\phi \rightarrow \overline{\x}\x}\,\int d\Pi_{\phi}
f^{\rm eq}_{\phi}(\vec{p}_1,T)\,, \nn \\
&\simeq& \int \Gamma^{\prime}_{\phi \rightarrow \overline{\x}\x}\,f^{\rm eq}_{\phi}(\vec{p}_1,T)
\dfrac{g_{\phi}\,d^3\vec{p}_1}{(2\pi)^3}\,,\nn \\ 
& \simeq & n^{\rm eq}_{\phi}\,\langle \Gamma \rangle_{\phi \rightarrow \overline{\x}\x} 
\,,
\label{BE-decay}
\eea 
where $\Gamma_{\phi \rightarrow \overline{\x}\x}$ and
$\Gamma^{\prime}_{\phi \rightarrow \overline{\x}\x}$
are partial decay widths of $\phi$ into $\overline{\x} \x$
final state in the rest frame of $\phi$ and in a
frame where $\phi$ is moving with a four momentum $P_1\,(E_1,\,\vec{p}_1)$
respectively and $\Gamma^{\prime}_{\phi \rightarrow \overline{\x}\x} = \dfrac{m_{\phi}}{E_1}
\,\Gamma_{\phi \rightarrow \overline{\x}\x}$. The quantity
$\langle \Gamma \rangle_{\phi \rightarrow \overline{\x}\x}$ is thermal
averaged decay width and it is defined as
\be
\langle \Gamma \rangle_{\phi \rightarrow \overline{\x}\x}
=m_{\phi}\dfrac{\int d\Pi_{\phi}\,\Gamma_{\phi \rightarrow
\overline{\x}\x}\,f^{\rm eq}_{\phi}(\vec{p}_1,T)}
{\int d\Pi_{\phi}\,E_1\,f^{\rm eq}_{\phi}(\vec{p}_1,T)}\,,
\ee 
while the equilibrium number density $n^{\rm eq}_{\phi}$ is defined
in Eq.\,(\ref{neq}).
Now, in order to proceed
further we have to choose a distribution function for the mother
particle $\phi$. As $\phi$ is in thermal equilibrium, if we consider
the Maxwell-Boltzmann distribution i.e. $f^{\rm eq}_{\phi}(\vec{p}_1,T) = \exp(-E_1/T)$
then both $\langle \Gamma \rangle_{\phi \rightarrow \overline{\x}\x}$
and $n^{\rm eq}_{\phi}$ reduce to respective well known forms
i.e. $\langle \Gamma \rangle_{\phi \rightarrow \overline{\x}\x} =
\dfrac{{\rm K_1}\left(\frac{m_{\phi}}{T}\right)}{{\rm K_2}\left(\frac{m_{\phi}}{T}\right)}
\Gamma_{\phi \rightarrow \overline{\x}\x}$
and $n^{\rm eq}_{\phi} = \dfrac{T}{2\pi^2}\,g_{\phi}\,m^2_{\phi}\,
{\rm K_2}\left(\frac{m_{\phi}}{T}\right)$. Using these two expressions,
one can further simplify the collision term as
\bea
\frac{dn_\x}{dt}+3Hn_\x\simeq \dfrac{g_{\phi}\,m^2_{\phi}\,
\Gamma_{\phi \rightarrow \overline{\x}\x}\,T}{2\pi^2}\,
{\rm K_1}\left(\frac{m_{\phi}}{T}\right)\,.
\eea
The left hand side of the above equation can be written as a
temperature variation of comoving number density $Y_{\x}$
following the procedure discussed in the previous section.
Therefore, in terms of $Y_{\x}$, the Boltzmann equation
for freeze-in via decay has the following form
\bea
\frac{dY_{\x}}{dT}&\simeq&-\frac{1}{sHT}\,
\dfrac{g_{\phi}\,m^2_{\phi}\,
\Gamma_{\phi \rightarrow \overline{\x}\x}\,T}{2\pi^2}\,
{\rm K_1}\left(\frac{m_{\phi}}{T}\right)\,,
\eea 
and
\bea
Y_{\x} &\simeq& -\dfrac{g_{\phi}\,m^2_{\phi}\,
\Gamma_{\phi \rightarrow \overline{\x}\x}}{2\pi^2}
\int_{T_{max}}^{T_{min}}\frac{dT}{sH}\,
{\rm K_1}\left(\frac{m_{\phi}}{T}\right) \,, \nn \\
&\simeq& -\dfrac{45\,g_{\phi}\,m^2_{\phi}\,M_{pl}\,
\Gamma_{\phi \rightarrow \overline{\x}\x}}{1.66\times 4\pi^4}
\int_{T_{max}}^{T_{min}}dT\,
\dfrac{{\rm K_1}\left(\frac{m_{\phi}}{T}\right)}{T^5\,
\sqrt{g_{\rho}}\,g_s}\,. 
\eea 
Here also we have used the expressions of $H$
and $s$ for radiation dominated era. $T_{max}$
and $T_{min}$ are initial and final temperatures. As the integrand
has maxima around $T\simeq m_{\phi}$, considering $T_{max}=\infty$
and $T_{min}=0$, one can finally get the following expression
of $Y_{\x}$,
\bea
Y_{\x} \simeq  \dfrac{135\,M_{pl}}{1.66\times 8\pi^3}\dfrac{g_{\phi}\,
\Gamma_{\phi \rightarrow \overline{\x}\x}}{m^2_{\phi}\,
\sqrt{g_{\rho}(m_{\phi})}\,g_s(m_{\phi})}\,.
\eea
\,

\section{Calculation of X-ray flux:}
\label{X-ray_line}
\begin{figure}[h!]
\includegraphics[height=4cm,width=12cm]{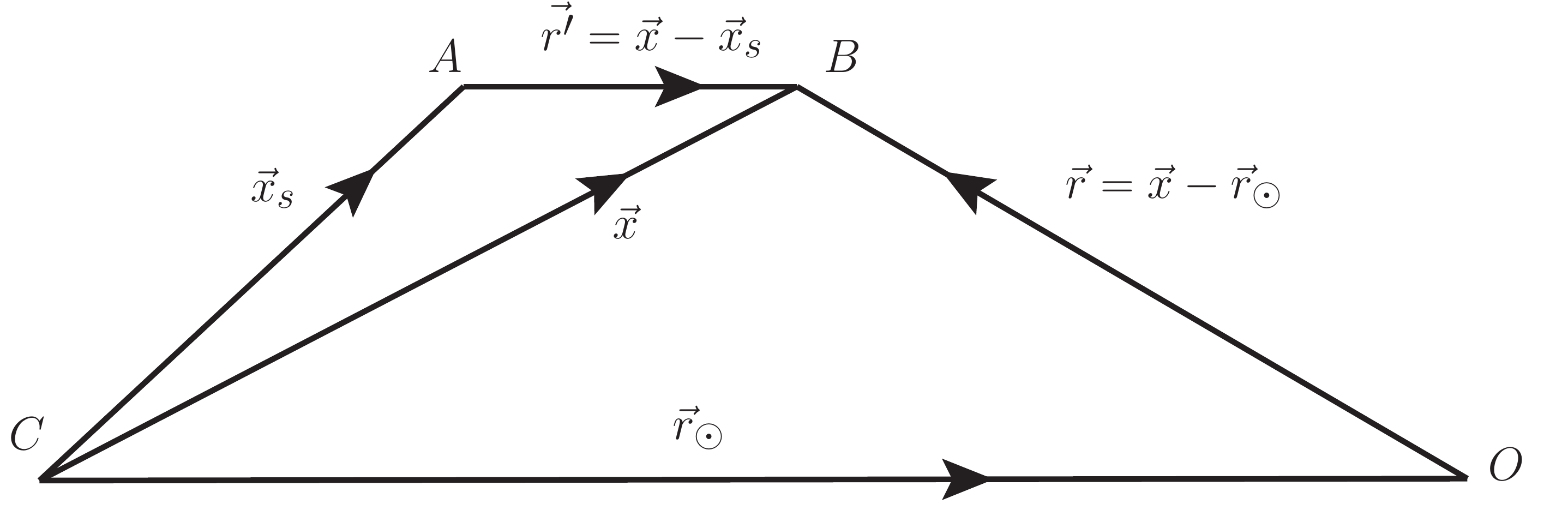}
\caption{Schematic diagram of dark matter annihilation
via a long lived mediator $\tf$.}
\label{Fig:flux_dia}
\end{figure}
Let us consider two dark matter particles annihilate into two $\tf$s and
each of them decays to two photons and further the corresponding decay of $\tf$ is not
instantaneous. This means $\tf$ travels a certain distance depending on this
lifetime before decaying into a pair of $\gamma$. This situation
has been illustrated in Fig.\,\ref{Fig:flux_dia}, where the point
$C$ represents the galactic centre. The annihilation of
dark matter occurs at $A$ while the decay of $\tf$
after travelling a distance $\vec{r^{\prime}}$ happens
at $B$. $O$ is the position of the earth with respect to the
galactic centre. Now, in such a situation, we want
to calculate the photon flux detected by a telescope
placed at $O$, which is produced at $B$ from the cascade
annihilations of $\x$. First, we derive the expression of $\gamma$-ray
flux for the present situation following \cite{Rothstein:2009pm, Chu:2017vao}
and then we discuss how it differs from the $\gamma$-ray
flux involving direct dark matter annihilation. 
As mentioned earlier, let two $\x$s annihilate at
$A$ to produce two $\tf$s and each of them decays
into two $\gamma$s at $B$ while the position vectors
of $A$ and $B$ with respect to the galactic centre $C$
are $\vec{x}_s$ and $\vec{x}$ respectively.
If total number density of dark matter at $\vec{x}_s$ is $n_{\chi}(\vec{x}_s)$ then
the annihilation rate per $\bar{\x}$ particle is given by,
\bea\label{eq12}
\Gamma_{ann}(\vec{x}_s)&=&\langle{\sigma {\rm v_{rel}}}
\rangle_{\bar{\x} \x\rightarrow\tf\tf}\,\,\dfrac{n_{\x}(\vec{x}_s)}{2}\nn\\
&=&\langle{\sigma {\rm v_{rel}}}
\rangle_{\bar{\x} \x \rightarrow\tf\tf}\,\,\dfrac{\rho_{\x}(\vec{x}_s)}{2m_{\x}}\,.
\eea
Here, the 2 factor arises because of the fact that
average number densities of $\x\,\,\text{and}\,\,\bar{\x}\,\,\text{are}\,\,\frac{n_{\x}}{2}$.
$\rho_{\x}(\vec{x}_s)$ is the dark matter density at
a distance $\vec{x}_s$ from the galactic centre while
$\sigmaVindirect$ is the annihilation cross section
for $\bar{\x} \x \rightarrow \tf \tf$ and it has the
following expression,
\bea
\sigmaVindirect = \dfrac{g^4}{16\,\pi\,m_{\x}}
\dfrac{\left(m^2_{\x}-m^2_{\tf}\right)^{3/2}}{\left(2\,m^2_{\x}
-m^2_{\tf}\right)^2} \,.
\label{sigmaVind}
\eea 
In the above we have assumed non-relativistic
nature of our dark matter particles during
their annihilations to $\tf$s. Now, an element
of volume $dV_{\vec{x}_s}$ at $A$ contains $\frac{\rho_{\x}(\vec{x}_s)}
{2\,m_{\x}} dV_{\vec{x}_s}$ number of $\bar{\x}$ particles. 
Hence the total annihilation rate\footnote{In case of Majorana fermion
the annihilation rate will be $$ \Gamma_{ann}^{total}(\vec{x}_s)=\dfrac{1}{2}\times\sigmaVindirect\, \left(\frac{\rho_{\x}(\vec{x}_s)}{\,m_{\x}}\right)^2 dV_{\vec{x}_s}\,\,.$$ 
The $\frac{1}{2}$ factor arises to avoid the overcounting since for
Majorana fermion particles and antiparticles are identical.} at
$\vec{x}_s$ for an element of volume $dV_{\vec{x}_s}$ is 
\begin{equation}\label{eq13}
\Gamma_{ann}^{total}(\vec{x}_s)=\sigmaVindirect\,\dfrac{\rho_{\x}(\vec{x}_s)}
{2m_{\x}} \times \frac{\rho_{\x}(\vec{x}_s)}{2\,m_{\x}} dV_{\vec{x}_s}\,.
\end{equation} 
 \\
Here from each annihilation of $\x$ and $\bar{\x}$ we get two pseudo scalars. Hence,
the production rate of $\tf$ in the elemental volume $dV_{\vec{x}_s}$ can be written
as
\bea
\Gamma^{\prime}_{\tf}(\vec{x}_s)&=&2\times\Gamma_{ann}^{total}(\vec{x}_s)\nn \\
&=& \dfrac{\sigmaVindirect}{2}\,\Bigg(\frac{\rho_{\x}(\vec{x}_s)}
{m_{\x}}\Bigg)^2 dV_{\vec{x}_s}\,, \nn \\
&=& \Gamma_{\tf}(\vec{x}_s)\,dV_{\vec{x}_s}\,,
\label{production_rate_phi}
\eea
where, $\Gamma_{\tf}(\vec{x}_s)$ in the right hand
side denotes the production rate of $\tf$ per
unit volume at $\vec{x}_s$.
Now, the probability that a particle $\tf$ will travel a
distance $r^{\prime}$ from its source point at $\vec{x}_s$ without decaying
is given by,
\begin{equation}\label{eq15}
P(r^{\prime})=e^{-\frac{r^\prime}{\lambda_{\tf}}}\,,
\end{equation}
where, $\lambda_{\tf}$ is the decay length of $\tf$, which is
the average distance travelled by a particle before decaying
and $\lambda_{\tf}$ can be expressed as,
\bea
\lambda_{\tf} &=& {\rm v}_{\tf}\, \tau_{\tf}\,, \nn \\ 
&=& {\rm v}_{\tf}\, \gamma_{\tf}\,\tau^0_{\tf}\,, \nn \\ 
&=& \dfrac{\sqrt{m^2_{\x}-m^2_{\tf}}}{m_{\tf}}\,\tau^0_{\tf}\,,\nn \\
&\simeq& \sqrt{0.2}\times 10^{7} \left(\dfrac{\delta}{10^{-5}}\right)^{1/2}
\,\left(\dfrac{\tau^0_{\tf}}{10^{20}\,{\rm s}}\right)\,\,{\rm kpc}\,,
\label{decay_length}
\eea  
where, ${\rm v}_{\tf}$ is the velocity of $\tf$ in the
laboratory frame, $\gamma_{\tf} \simeq \frac{m_{\x}}{m_{\tf}}$
(considering non-relativistic $\x$, i.e. $E_{\x}\simeq m_{\x}$).
Here, $\tau^0_{\tf}=1/\Gamma_{\tf\rightarrow \gamma\gamma}$
is the lifetime of $\tf$ at rest and the corresponding decay width is given by
\begin{equation}
{\Gamma_{\tf\rightarrow \gamma\gamma}\,=\, \frac{m_{\tf}^3}{16\pi\Lambda^2}} \,\,\,.
\label{GFF}
\end{equation}
The quantity $\delta$ is already
defined in the previous section, expressing the mass
splitting between $\x$ and $\tf$. 
Now to calculate how many $\tf$ will reach at $B$ starting from $A$, we need to
know the phase space distribution of $\tf$.

Number of $\tf$ which are produced within an elementary volume $dV_{\vec{x}_s}$ at $\vec{x}_s$ will
cross an elementary area $dA^{\prime}$ at $\vec{x}$ in time $dt$ is given by 
\bea
dN_{\tf}(\vec{x},\vec{l})=dV_{\vec{x}_s}
\Bigg(\Gamma_{\tf}(\vec{x}_s)dt\Bigg)
\Bigg(\frac{dA^{\prime}}{4\pi |\vec{x}-\vec{x}_s|^2}\Bigg)
\exp\left[-\frac{|\vec{x}-\vec{x}_s|}{\lambda_{\tf}}\right]\,,
\eea
where, $\vec{l}$ represents the direction of propagation
of $\tf$ which is along $\vec{r^\prime} = \vec{x}-\vec{x}_s$. Moreover,
without loss of any generality we have considered both $\vec{l}$
and $\vec{dA^{\prime}}$ are along the same direction. The
differential flux $d\Phi_{\tf}$ at $\vec{x}$ is defined as
\bea
d\Phi_{\tf} (\vec{x},\vec{l}) =
\dfrac{d^2N_{\tf}(\vec{x},\vec{l})}{dA^{\prime}\,dt} \,,
\label{flux-def}
\eea
and, therefore, the flux of $\tf$ at $\vec{x}$ is given by,
\bea
\Phi_{\tf}(\vec{x}) = 
\dfrac{1}{4\pi}\int dV_{\vec{x}_s}
\,{\Gamma_{\tf}(\vec{x}_s)} 
\dfrac{\exp\left[-\frac{|\vec{x}-\vec{x}_s|}{\lambda_{\tf}}\right]}{|\vec{x}-\vec{x}_s|^2}\,.
\eea
Finally, we have defined a quantity $f_{\tf}(\vec{x})$,
called the density function of $\tf$
at $\vec{x}$ as
\bea
f_{\tf}(\vec{x}) &=& 
\dfrac{\Phi_{\tf}(\vec{x})}{{\rm v}_{\tf}}\,,\nn \\
&=& \dfrac{1}{4\pi\,{\rm v}_{\tf}}\int dV_{\vec{x}_s}
\,{\Gamma_{\tf}(\vec{x}_s)} 
\dfrac{\exp\left[-\frac{|\vec{x}-\vec{x}_s|}{\lambda_{\tf}}\right]}{|\vec{x}-\vec{x}_s|^2}\,.
\label{disfunc_phi}
\eea

The angular distribution of photon is isotropic in the rest frame of
$\tf$ (since the scalar is a spin zero object). However, in the galactic frame the emitted
photons have some angular distribution. Let $\vec{k}$ be the direction of emitted
photons (i.e.\,\,along the position vector $-\vec{r}$
in Fig.\,\ref{Fig:flux_dia}) then the angular distribution of photons will depend on the angle
between $\vec{l}$ and $\vec{k}$. If $dN_{\gamma}$ is the number of photons emitted
along the direction $\vec{k}$ within a solid angle $d\Omega_{\gamma}$ and
having energy between $E_{\gamma}$ and $E_{\gamma}+dE_{\gamma}$, then
let us define a function $\mathcal{G}(\vec{k}-\vec{l},E_{\gamma})$ as
\begin{equation}
\mathcal{G}(\vec{k}-\vec{l},E_{\gamma}) =
\frac{d^2N_\gamma}{dE_{\gamma }\,d\Omega_{\gamma}}\,.
\label{gfunction}
\end{equation}
Therefore, the number of photons coming towards the earth
in time $dt$ along the direction $\vec{k}$ from an elementary
volume $dV$ situated at
${\vec{x}}$
with respect to the galactic centre is 
\begin{equation}\label{eq20}
dN_{\gamma}=dV\Bigg(\frac{dt}{\tau_{\tf}} f_{\tf}
(\vec{x}) \Bigg)\Bigg(\mathcal{G}(\vec{k}-\vec{l},
E_{\gamma})\,dE_{\gamma}\,\frac{dA}{|\vec{x}-\vec{r}_\odot|^2}\Bigg)\,,
\end{equation}
where $dA$ is an elementary area placed on the earth (i.e.\,at
$O$, the position of telescope) and similar to the previous
case, here also we have considered that the photons are coming
along the normal to the area $dA$. Furthermore, from Fig.\,\ref{Fig:flux_dia} we can write 
$\vec{r}=\vec{x}-\vec{r}_\odot$ where $\vec{r}_\odot$ describes the
position of the earth with respect to the galactic centre.
Now, using
the definition of differential flux given in Eq.\,(\ref{flux-def}),
one can easily calculate the total photon flux from the decay of $\tf$ as
\bea
\Phi_{\gamma}=\int  \,\dfrac{f_{\tf}(\vec{x})}{\tau_{\tf}}
\mathcal{G}(\vec{k}-\vec{l},
E_{\gamma})\,\frac{dE_{\gamma}}{|\vec{x}-\vec{r}_\odot|^2}\, dV.
\eea
The differential photon flux on the earth can be obtained using $\vec{r}=\vec{x}-\vec{r}_\odot$ as 
\begin{equation}\label{eq21}
\frac{d^2\Phi_{\gamma}}{dE_{\gamma}\,d\Omega}=
\frac{1}{\tau_{\tf}}\int_{l.o.s}dr\,f_{\tf}(\vec{x})\,
\mathcal{G}(\vec{k}-\vec{l},E_{\gamma})\,,
\end{equation}
where $\int d\Omega=\Delta\Omega$ is the field of view of the telescope placed at O.
The above integration over $r$ is along the line of sight ($l.o.s$) distance.
Now substituting the expression of $f_{\tf}(\vec{x})$ in
the above we get,
\begin{equation}\label{eq22}
\frac{d^2\Phi_{\gamma}}{dE_{\gamma}\,d\Omega}
=\frac{1}{4\pi {\rm v}_{\tf}
\tau_{\tf}}\int_{l.o.s}dr
\int dV_{\vec{x}_s}
{\Gamma_{\tf}(\vec{x}_s)} \dfrac{\exp\left[-\frac{|\vec{x}-\vec{x}_s|}{\lambda_{\tf}}\right]}{|\vec{x}-\vec{x}_s|^2}
\,\mathcal{G}(\vec{k}-\vec{l},E_{\gamma})\,.
\end{equation}
Finally, using Eqs.\,(\ref{production_rate_phi},
\ref{decay_length}) the expression
of photon flux is 
\bea
\frac{d^2\Phi_{\gamma}}{dE_{\gamma}\,d\Omega}
&=&\frac{1}{4\pi\lambda_{\tf}}\int_{l.o.s}
dr \int dV_{\vec{x}_s}
\dfrac{\sigmaVindirect}{2}\,\Bigg(\frac{\rho_{\x}(\vec{x}_s)}
{m_{\x}}\Bigg)^2 \dfrac{\exp\left[-\frac{|\vec{x}-\vec{x}_s|}{\lambda_{\tf}}\right]}
{|\vec{x}-\vec{x}_s|^2}\,\mathcal{G}(\vec{k}-\vec{l},E_{\gamma})\,,\nn \\
&=& \dfrac{\sigmaVindirect}{2\,m^2_{\x}}
\int_{l.o.s}
dr\,\rho^2_{\rm eff}(x)\,
\mathcal{G}(\vec{k}-\vec{l},E_{\gamma})\,,
\eea
where 
\bea
\rho^2_{\rm eff}(x) = \int dV_{\vec{x}_s}
\frac{\rho^2_{\x}(\vec{x}_s)}{4\pi\lambda_{\tf}}
\dfrac{\exp\left[-\frac{|\vec{x}-\vec{x}_s|}{\lambda_{\tf}}\right]}
{|\vec{x}-\vec{x}_s|^2}\,,
\label{rhoeff}
\eea
is the square of effective dark matter density
at point B. In the present case,
dark matter candidate $\x$ is non-relativistic
and the mass of $\x$ and $\tf$ are required to be
almost degenerate to generate a {\it line}-like
photon spectrum. Consequently, the intermediate particles
$\tf$ are also non-relativistic and hence
the photon spectrum will appear isotropic
to the observer at the earth as well. Therefore,
one can take $\mathcal{G}(\vec{k}-\vec{l},E_{\gamma})
\simeq\dfrac{1}{4\pi}\dfrac{dN_{\gamma}}{dE_{\gamma}}$
and the differential photon flux from the cascade
annihilation of dark matter reduces to the familiar
form 
\bea
\frac{d\Phi_{\gamma}}{dE_{\gamma}}
&=&2\times \dfrac{1}{4}\,\dfrac{r_{\odot}}{4\pi}
\left(\dfrac{\rho_{\odot}}{m_{\x}}\right)^2
\sigmaVindirect\,\dfrac{dN_{\gamma}}{dE_{\gamma}}\,
J_{\rm eff}\Delta{\Omega}\,.
\eea
\bibliographystyle{jhep}
\bibliography{UV-IR-freezein}
\end{document}